\documentclass[iop]{emulateapj} 

\usepackage[caption=false]{subfig} 
\usepackage{rotating}
\usepackage[usenames,dvipsnames]{xcolor}
\usepackage{color}

\newcommand{\lya}{Ly$\alpha$}
\newcommand{\halpha}{H$\alpha$}
\newcommand{\hbeta}{H$\beta$}
\newcommand{\hi}{H{\sc i}}

\newcommand{\lstar}{$L^\star$}
\newcommand{\mstar}{$M^\star$}

\newcommand{\llya}{$L_{\mathrm{Ly}\alpha}$}

\newcommand{\lfuv}{$L_\mathrm{FUV}$}
\newcommand{\ewlya}{$W_{\mathrm{Ly}\alpha}$}
\newcommand{\ewha}{$W_{\mathrm{H}\alpha}$}

\newcommand{\ergsec}{erg~s$^{-1}$}
\newcommand{\ergsecaa}{erg~s$^{-1}$~\AA$^{-1}$}
\newcommand{\ergseckpc}{erg~s$^{-1}$~kpc$^{-2}$}
\newcommand{\ergsecaakpc}{erg~s$^{-1}$~\AA$^{-1}$~kpc$^{-2}$}
\newcommand{\ergseccmarcsec}{erg~s$^{-1}$~cm$^{-2}$~arcsec$^{-2}$}

\newcommand{\ergseccm}{erg~s$^{-1}$~cm$^{-2}$}
\newcommand{\msun}{M$_\odot$}
\newcommand{\msunyr}{M$_\odot$~yr$^{-1}$}
\newcommand{\fesclya}{$f_\mathrm{esc}^{\mathrm{Ly}\alpha}$}
\newcommand{\ebv}{$E_{B-V}$}
\newcommand{\ebvneb}{$E_{B-V}^\mathrm{neb}$}
\newcommand{\ebvstel}{$E_{B-V}^\mathrm{stel}$}
\newcommand{\kms}{km~s$^{-1}$}

\newcommand{\oII}{[O{\sc ii}]}
\newcommand{\oIII}{[O{\sc iii}]}
\newcommand{\oI}{[O{\sc i}]}
\newcommand{\sII}{[S{\sc ii}]}

\newcommand{\nII}{[N{\sc ii}]}

\slugcomment{Submitted to ApJ, accounted for first referee's comments}

\shorttitle{LARS II: HST Imaging Results}
\shortauthors{M. Hayes et al.}

\begin{document}

\title{the Lyman alpha Reference Sample: II. HST imaging results,
integrated properties and trends\altaffilmark{1}}

\author{Matthew Hayes\altaffilmark{2,3,4},
G\"oran \"Ostlin\altaffilmark{4}, 
Florent Duval\altaffilmark{4}, 
Andreas Sandberg\altaffilmark{4}, 
Lucia Guaita\altaffilmark{4}, 
Jens Melinder\altaffilmark{4}, 
Angela Adamo\altaffilmark{5}, 
Daniel Schaerer\altaffilmark{6,2}, 
Anne Verhamme\altaffilmark{6}, 
Ivana Orlitov{\'a}\altaffilmark{6,7}, 
J. Miguel Mas-Hesse\altaffilmark{8}, 
John M. Cannon\altaffilmark{9}, 
Hakim Atek\altaffilmark{10}, 
Daniel Kunth\altaffilmark{11}, 
Peter Laursen\altaffilmark{12}, 
H\'ector Ot\'i-Floranes\altaffilmark{13,8,14}, 
Stephen Pardy\altaffilmark{9}, 
Th{\o}ger Rivera-Thorsen\altaffilmark{4},
E. Christian Herenz\altaffilmark{15} 
}

\email{matthew@astro.su.se}
\altaffiltext{1}{Based on observations made with the NASA/ESA Hubble 
Space Telescope, obtained at the Space Telescope Science Institute, 
which is operated by the Association of Universities for Research in 
Astronomy, Inc., under NASA contract NAS 5-26555. These observations are
associated with program \#12310.}
\altaffiltext{2}{Universit\'e de Toulouse; UPS-OMP; IRAP; Toulouse, France}
\altaffiltext{3}{CNRS; IRAP; 14, avenue Edouard Belin, F-31400 Toulouse, France}
\altaffiltext{4}{Department of Astronomy, Oskar Klein Centre, Stockholm University, AlbaNova University Centre, SE-106 91 Stockholm, Sweden}
\altaffiltext{5}{Max Planck Institute for Astronomy, K\"onigstuhl 17, D-69117 Heidelberg, Germany}
\altaffiltext{6}{Geneva Observatory, University of Geneva, 51 Chemin des Maillettes, CH-1290 Versoix, Switzerland}
\altaffiltext{7}{Astronomical Institute, Academy of Sciences of the Czech Republic, Bo\v cn{\'\i} II, CZ-14131 Prague, Czech Republic}
\altaffiltext{8}{Centro de Astrobiolog\'ia (CSIC--INTA), Departamento de Astrof\'isica, POB 78, E–28691 Villanueva de la Ca\~nada, Spain}
\altaffiltext{9}{Department of Physics and Astronomy, Macalester College, 1600 Grand Avenue, Saint Paul, MN 55105, USA}
\altaffiltext{10}{Laboratoire d’Astrophysique, \'Ecole Polytechnique F\'ed\'erale de Lausanne (EPFL), Observatoire, CH-1290 Sauverny, Switzerland.}
\altaffiltext{11}{Institut d'Astrophysique de Paris, UMR 7095 CNRS \& UPMC, 98 bis Bd Arago, 75014 Paris, France.}
\altaffiltext{12}{Dark Cosmology Centre, Niels Bohr Institute, University of Copenhagen, Juliane Maries Vej 30, 2100 Copenhagen, Denmark}
\altaffiltext{13}{Instituto de Astronom\'ia, Universidad Nacional Aut\'onoma de M\'exico, Apdo. Postal 106, Ensenada B. C. 22800 Mexico}
\altaffiltext{14}{Dpto. de F\'isica Moderna, Facultad de Ciencias, Universidad de Cantabria, 39005 Santander, Spain}
\altaffiltext{15}{Leibniz-Institut f\"ur Astrophysik (AIP), An der Sternwarte 16, D-14482 Potsdam, Germany.}

\begin{abstract}
We report upon new results regarding the \lya\ output of galaxies, 
derived from the \emph{Lyman alpha Reference Sample} (LARS), focusing 
on Hubble Space Telescope imaging. For 14 galaxies we present intensity 
images in \lya, \halpha, and UV, and maps of \halpha/\hbeta, \lya\ 
equivalent width (EW), and \lya/\halpha. We present \lya\ and UV 
light profiles and show they are well-fitted by S\'ersic profiles, but
\lya\ profiles show indices systematically lower than those of the UV 
($n\approx 1-2$ instead of $\gtrsim 4$). This reveals a general lack of 
the central concentration in \lya\ that is ubiquitous in the UV. 
Photometric growth curves increase more slowly for \lya\ than the FUV, 
showing that small apertures may underestimate the EW.  For most 
galaxies, however, flux and EW curves flatten by radii $\approx 10$~kpc,
suggesting that if placed at high-$z$ , only a few of our galaxies 
would suffer from large flux losses. We compute global properties 
of the sample in large apertures, and show total luminosities to 
be independent of all other quantities. Normalized \lya\ throughput, 
however, shows significant correlations: escape is found to be higher 
in galaxies of lower star formation rate, dust content, mass, and several 
quantities that suggest harder ionizing continuum and lower metallicity.
Eight galaxies could be selected as high-$z$ \lya\ emitters, based upon 
their luminosity and EW. We discuss the results in the context 
of high-$z$ \lya\ and UV samples. A few galaxies have EWs above
50~\AA, and one shows \fesclya\ of 80\%; such objects have not 
previously been reported at low-$z$.
\end{abstract}

\keywords{cosmology: observations --- galaxies: evolution --- galaxies: formation --- galaxies: starburst --- radiative transfer}

\section{Introduction}

The \emph{Lyman-alpha Reference Sample} (LARS; \"Ostlin et al, 2013, 
hereafter paper I) is a sample of 14 nearby star-forming galaxies, 
selected for observation with the Hubble Space Telescope and an array
of other telescopes. The primary goals are to return 
detailed observations -- both in images and spectra -- of the \hi\ 
\lya\ emission line, and to do so in a sample that is simultaneously 
as free from bias as possible, statistically meaningful enough to 
observe trends within the sample, and comparable in selection to 
galaxies observed in the high-$z$ universe. With as much 
information about the emitted \lya\ as possible, the data-set is then
completed by amassing as much information on the stars, and various 
phases of the interstellar medium as possible; from these we measure
and map the properties of the galaxies, their intrinsic \lya, and the 
properties of the medium through which the \lya\ must travel in order 
to escape and reach the observer.

The astrophysical motivation for LARS is straightforward: \lya\ is 
intrinsically the most luminous feature in the spectrum of a hot source 
\citep{Schaerer2003,Raiter2010}, and was recognized almost half a century ago as a
potential beacon through which to study the distant universe 
\citep{Partridge1967}. However while intrinsically very luminous, the 
fact that it scatters wherever it encounters \hi\ gas means that it 
may encode not only information about the nebulae in which it was 
produced, but also the interstellar, circumgalactic, and intergalactic
media through which it must have transferred. This radiative transport 
modifies the spectral distributions of the \lya\ radiation,
typically resulting in the observed asymmetric line profiles 
\citep{Kunth1998,Rhoads2003,Shapley2003,Shimasaku2006,Tapken2007,Vanzella2010,Lidman2012,Wofford2013}
or double peaks \citep{Shapley2006,Quider2009,Heckman2011,Kulas2012,Yamada2012}, 
and spatially extended \lya\ emission \citep{Fynbo2001,Steidel2003,Steidel2011,Hayes2005,Hayes2007,Ostlin2009,Matsuda2012}. 

The \lya\ output from galaxies has been shown to evolve strongly with 
redshift \citep{Hayes2011evol,Stark2011,Schaerer2011frac,Blanc2011,Dijkstra2013} in a 
fashion that, in contrast to the star formation rate (SFR) density 
\citep[e.g.][]{Hopkins2006}, is completely monotonic. Naturally this phenomenon 
must reflect a fundamental change in the galaxy population, but 
currently the \lya\ emissivity is little more than an indicator. If we 
are to use \lya\ to study the high-$z$ galaxy population, and for 
example as a probe of cosmic reionization 
\citep[e.g.]{Haiman1999,Santos2004,Malhotra2004,Kashikawa2006,Dijkstra2007},
we need to understand under what 
conditions galaxies become strong \lya\ emitters. In doing so, we must
simultaneously try to account for aperture losses that may result from 
\lya\ scattering, as low surface brightness haloes may, or may not 
\citep{Feldmeier2013} be ubiquitous. This can only be achieved by
assembling a data-set that comprises a sufficiently large sample of 
galaxies, and observations that measure all the quantities thought to
be complicit in the \lya\ transport process. This second requirement 
also sets a limit on redshift: it must be low enough to enable us to 
study galaxies in detail, and probe the scattering medium directly.
It was precisely to this end that the LARS survey was designed  --
Paper I gives a complete overview of the survey, motivations, and 
sample selection.

The backbone of LARS is a HST ultraviolet (UV) and optical imaging 
campaign, carried out with the \emph{Advanced Camera for Surveys} (ACS)
and \emph{Wide Field Camera 3} (WFC3). The observational parameters of 
the survey are described in Paper I. This enables us to map the \lya\ emission
and absorption on spatial scales as low as 30~pc, and
also determine the properties of the stellar population and nebular gas, 
including the dust attenuation in both phases on the same scales.
Additional campaigns include HST UV spectroscopy in \lya\ and the 
nearby continuum, 21~cm line observations of the \hi, 3D spectroscopy 
and narrowband imaging in more nebular emission lines, 
and all of these data-sets will provide the subject of forthcoming 
papers. 

In this paper we present the first set of `global' results that can
be obtained from the HST imaging observations. In \citet{Hayes2013} we 
showed the first images in \lya, \halpha, and the UV continuum, and 
quantified the extent of \lya\ emission and compared it with the other 
hot stellar and nebular radiation. In this paper we take that analysis 
a lot further and present a library of individual intensity images, 
maps of line ratios and equivalent widths, and radial light-profiles 
of the same types of radiation. We also present aperture growth curves
to examine the recovered quantities as a function of aperture size, 
and global properties of the sample, computed in homogeneously defined 
apertures, in order to examine global trends in the \lya\ throughput.
Generally, Sections~\ref{sect:data}--\ref{sect:globals} describe the 
data, while  Sections~\ref{sect:individual}--\ref{sect:corr} discuss 
its astrophysical meaning. 

In Section~\ref{sect:data} we present a brief overview of the imaging
data.
In Section \ref{sect:images} we present images in \lya\ and other 
relevant wavelengths (\halpha\ and FUV continuum) and discuss line 
ratios and equivalent width maps. 
In Section~\ref{sect:spatdist} we show the spatial distribution of 
\lya\ and related observables, both as radial surface brightnesses and 
the integrated quantities in collapsed apertures. 
In Section~\ref{sect:globals} we measure some global properties, and 
show how the \lya\ output of the sample may depend upon various 
inferable stellar and nebular quantities, and 
in Section~\ref{sect:individual} we point out interesting and relevant 
features in the individual targets, and discuss their properties. 
Section~\ref{sect:select} describes the `analog' nature of the sample 
and discusses how LARS galaxies would conform to high-$z$ UV and \lya\
selection functions and be detected/selected in survey data. 
In Section~\ref{sect:corr} we discuss how various properties affect the
\lya\ output, and under what conditions galaxies may emit their \lya. 
Finally in Section~\ref{sect:conc} we present a summary of  our 
findings.  Throughout we assume a cosmology of 
$(H_0, \Omega_\mathrm{M}, \Omega_\Lambda) = 
(70~\mathrm{km~s}^{-1}~\mathrm{Mpc}^{-1}, 0.3, 0.7)$.

\section{Overview of the Data}\label{sect:data}

A detailed description of the data and its reduction, binning, and 
processing can be found in Paper I, and here we present 
only a brief overview. All 14 galaxies were imaged with HST in three 
configurations to isolate the emission lines: optical narrowband filters
for \halpha\ and \hbeta\ with either the Wide Field Camera 3 (WFC3) or 
Advanced Camera for Surveys (ACS), and a UV long-pass pair-subtraction 
method to isolate a clean well-defined intermediate bandpass that 
captures \lya\ using the Solar Blind Channel (SBC) of ACS (see Paper I
and \citealt{Hayes2009}).  Further to
this we obtained FUV continuum imaging (SBC) and optical broadband 
imaging in approximately the $U-$, $B-$, and $I-$bands (ACS and/or 
WFC3); exactly which filter depends on redshift (Paper I). Imaging data 
were reduced with their standard HST {\tt CALACS} and {\tt CALWF3} 
pipelines, and drizzled onto a common pixel scale of 0.04 arcsec using 
the {\tt MULTIDRIZZLE} task\footnote{Fruchter, A. and Sosey, M. et al. 
2009, "The MultiDrizzle Handbook", version 3.0, (Baltimore, STScI)}. 
Finally reduced images were matched in point spread function (PSF) 
using the image with the broadest PSF as reference.  ACS/SBC 
photometric zero-points were corrected following the method described 
in Paper I.

Science frames were adaptively binned using a Voronoi 
tessellation algorithm and the weight maps output from 
{\tt MULTIDRIZZLE}. 
In this article we present results derived from both binned 
and un-binned frames, depending on the application. For this we used 
the WVT binning algorithm by \citet{Diehl2006}, which is a 
generalization of the Voronoi binning algorithm of 
\citet{Cappellari2003}. For the reference bandpass we use the FUV 
continuum image and  require a threshold  signal-to-noise ratio of 10 per 
bin, but do not permit bin sizes to exceed 1 square arcsecond (625
pixels at the native scale of WFC3/UVIS). 

\begin{figure*}[t!]
\centering 
\includegraphics[scale=0.45, clip=true, trim=02mm 11mm 6mm 1mm]{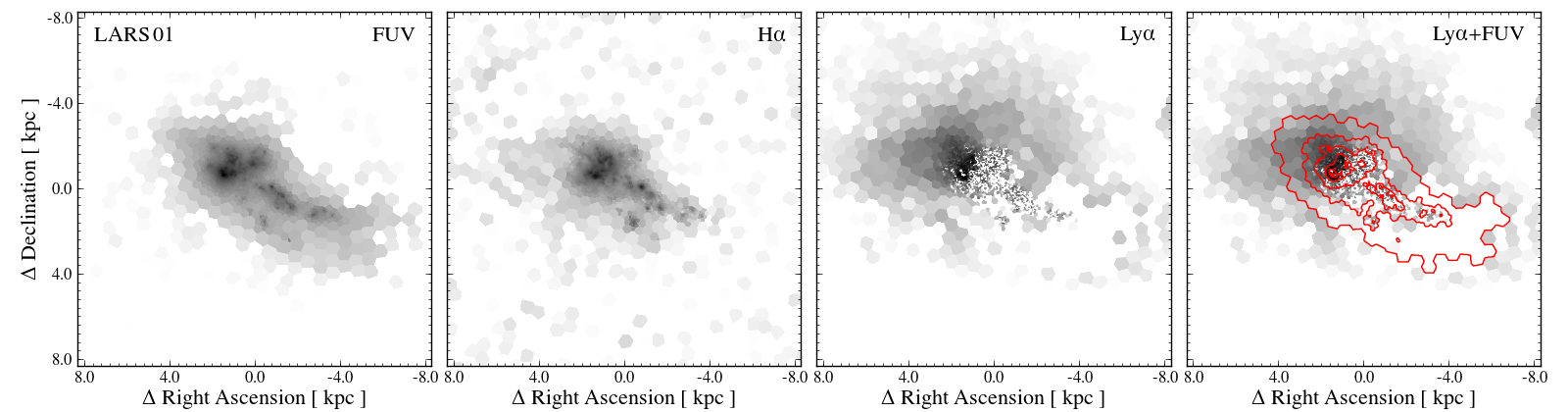}
\includegraphics[scale=0.45, clip=true, trim=02mm 11mm 6mm 1mm]{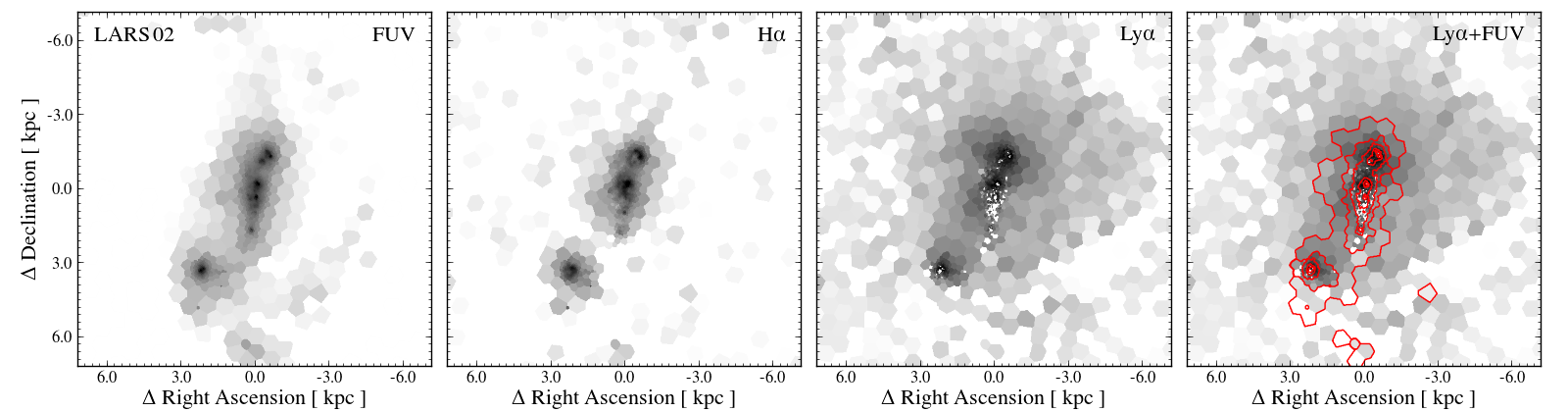}
\includegraphics[scale=0.45, clip=true, trim=02mm 11mm 6mm 1mm]{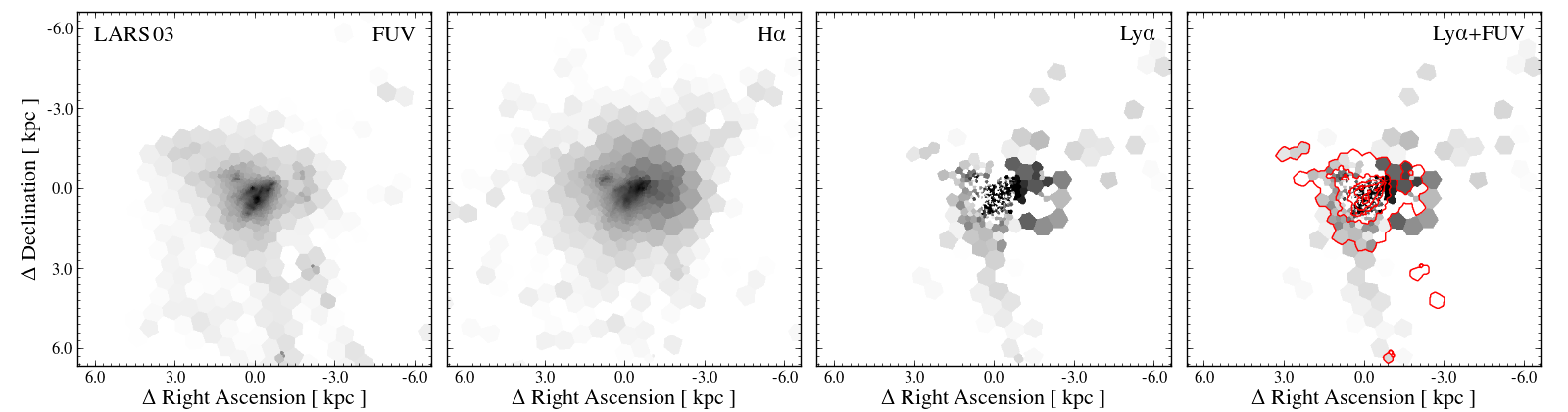}
\includegraphics[scale=0.45, clip=true, trim=02mm 11mm 6mm 1mm]{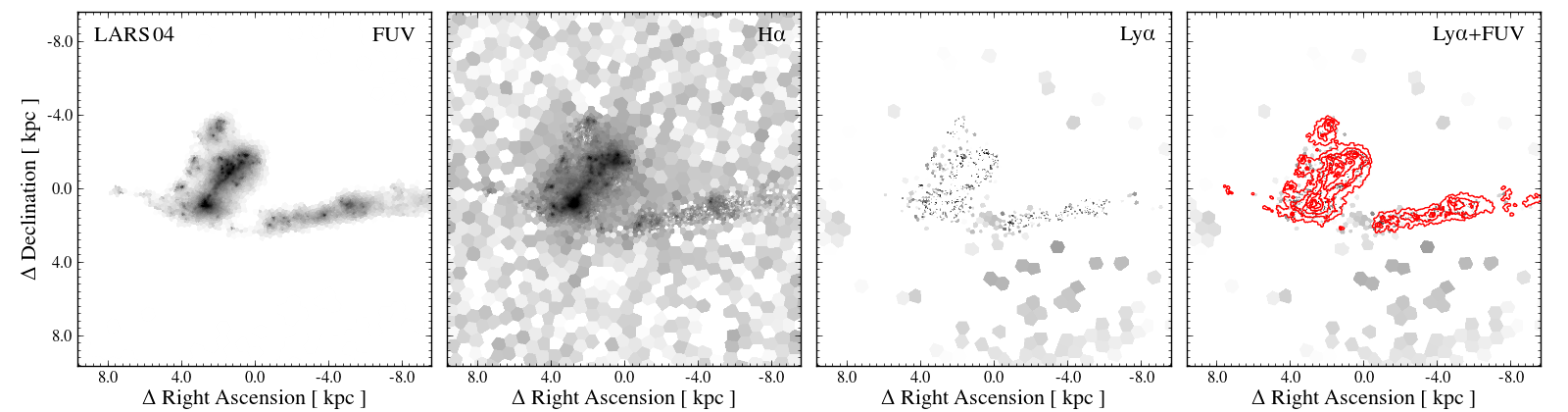}
\includegraphics[scale=0.45, clip=true, trim=02mm 01mm 6mm 1mm]{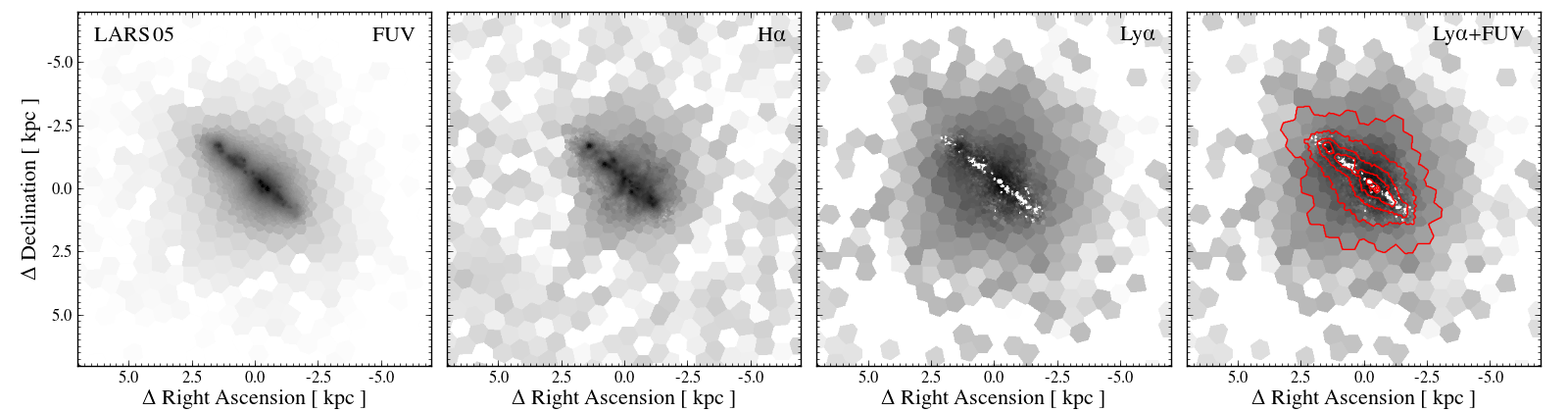}

\caption{HST imaging of the LARS galaxies. \emph{Left most} panels show 
the FUV continuum at $\lambda \approx 1500$\AA, which traces massive 
star-forming regions that are not obscured by dust. \emph{Center left} 
panels show continuum-subtracted \halpha, which traces the nebulae
that result from star formation. The \emph{Center Right} panels show 
continuum-subtracted \lya\ maps, and the \emph{Right most} images are 
the same but with FUV contours overlaid in red.  All images have been 
adaptively binned using Voronoi tessellation, and are shown in a 
negative logarithmic intensity scaling in which the cut levels are set 
to show detail.  \lya\ frames are continuum-subtracted, and hence these
frames in particular can contain negative flux; in logarithmic scaling 
this is undefined so \lya\ absorbtion along the line-of-sight is often
seen as white speckles.  Each cutout is scaled to show the overall 
galaxy morphology, so each object is shown on its own physical scale; 
the size in kpc is labeled for each object.  North is up and east to 
the left.  \emph{Top} to \emph{Bottom} shows galaxies LARS\,01, 02, 03, 
04, and 05.}
\label{fig:maps}
\end{figure*}

\begin{figure*}[t!]
\ContinuedFloat

\centering 
\includegraphics[scale=0.45, clip=true, trim=02mm 11mm 6mm 1mm]{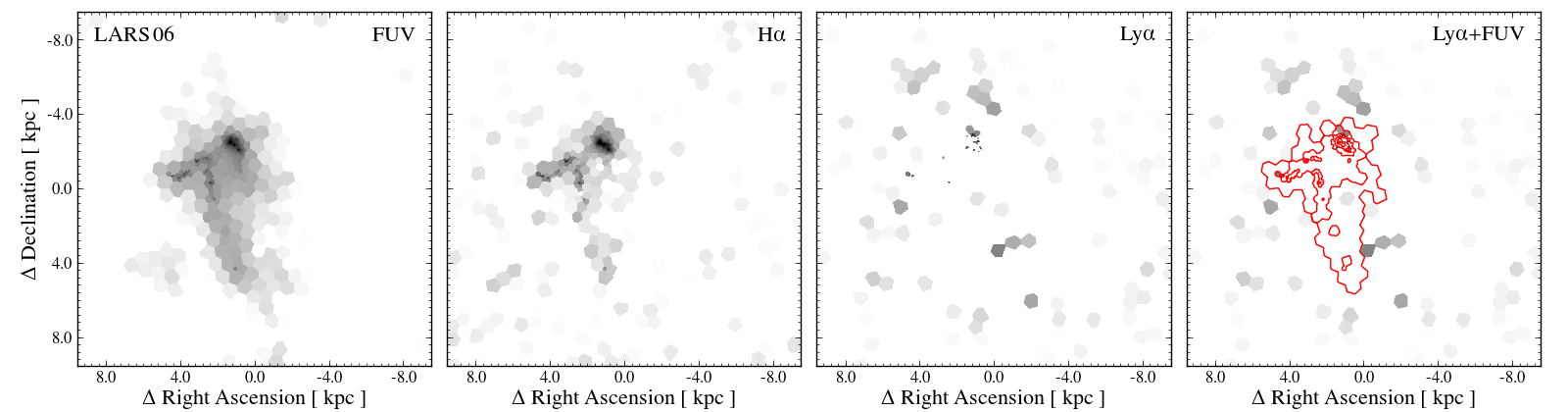}
\includegraphics[scale=0.45, clip=true, trim=02mm 11mm 6mm 1mm]{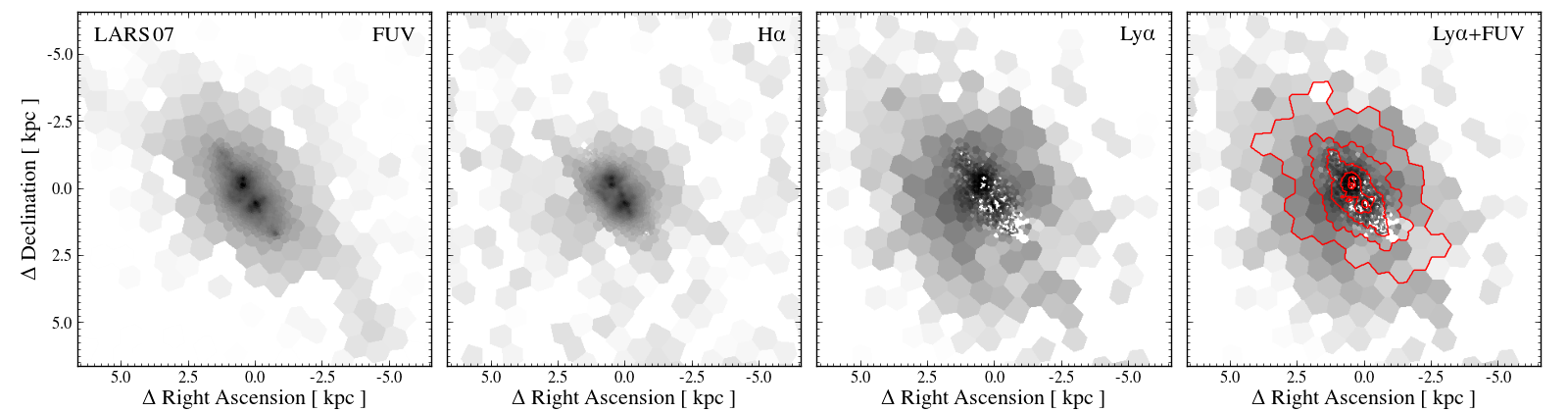}
\includegraphics[scale=0.45, clip=true, trim=02mm 11mm 6mm 1mm]{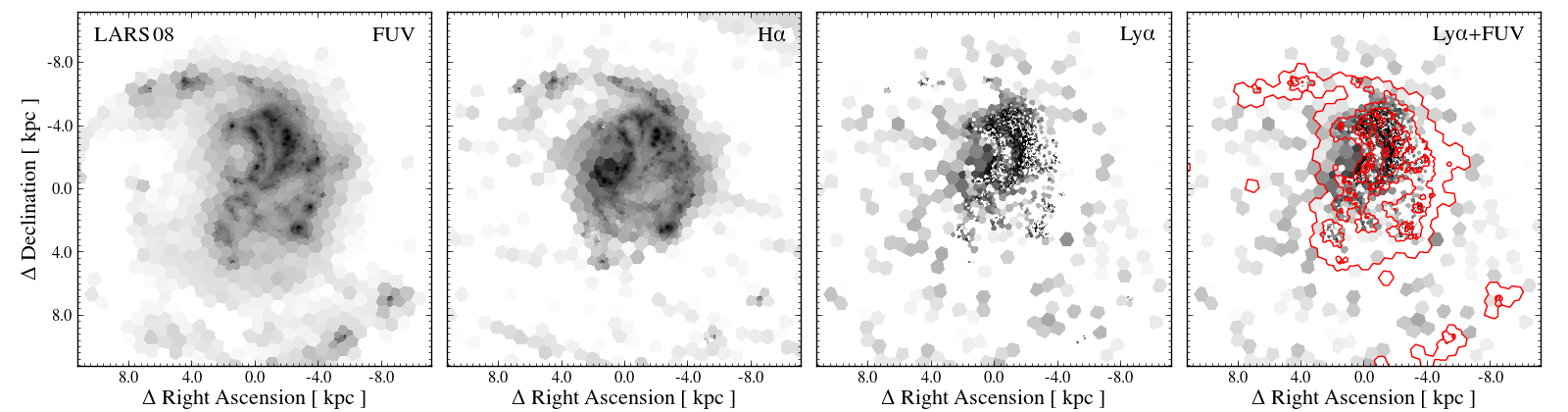}
\includegraphics[scale=0.45, clip=true, trim=02mm 11mm 6mm 1mm]{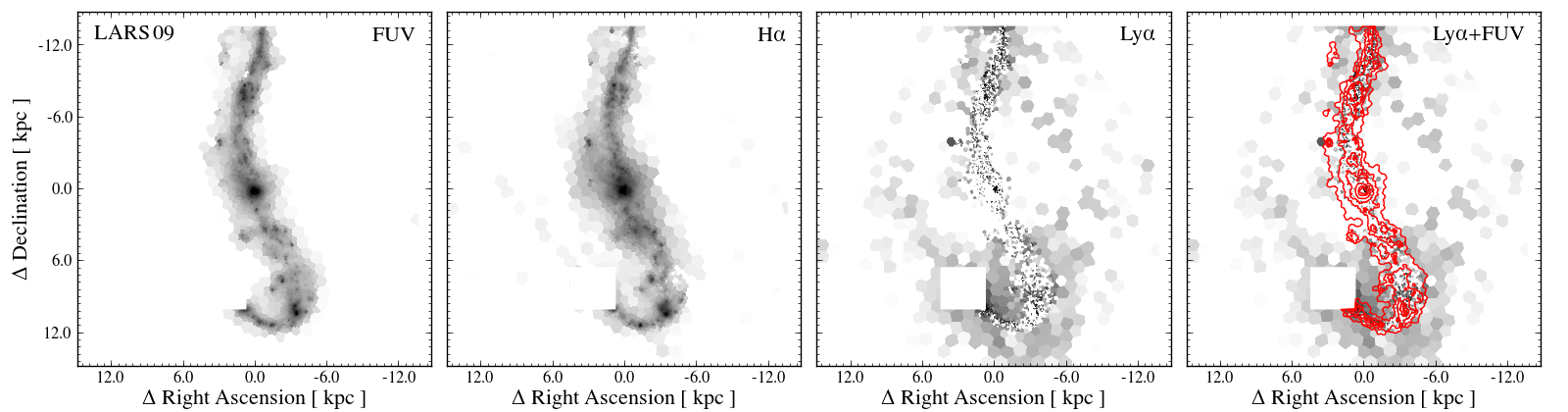}
\includegraphics[scale=0.45, clip=true, trim=02mm 01mm 6mm 1mm]{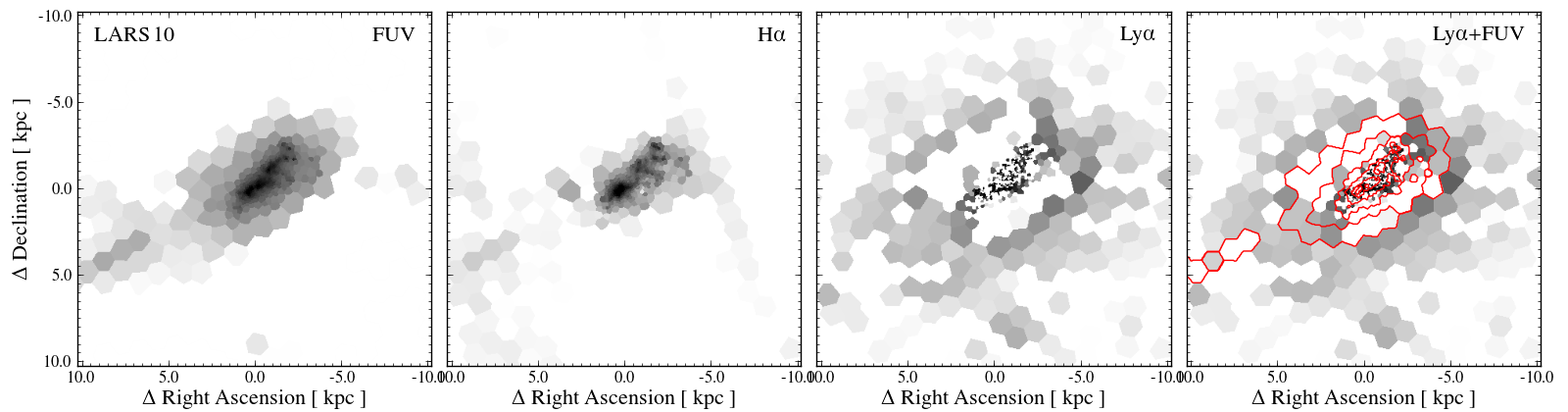}

\caption{{\bf --- continued ---} \emph{Top} to \emph{Bottom} shows 
galaxies LARS\,06, 07, 08, 09, and 10.  In LARS\,09 a field star has 
been masked with a square box.} 
\end{figure*}

\begin{figure*}[t!]
\ContinuedFloat
\includegraphics[scale=0.45, clip=true, trim=02mm 11mm 6mm 1mm]{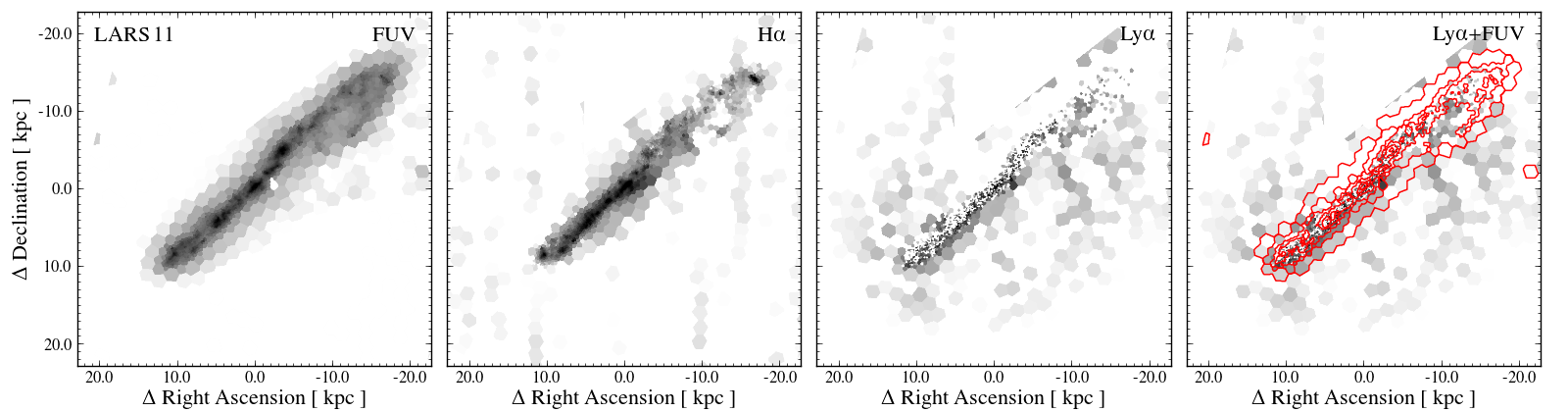}
\includegraphics[scale=0.45, clip=true, trim=02mm 11mm 6mm 1mm]{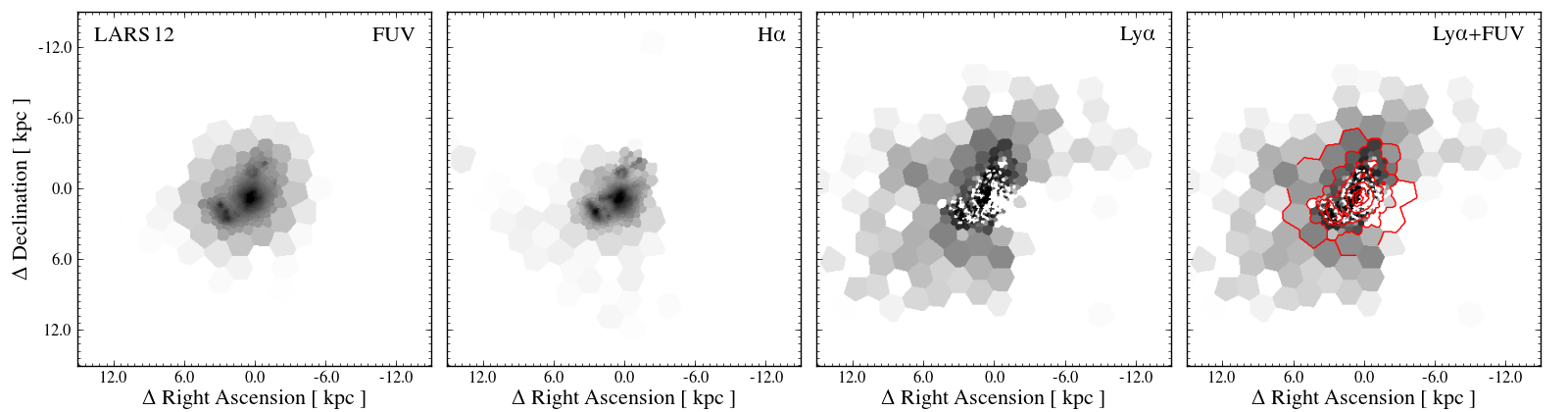}
\includegraphics[scale=0.45, clip=true, trim=02mm 11mm 6mm 1mm]{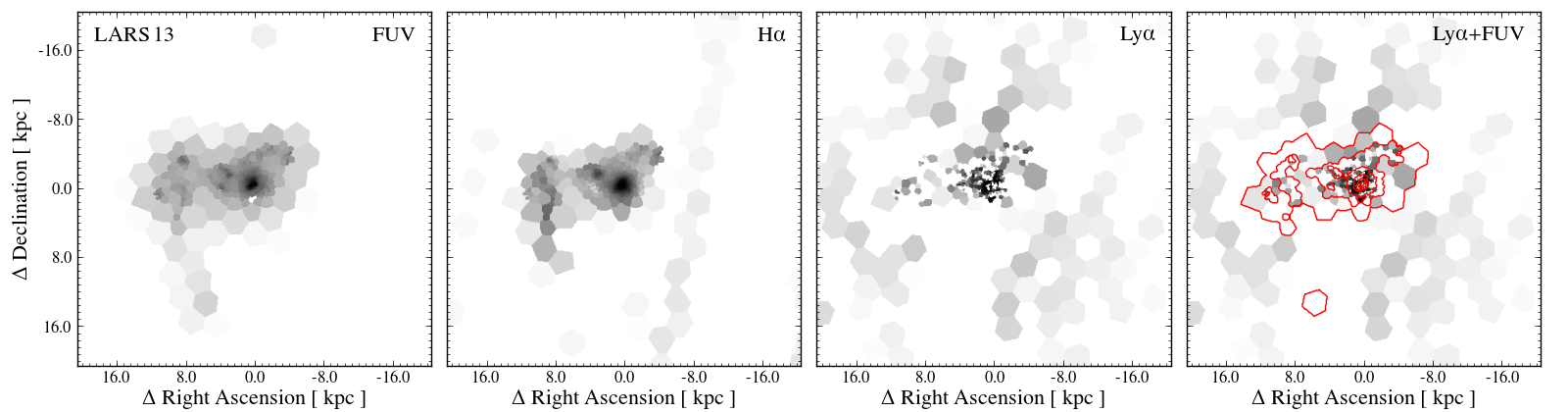}
\includegraphics[scale=0.45, clip=true, trim=02mm 01mm 6mm 1mm]{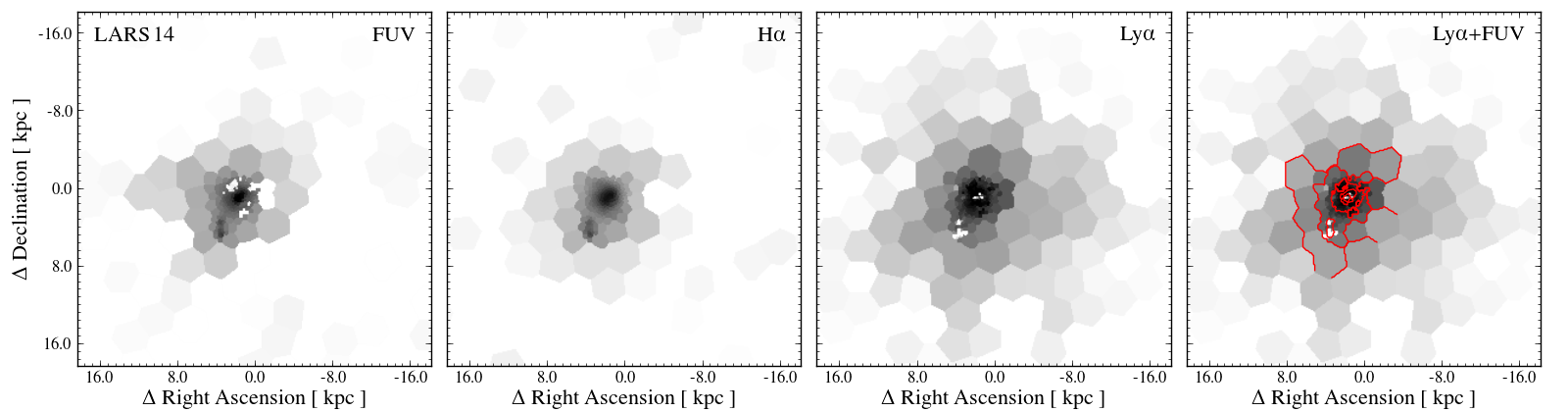}

\centering 

\caption{{\bf --- continued ---} \emph{Top} to \emph{Bottom} shows 
galaxies LARS\,11, 12, 13, and 14.  In LARS\,11 a nearby projected 
spiral galaxy has been masked with the triangular shape.} 
\end{figure*}

The \lya-transmitting UV filter was continuum-subtracted using the 
\emph{Lyman-alpha eXtraction Software} ({\tt LaXs}), which is an 
extension of the method presented in \citet{Hayes2009}, where full 
control of the nebular spectrum is obtained through narrowband imaging
in \halpha\ and \hbeta. {\tt LaXs} also handles the iterative continuum
subtraction of \halpha\ and \hbeta, and estimates the stellar 
absorption under \lya\ \citep[see][]{Valls-Gabaud1993} and each 
Balmer line \citep[e.g.][]{Gonzalez-Delgado1999}. 
The continuum-subtraction of \halpha\ is also corrected for the 
contribution of both \nII$\lambda\lambda6548$,6584\AA\ lines at the
appropriate transmission level, where the \nII/\halpha\ ratio (N2) 
is measured in the SDSS spectrum. Thus it accounts for no spatial 
variation, although this effect is shown not to be strong in most local 
starbursts. 

We caution again that because Lyman series transitions occur to the 
ground-state, interstellar \hi\ absorption in the \lya\ resonance
can be very strong. This absorption affects not only nebular \lya, but
also $\lambda=1216$\AA\ radiation in the stellar continuum which,
particularly along sightlines to the brightest clusters, can be 
substantial. Furthermore, the associated
scattering and frequency redistribution of these 
photons may result in complicated, extended, and P\,Cygni-like line 
profiles \cite[e.g.][]{Mas-Hesse2003}, the absorption components of 
which may extend hundreds of \kms\ bluewards of line center. 
Imaging methods have no capability to correct for interstellar 
absorption nor disentangle the red and blue components of a P\,Cygni
profile.
What we refer to as \lya\ in this paper is any deviation from the 
predicted stellar UV continuum, integrated over the synthetic narrow
bandpass. However we note also that if this 
constitutes a `problem' it is not one that is unique to low-$z$
observations, as high-$z$ narrowband imaging (and also low-resolution 
spectroscopy) will see the same blend of absorption and emission. 
Indeed the high spatial resolution afforded by HST imaging at low-$z$
enables us to see precisely where emission and absorption dominate in 
the ISM, and simulations of LARS galaxies observations at the 
degraded spatial sampling of higher redshifts are ongoing (Guaita et 
al.  in preparation).

The output of \texttt{LaXs}
includes not only emission line and line-free continuum images, but 
also maps of the stellar age and mass, and attenuation seen by the 
stellar continuum; several of these additional outputs are also
used for analysis in this paper.

\section{Images and Line Maps}\label{sect:images}

\subsection{Line Intensity Maps}

We present our first results in Figure~\ref{fig:maps}. 
Interloping objects, such as a field star in the 
frame of LARS\,09 and a chance projection of a nearby galaxy in the 
frame of LARS\,11, have been masked in these frames. The \emph{far left}
column  shows the observed intensity of 
radiation in the far ultraviolet, which traces the unobscured massive 
stars, and roughly 
incorporates the sites that produce the ionizing photons (LyC). In the 
\emph{center-left} column we show continuum-subtracted \halpha, which 
traces the nebulae in which the aforementioned ionizing photons are 
absorbed. Morphologically the FUV and \halpha\ emission maps are very 
similar, which is to be expected if LyC photons do not travel great 
distances; the flux emitted at one wavelength roughly results from the 
other. Discrepancies 
between FUV continuum and \halpha\ may result from dust attenuation
(which would obscure the UV more than \halpha, at least for the 
approximation of a dust screen), an evolved stellar 
population (e.g. an A-star dominated region will not produce ionizing 
photons, but will remain bright in the FUV), or the fact that we can 
spatially resolve the nebulae from their ionizing sources. Some 
galaxies, for example LARS\,03 and 05, do show somewhat extended 
\halpha, whereas in others such as LARS\,01 and 06 \halpha\ appears 
more compact.  While we have not yet dust-corrected 
the \halpha\ frame, its unobscured intensity traces the underlying 
production of \lya. The continuum subtracted \lya\ observation is shown
in the \emph{center right} panel, and in the \emph{far right} column 
with UV contours. Immediately it can be seen that these 
\lya\ morphologies bear only limited resemblance to their counterparts
in the FUV and \halpha: in some cases \lya\ appears to be almost 
completely absent, whereas in others smooth, large-scale structures are
visible that are not seen at other wavelengths \citep[see also][]{Hayes2013}. 

The intensity scaling of all the 
images is logarithmic, and the cut levels are set to show the maximum
of structure and the level at which the faintest features fade into 
the background noise. Should the reader wish to compare galaxies in 
absolute surface intensity, they are referred to 
Section~\ref{sect:spatdist} and Figure~\ref{fig:radprof}. 
\lya\ can appear in absorption as well as 
emission but these logarithmic images do not display any relative 
information of \lya\ intensity below zero. Despite this tuning to show
detail in emission, some galaxies are almost 
invisible in \lya: LARS\,04 and 06 in particular show only small hints 
of weak patchy \lya\ emission but otherwise little sign of contiguous 
emission regions. We note that every galaxy in the sample shows 
\emph{some} \lya\ emission, although that does not necessarily qualify
every galaxy as a `Lyman-alpha Emitter' in the common sense of the 
term.  Section~\ref{sect:select} is dedicated to the discussion of this.

A general trend is
that while the FUV and \halpha\ images show a great deal of central 
structure, not much of that structure remains in the \lya\ morphology.
Many of the complexes of star-forming knots shown in \halpha\ give way
to much smoother and featureless \lya\ images; LARS 05 and 08 are some 
particularly striking examples of this. The extended halo emission 
discussed in \citet{Hayes2013} is clearly visible, particularly in 
LARS\,01, 02, 05, 07, 10, 12, 14.

\begin{figure*}[t!]
\centering 
\includegraphics[scale=0.4, clip=true, trim=00mm 08mm 10mm 02mm]{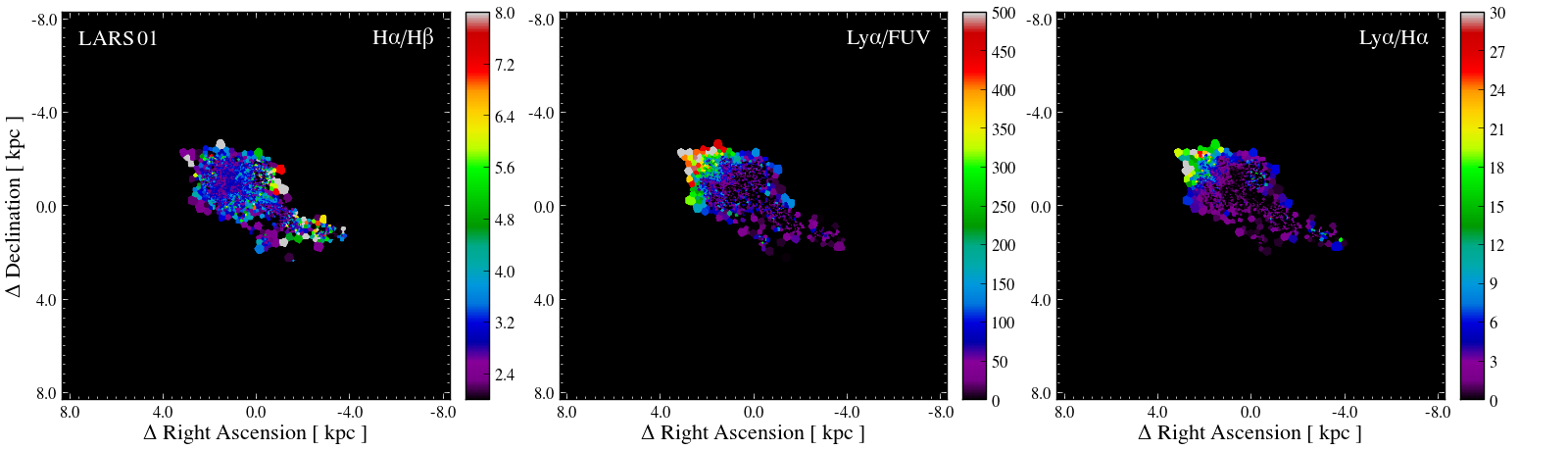}
\includegraphics[scale=0.4, clip=true, trim=00mm 08mm 10mm 02mm]{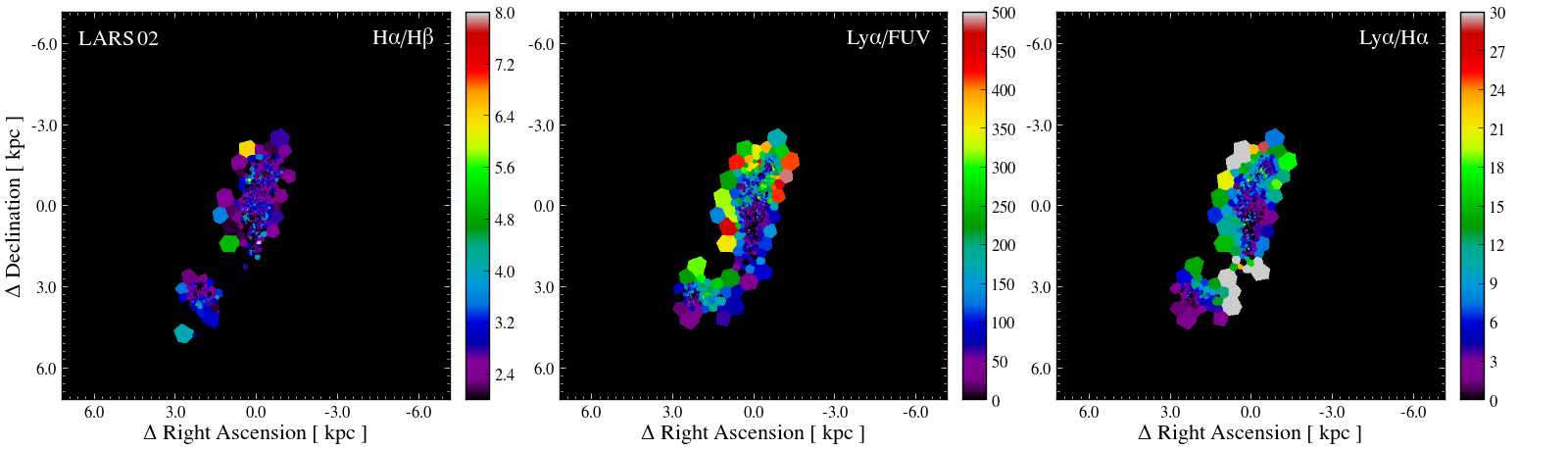}
\includegraphics[scale=0.4, clip=true, trim=00mm 08mm 10mm 02mm]{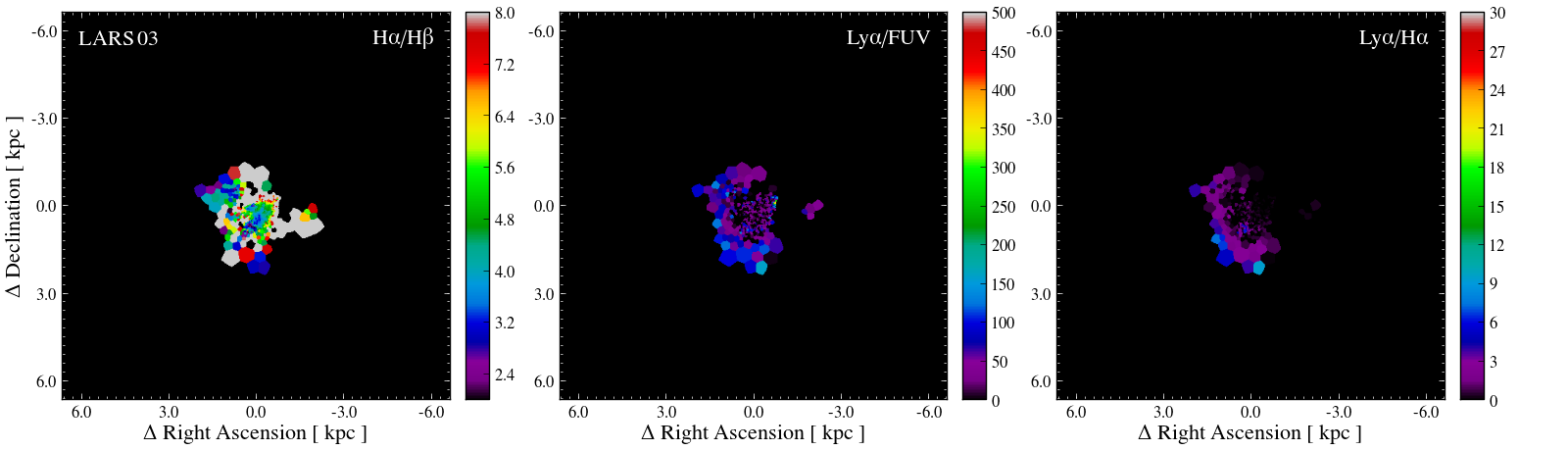}
\includegraphics[scale=0.4, clip=true, trim=00mm 08mm 10mm 02mm]{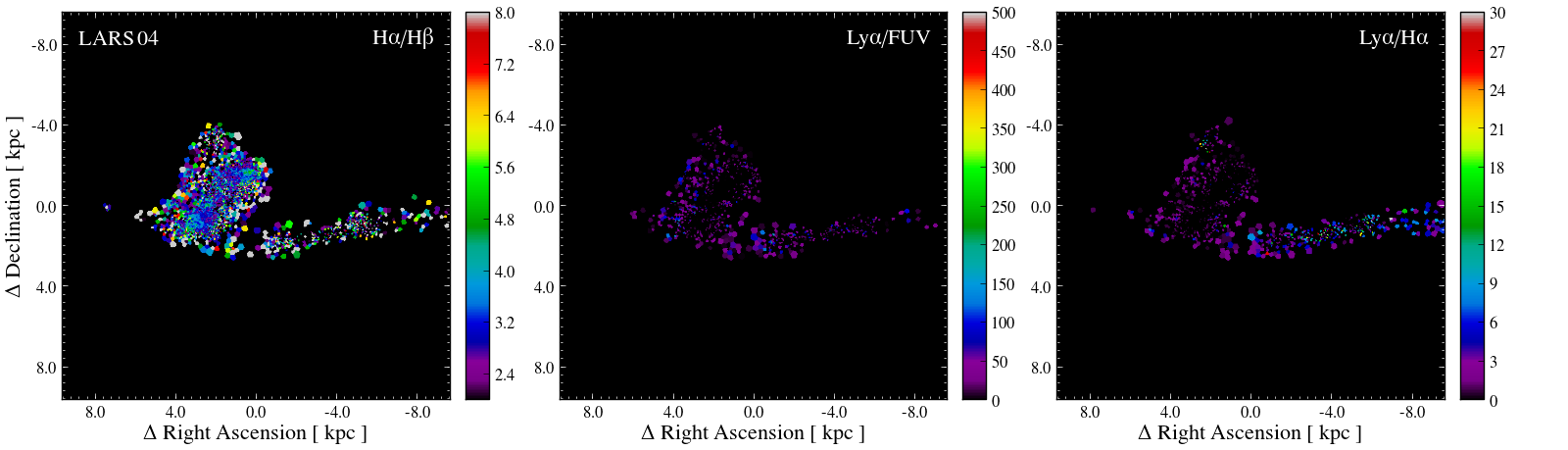}
\includegraphics[scale=0.4, clip=true, trim=00mm 00mm 10mm 02mm]{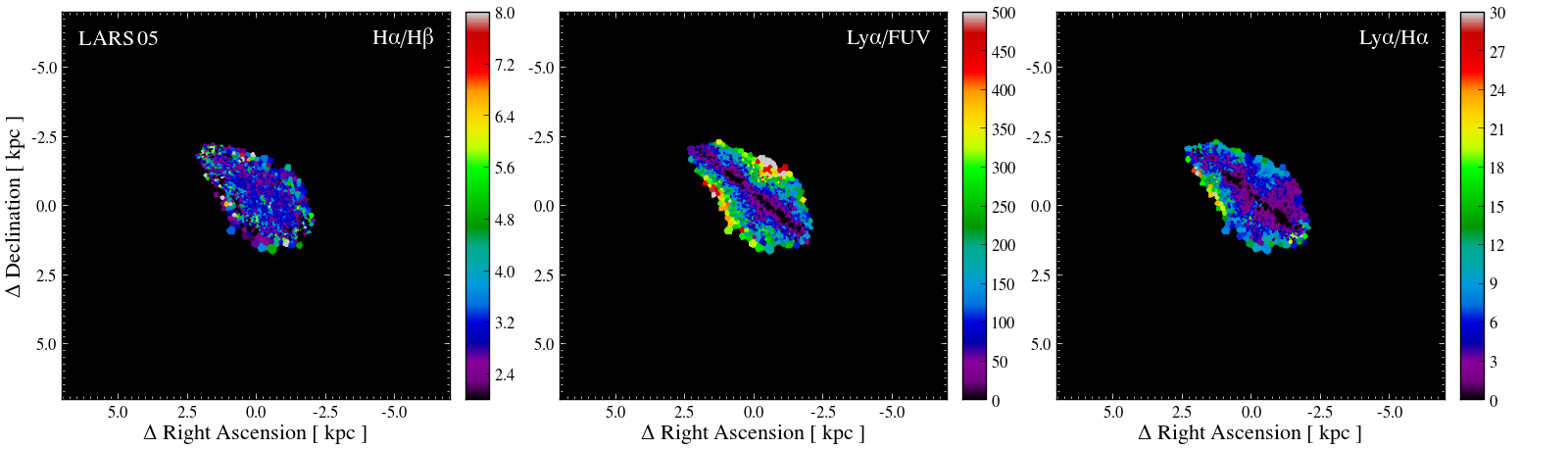}

\caption{Line ratios for LARS galaxies 01, 02, 03, 04, and 05. \emph{Left} panels 
show \halpha/\hbeta, which approximates the attenuation due to dust in 
the nebular phase. For reference, note that \halpha/\hbeta\ for $10^4$K 
gas is 2.86 \citep{Osterbrock1989}. \emph{Center} panels show the \lya\ EW -- 
this takes a value of $\approx 100$~\AA\ for star formation at 
equilibrium (after roughly 100~Myr) and up to $\approx 250$~\AA\  for 
the youngest bursts \citep{Schaerer2003}. The \emph{Right} panels show the 
\lya/\halpha\ ratios, which at normal temperatures and densities takes 
a value of $\approx 8.7$ \citep{Hummer1987}. Scales are the same as in 
Figure~\ref{fig:maps}.
\vspace{5mm}
}
\label{fig:ratios}
\end{figure*}

\begin{figure*}[t!]
\ContinuedFloat

\centering 
\includegraphics[scale=0.4, clip=true, trim=00mm 08mm 10mm 02mm]{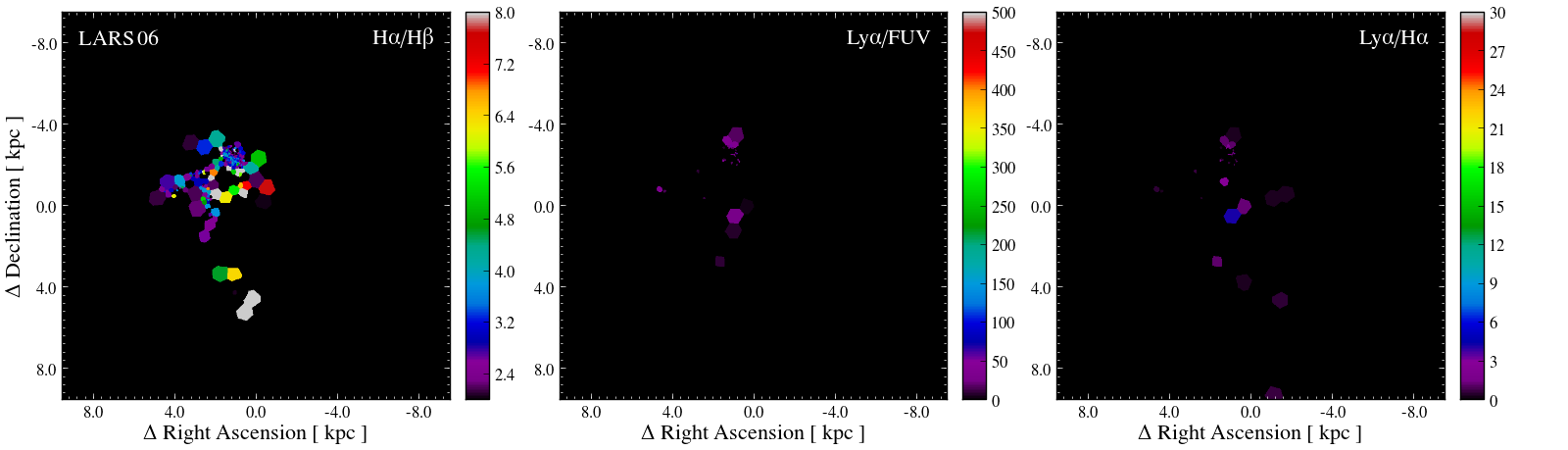}
\includegraphics[scale=0.4, clip=true, trim=00mm 08mm 10mm 02mm]{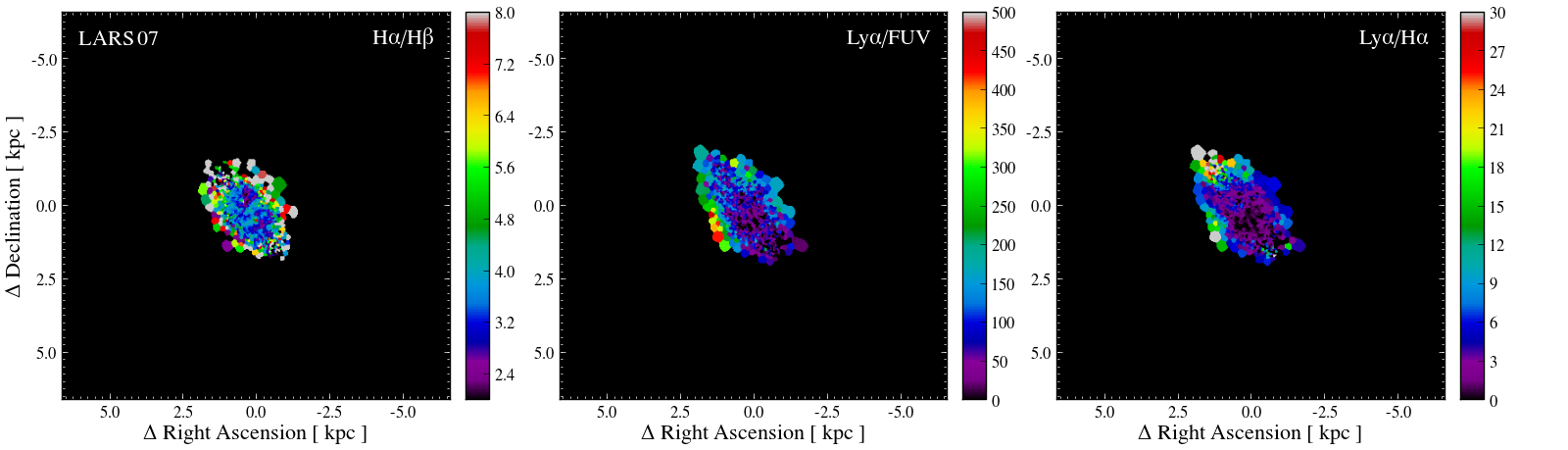}
\includegraphics[scale=0.4, clip=true, trim=00mm 08mm 10mm 02mm]{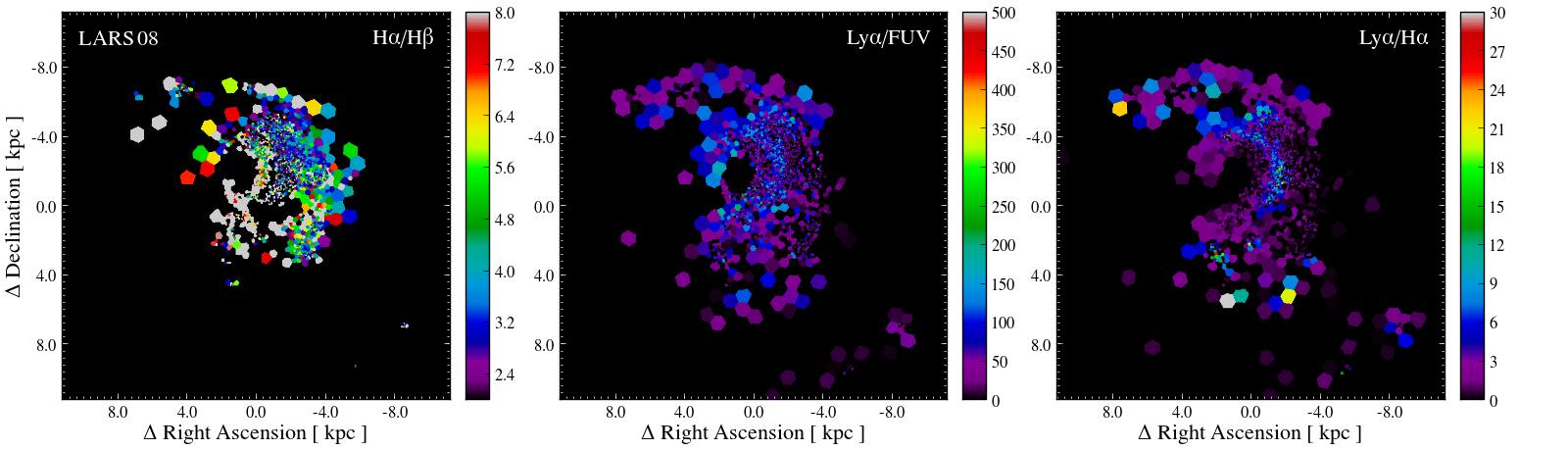}
\includegraphics[scale=0.4, clip=true, trim=00mm 08mm 10mm 02mm]{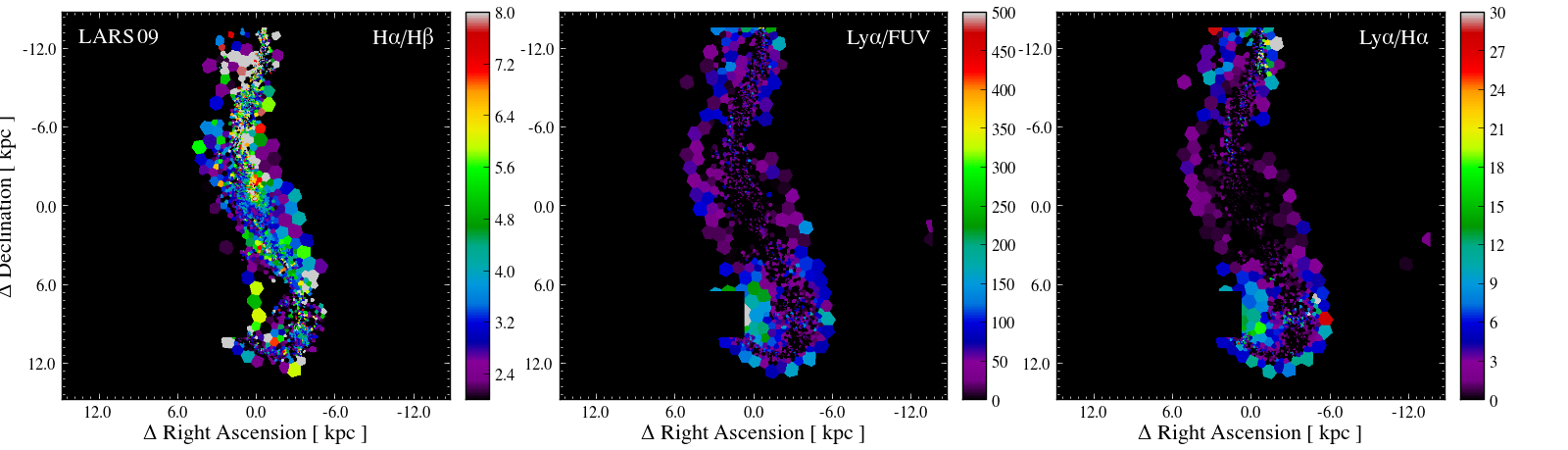}
\includegraphics[scale=0.4, clip=true, trim=00mm 00mm 10mm 02mm]{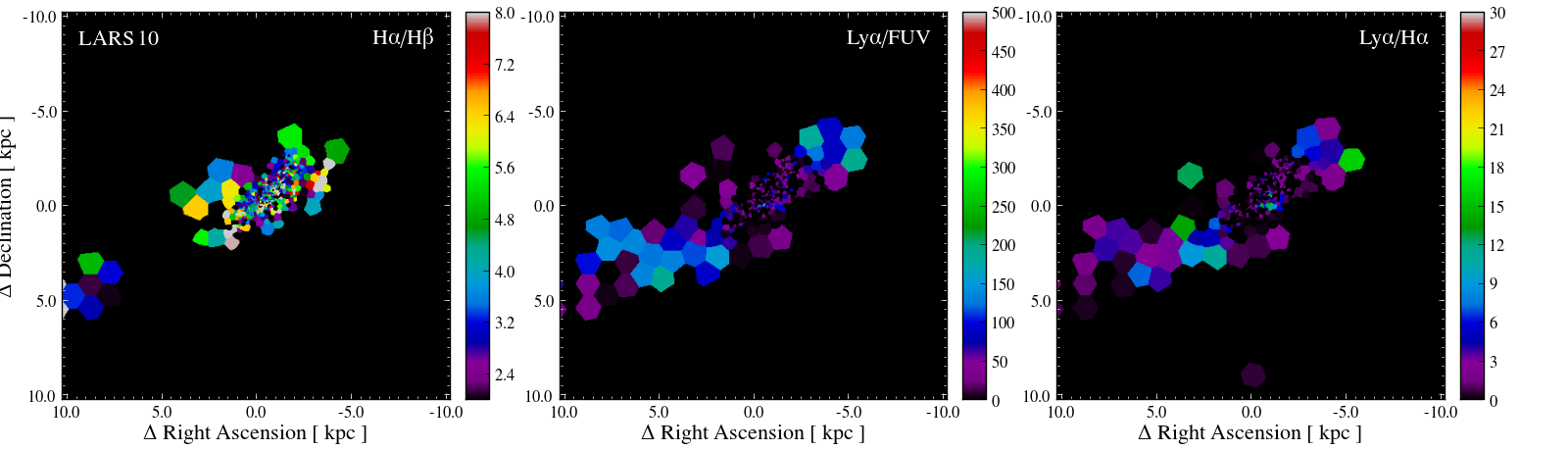}

\caption{{\bf --- continued ---} Galaxies: LARS 06, 07, 08, 09, and 10. 
Note that a star has been masked in the image of LARS 09.}
\end{figure*}

\begin{figure*}[t!]
\ContinuedFloat

\centering 
\includegraphics[scale=0.4, clip=true, trim=00mm 08mm 10mm 02mm]{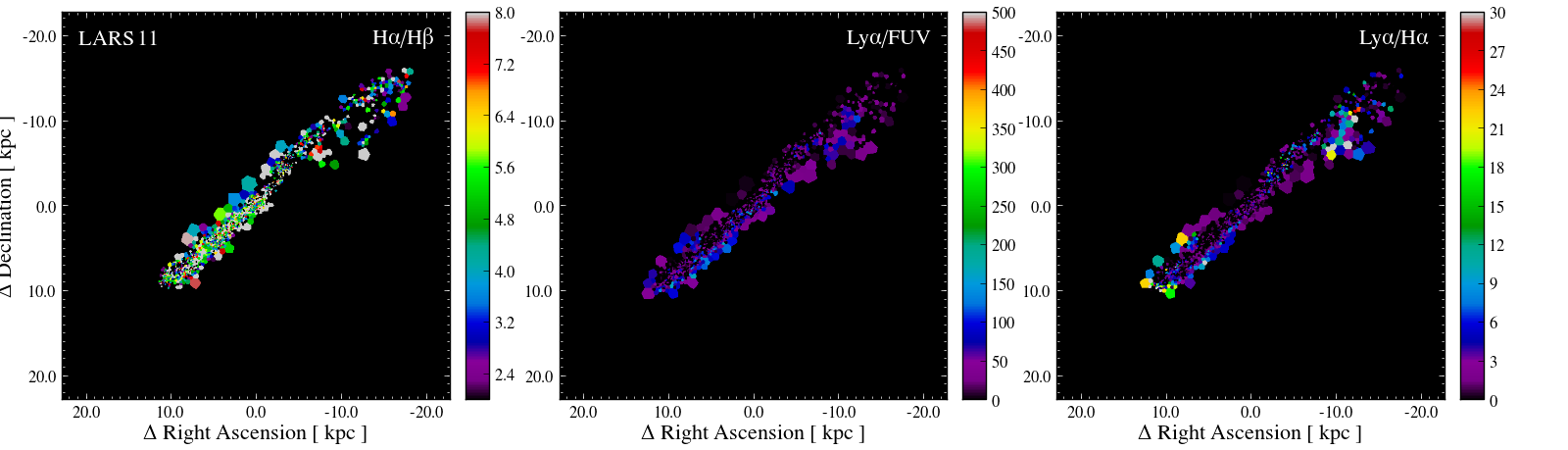}
\includegraphics[scale=0.4, clip=true, trim=00mm 08mm 10mm 02mm]{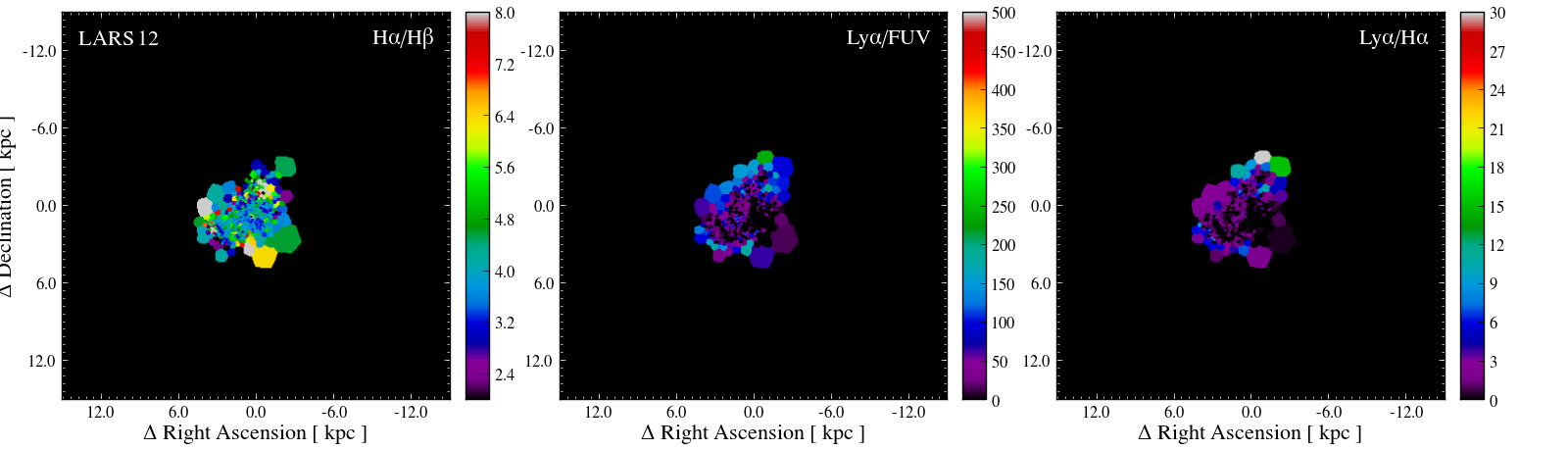}
\includegraphics[scale=0.4, clip=true, trim=00mm 08mm 10mm 02mm]{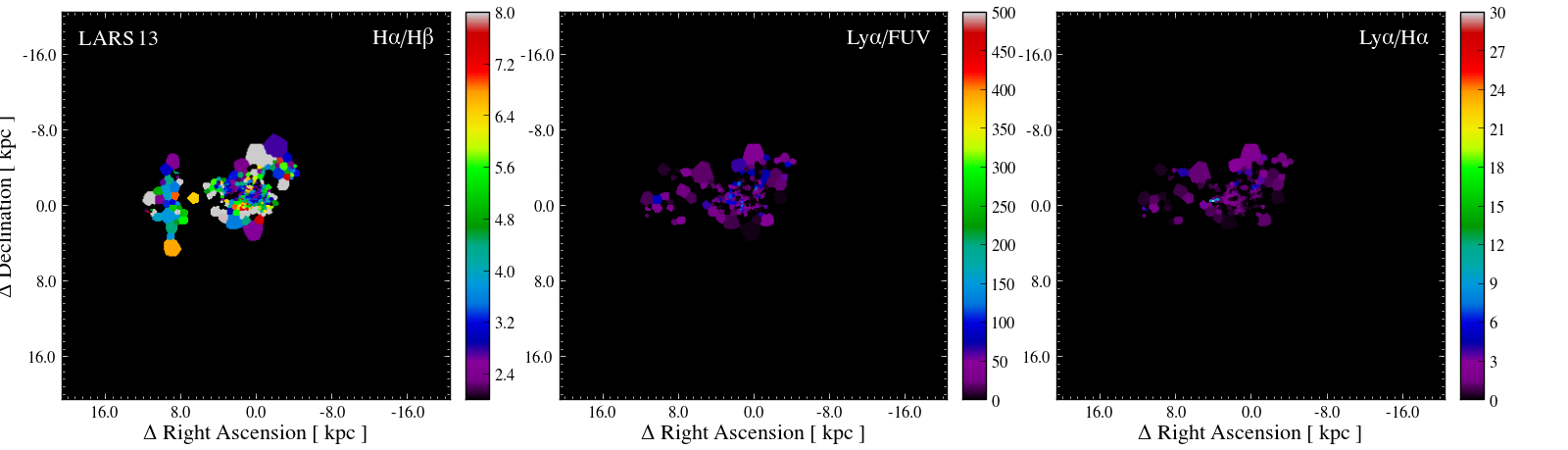}
\includegraphics[scale=0.4, clip=true, trim=00mm 00mm 10mm 02mm]{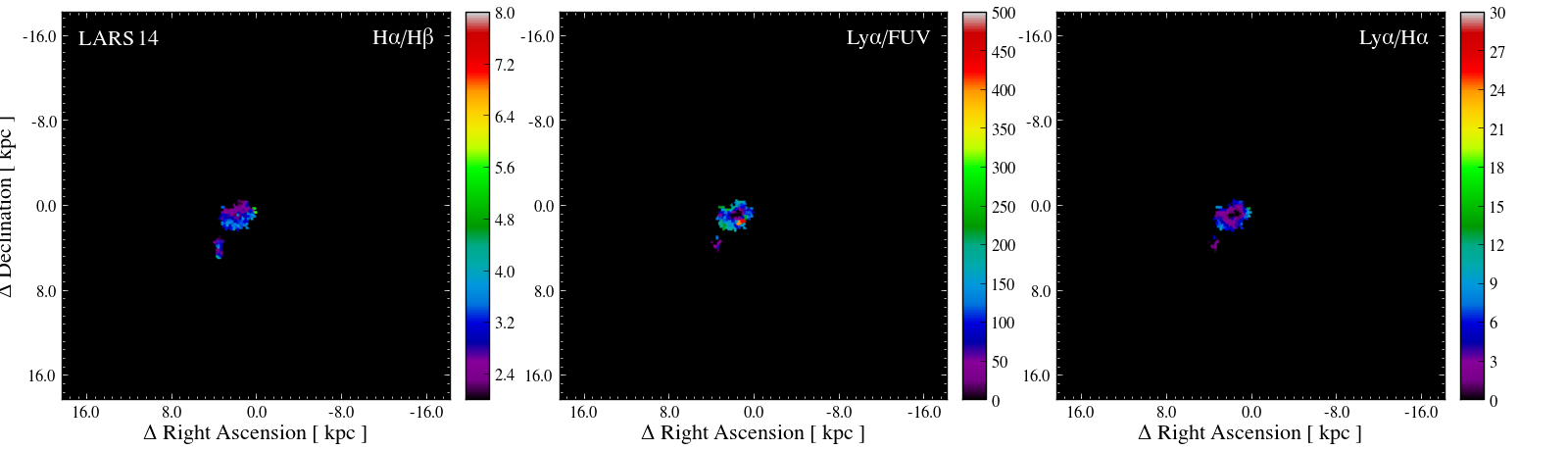}

\caption{{\bf --- continued ---} Galaxies: LARS 11, 12, 13, and 14.
Note that a galaxy has been masked in the image of LARS 11.}
\end{figure*}

ACS/SBC, through which our FUV images are 
obtained, is known to have a dark current that is centrally peaked, 
causing a pattern in the image that may be somewhat resemblant of our
low surface brightness \lya\ structures. However, we are convinced that 
none of the observed diffuse component is the result of this 
instrumental signature. Firstly, and as discussed in Paper I, the SBC 
dark current increases as a strong function of operating 
temperature, which in turn grows with the time since the SBC amplifiers
have been operating -- the dark current is constant and negligible at 
temperature below 28~$^\circ$C, which corresponds to about four hours of
operation\footnote{Gonzaga, S., et al., 2011, ACS Data Handbook, 
Version 6.0, (Baltimore: STScI).}. Our SBC observations were executed in
visits of just two orbits, which corresponds to no more than three 
hours of operation.
Furthermore	we have been careful to make sure that our SBC visits are 
only executed when the camera has not been used in the preceding visits 
and the amplifiers have been switched off, thereby ensuring a cold 
detector when the observations begin. Secondly, we adopted a strategy of 
interleaving on- and off-line observations, such that a dark current 
signature should it somehow be present, would be evenly distributed 
between the filters. Furthermore the fact that the extended halo
feature is absent in seven of the 
14 galaxies (LARS\,03, 04, 06, 08, 09, 11, 13), and 
that galaxies show differing morphologies, 
asymmetries, and intensities in the diffuse component all lead us to 
confidence about its reality.

\subsection{Line Ratio Maps}

In Figure~\ref{fig:ratios} we show maps of the following ratios: 
\halpha/\hbeta, which quantifies the dust attenuation in the nebular 
phase (\emph{Left}); \lya/FUV continuum, the equivalent width 
(\ewlya, \emph{Center}); and \lya/\halpha\ (\emph{Right}). Pixel-to-pixel 
noise in some of these images may be high, and usually comes from low 
signal in either the \lya\ or \hbeta\ image. To remove this from the 
presented images we restrict our display to the isophotal level in 
the UV that corresponds to twice the Petrosian radius, defined by 
$\eta=0.2$ \citep{Petrosian1976}. This is the 
isophotal version of the circular apertures used to compute the global
quantities in Section~\ref{sect:globals}. 

Galaxies show varying degrees of structure in all the intensity ratios.
In \halpha/\hbeta\ some objects (e.g. LARS\,03) show smooth 
spatial gradients in their extinction, while others show very little 
well-defined structure (e.g. LARS\,02). Spatially  \halpha/\hbeta\ 
ranges between values as low as $\sim 2$ (notably in LARS\,02), to 
galaxies that show heavily 
extinguished nuclei with \halpha/\hbeta\ above 8 (LARS\,03 and 08). 
The \lya\ EW, and \lya/\halpha\ maps are rich with features that 
reflect the complex \lya\ astrophysics. The \lya\ EW, shown in the 
central panels 
in Figure~\ref{fig:ratios}, is one of the most common high-$z$ 
observables, and the main selection criterion for most \lya\ surveys.
At high-$z$, of course, the EW is always measured in collapsed 
apertures. For the color-scaling in Figure~\ref{fig:ratios}, black 
shows an absence of \lya\ (or absorption), a blue color encodes 
a \lya\ EW of 100~\AA, corresponding to the expected \lya\ EW of 
star formation at equilibrium, and mid green (250~\AA) to the approximate
value that can be attained for very young star formation episodes
\citep{Charlot1993}. All yellow, red and white resolution elements 
exceed this value. For the \lya/\halpha\ maps, the intrinsic ratio for 
Case B recombination is 8.7 ($\approx 11$ for Case A, 
\citealt{Hummer1987}). In the central regions of all galaxies,
where the UV surface brightness is high, the overwhelming majority of 
the displayed pixels are in the black--purple range, showing EW below
a few tens of \AA, and \lya/\halpha\ in the range of $<0$ to $\sim3$. The 
low EWs can be explained if the star formation episode is in the 
process of turning off, but the \lya/\halpha\ ratio cannot: the 
black-purple pixels in the \lya/\halpha\ map can only be explained by a
stronger suppression of \lya\ than \halpha. This could be because of 
dust attenuation or because \lya\ photons are being resonantly 
scattered out of the line-of-sight by neutral hydrogen atoms.  In many 
cases, however, the values of \halpha/\hbeta\ in these regions imply 
nebular attenuations of around zero, which would suggest that 
scattering is the dominant mechanism of removing photons from the 
sightline. 

Another feature seen in most of the galaxies is that as we look away 
from the highest surface brightness regions, both the \lya\ EW and 
\lya/\halpha\ ratios increase. In examples like LARS 05, a thin column 
of low ratios runs across the body of the galaxy, but rapidly increases
away from the plane. In this example the EW exceeds 500~\AA\ (twice the 
maximum value for a zero-aged star formation episode), and 
\lya/\halpha\ exceeds 30. One may certainly expect EWs to be high on
small scales simply because the stars and the nebulae can be spatially 
resolved, but this cannot explain the high local line ratios. 
In \citet{Oti-Florannes2012} we showed that we could not easily produce
such line ratios in collisionally excited plasmas, and these local 
enhancements must almost certainly result from the scattering of \lya\
into the line-of-sight. 

One natural consequence of the larger EW at larger radius is that the 
integrated EW measured by matched aperture on-line/off-line photometry 
-- as done for instance by {\sc SExtractor} \citep{Bertin1996} in 
dual-image mode -- will be strongly dependent upon the size of the 
aperture (see Section~\ref{sect:globals}).

\section{The Spatial Distribution of Light}\label{sect:spatdist}

We now quantitatively examine and compare the spatial 
distribution of quantities. In a high-$z$ narrowband survey all 
galaxies will lie at effectively the same redshift, so the whole sample
will have the same physical sampling scale and luminosity distance.  
This is not true for the LARS sample, where redshifts differ by a 
factor of six (0.028 to 0.18). Hence to compare galaxies equally 
we present all quantities in restframe luminosities (instead of 
flux) and converted onto the physical scale in kpc (instead
of the observed angular scale in arcsec) -- i.e. when we refer to 
line `surface brightness', the units are \ergseckpc\ (instead of 
\ergseccmarcsec). For reference a galaxy at $z=3$ with an observed 
surface brightness of $10^{-18}$~\ergseccmarcsec\ would have an 
intrinsic restframe brightness of $1.30\times10^{39}$~\ergseckpc\ for 
our assumed cosmology. 

The galaxies have various morphologies and the inclinations of the 
underlying starburst hosts are currently unknown. Hence for radial work we 
attempt no deprojection and use simple circular annuli; when we have 
obtained deep optical/NIR imaging for the sample we will study radial 
profiles and inclinations in more detail.

\begin{figure}[t]
\centering 
\includegraphics[scale=0.57, clip=true, trim=1mm 1mm 3mm 2mm]{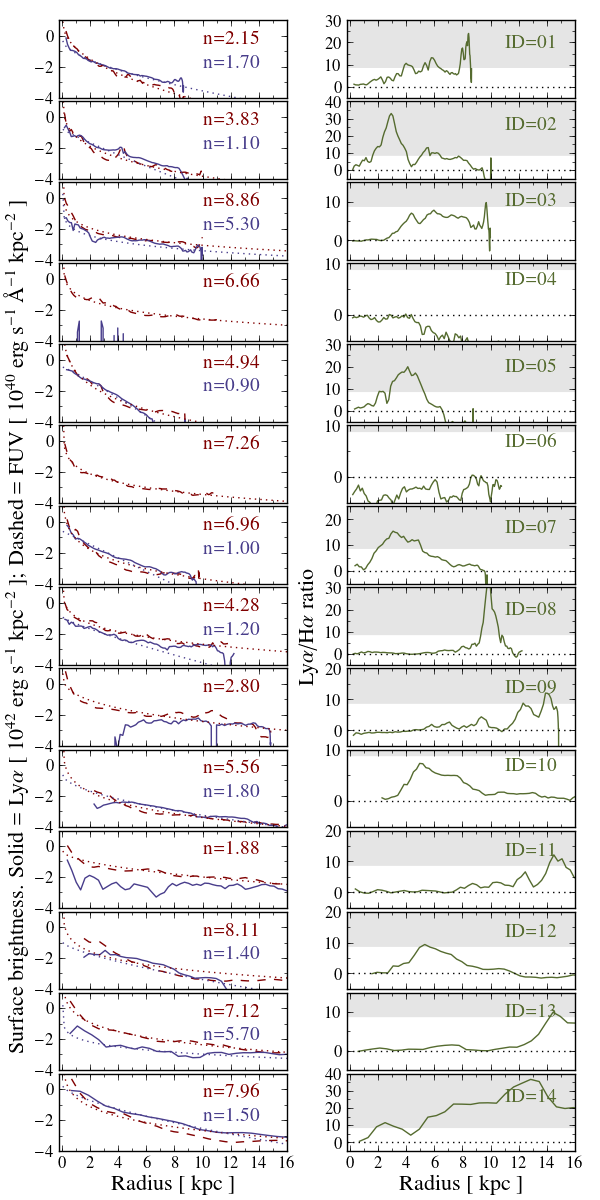}
\caption{\emph{Left} panels show the radial (circular annuli) light 
profiles computed from the \lya\ (blue solid) and FUV (red dashed) 
images. For comparison, profiles are scaled from surface brightness 
to restframe emissivities per areal unit, and units of the ordinate 
axis are $10^{42}$~\ergseckpc\ for \lya\ and $10^{40}$~\ergsecaakpc\ 
the continuum. Best-fitting S\'ersic profiles are shown with dotted 
lines of the corresponding colors, and the recovered S\'ersic index, 
$n$, is listed for each. 
\emph{Right} panels show the local \lya/\halpha\ ratios in the same 
annuli. The grey shading shows regions for which the intrinsic Case B 
value for this line ratio is exceeded.}\label{fig:radprof}
\end{figure}

\begin{figure*}[t!]
\centering 
\includegraphics[scale=0.4, clip=true, trim=0mm 0mm 0mm 5mm]{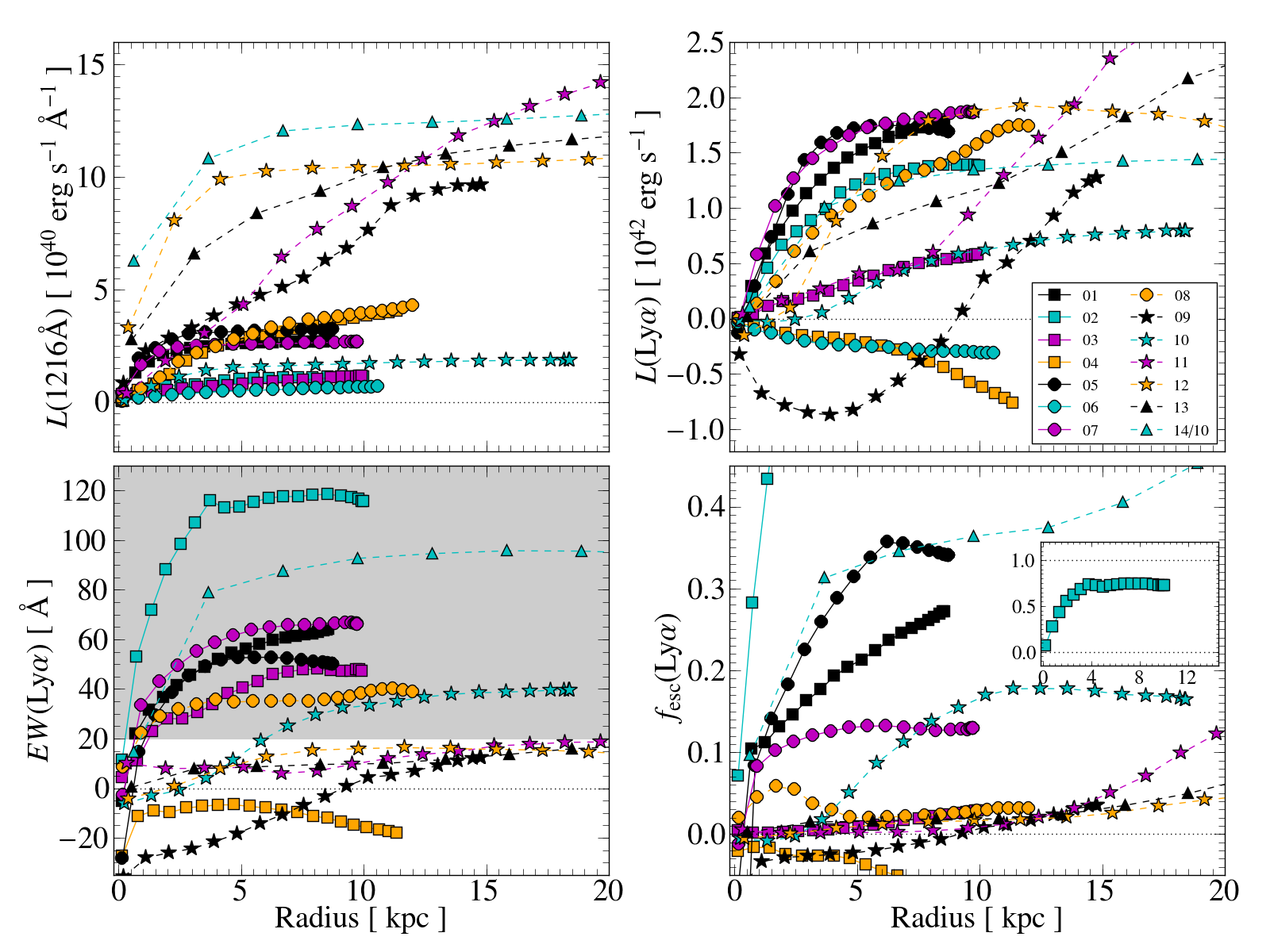}
\caption{Radial photometric curves-of-growth. I.e. the same as Figure 
\ref{fig:radprof} but instead of averaging over the differentially 
masked region, we integrate inside the total aperture. We show 
the FUV luminosity (\emph{upper left}), 
\lya\ luminosity (\emph{upper right}), 
\lya\ EW (\emph{lower left}), 
\fesclya\ (\emph{lower right}).
In the plot of \lya\ luminosity, the curve for LARS\,14 has been 
divided by 10 so that the galaxy may be visualized. In the figure of 
\ewlya, the shaded region shows \ewlya$\ge 20$~\AA, the common 
selection criterion used to select \lya-emitters at high redshift. }
\label{fig:growcurve}
\end{figure*}

We describe the light distribution in two ways: firstly we present the 
radial profiles of quantities measured locally within 
annular apertures (Section~\ref{sect:radprof}), which describes how the 
emergent light is 
actually distributed as a function of radius, and is common in detailed 
spatial studies of low-$z$ galaxies. Secondly, we show the summed 
quantity within each aperture (same as above) in 
Section~\ref{sect:growcurve}.  This is more comparable to the 
integrated measurements that would be made at high-$z$.  In 
determining the center we simply adopt the brightest pixel in the UV 
continuum image.  This is the same as in \citet{Hayes2013} and, while 
it is possible to make the case for other centroids, is the same as will
be used for isophotal light profiles used in upcoming studies.

\subsection{Radial Profiles}\label{sect:radprof}

Figure~\ref{fig:radprof} shows the radial distribution profiles of 
\lya\ an the FUV continuum, which are presented logarithmically to 
simultaneously contrast low and high surface intensities. The cost is
that regions where \lya\ is on average absorbed cannot be visualized. 
Typically the profiles are more strongly peaked towards the center in 
the continuum than in \lya: in the inner regions \lya\ profiles run
systematically flatter than their FUV counterparts. To quantify this 
central peaking, we fit S\'ersic light 
profiles\footnote{$\mu(r) = \mu_0 \exp[-(r/r_0)^{(1/n)}]$} to both line 
and continuum, finding the profiles adequately approximate reality in 
most cases. Recovered indices, $n$, are found to be high in the 
continuum -- usually between 4 and 8 -- which are typical of the very
compact nature of the UV. \lya\ profiles, however, typically show 
indices around $n=1-2$, and much closer to the exponential profile seen
in disk galaxies.

In the \emph{right} panels of Figure~\ref{fig:radprof} we present 
radial profiles of the \lya/\halpha\ ratio in the same annuli. Values
of \lya/\halpha$>8.7$ are shaded grey to indicate where the intrinsic
value of the line ratio is exceeded, and it is clear that this happens
at some radius in most galaxies in the sample. Most notably LARS 01, 02, 
05, 07, 08, and 14 show \lya\ to greatly exceed the intensity that is 
predicted by the \halpha\ line, and ratios three times the intrinsic 
value are easily obtained in some galaxies.

\subsection{Photometric Curves-of-Growth}\label{sect:growcurve}

In Figure~\ref{fig:growcurve} we show the total aperture-integrated 
FUV luminosity, and \lya\ luminosity, EW, and \fesclya\ as a function of 
aperture radius. The {\em Upper Left} panel shows the cumulative FUV 
luminosity; naturally, with a strong central peaking of this light 
profile (Figure~\ref{fig:radprof}), this quantity increases monotonically
and rapidly within the central few kpc. For all galaxies except 
one (LARS\,11) these growth curves have essentially flattened by radii 
of $\approx 10$~kpc, and all the more nearby ones (LARS 01 to 07),
flatten by substantially smaller radii still (3--5~kpc), suggesting 
that all the light has been captured by the aperture. \lya\ growth 
curves, shown to the {\em Upper Right} behave somewhat similarly, and 
most of these have also flattened by $r\sim 10$~kpc. However some 
(LARS 08, 09, 11, 13 at least) show profiles that are still growing 
at radii that correspond to the full extent of the SBC detector. 
LARS 04 and 06 are strong \lya\ absorbing galaxies, for which as radii
become larger more \lya\ flux is absorbed. Similar behaviour is seen in
LARS 09 at small radii, but at $r\gtrsim 4$ arcsec positive \lya\ flux 
starts to be captured, the curve increases, and LARS\,09 becomes a 
global emitter by about 10~kpc. 

The \emph{Lower left} panel shows aperture growth of \ewlya, or simply 
the ratio of the \emph{Upper Right} and \emph{Upper Left} plots. The 
positive gradient in nearly all of the lines shows that the UV light is
more strongly centrally concentrated than \lya\ 
(c.f. Section~\ref{sect:radprof} and Figure~\ref{fig:radprof}). 
Some of these curves 
converge at rather high \ewlya, above 40\AA, and two galaxies show 
\ewlya\ around 100\AA. The 
gray shaded area shows \ewlya\ above 20~\AA, the canonical value by 
which LAEs are defined at high-$z$, and we see that eight of our 
galaxies enter this region at some radius to which our HST observations
are sensitive. However between radii of $\approx 10$ and 20~kpc, the 
\ewlya\ of four galaxies is still slowly growing, reaching almost
this threshold value by 20~kpc ($\approx 2.5$ arcsec aperture at 
$z\gtrsim 2$). Whether they would become canonical LAEs using larger 
apertures we cannot say. As we will discuss in 
Section~\ref{sect:select} these curves of \llya\ and \ewlya\ completely
determine at what radius a high-$z$ galaxy may be selected as an LAE.

The {\em Lower Right} panel shows the measured \lya\ escape fraction, 
\fesclya, which is defined here and throughout as the observed \lya\ 
luminosity divided by the intrinsically produced luminosity. This 
value is derived from the \halpha\ luminosity that has been corrected 
for nebular dust 
attenuation using \halpha/\hbeta\ and following \citet{Hayes2005}:
\begin{eqnarray}
f_\mathrm{esc}^{\mathrm{Ly}\alpha} &=& L_{\mathrm{Ly}\alpha}^\mathrm{obs} / L_{\mathrm{Ly}\alpha}^\mathrm{int}      \nonumber \\
   &=& L_{\mathrm{Ly}\alpha}^\mathrm{obs} / ( 8.7 \times L_{\mathrm{H}\alpha}^\mathrm{obs} \times 10^{0.4\cdot E_{B-V} \cdot k_{6563}} ).
\label{eq:fesc}
\end{eqnarray}
\ebv\ in this case is the nebular value derived from the ratio of 
\halpha/\hbeta, where we adopt the intrinsic ratio of 2.86 
\citep{Osterbrock1989} and use the extinction curve derived in the 
Small Magellanic Cloud \citep{Prevot1984}. Qualitatively the picture 
does not change if we adopt other extinction laws. The way \fesclya\ 
has been defined permits it to take negative values in the case of 
\lya-absorbing galaxies, and represents the additional fraction of 
continuum \lya\ absorbed relative to the intrinsic nebular luminosity. 
We elect not to use global laws of effective attenuation such as that 
of \citet{Calzetti1994} or \citet{Charlot2000} because the scales we 
probe are very different from those over which such laws were computed 
(tens of pc vs several kpc). 

Generally most galaxies 
exhibit low escape fractions, below a few per cent, while five 
objects show \fesclya\ above 10 per cent (and growing) at the largest
radii. The slopes of these curves differ markedly, with some rising
sharply (e.g. LARS 02, 05, 14) and some much more slowly. Convergence
also differs greatly: LARS 02 (\emph{inset}) flattens at 
\fesclya$\approx 75$~\% within $\approx 4$~kpc, but in LARS 11 and the 
others with slowly rising \ewlya, \fesclya\ continues to grow until 
radii of at least 20~kpc.

\section{Aperture-Integrated Properties}\label{sect:globals}

Next we examine how total \lya-related quantities compare with physical
properties that can be derived from our HST imaging data, or archival 
survey data that have been previously obtained (SDSS spectroscopy). 
Properties derived from other data-sets, e.g. \hi\ masses and 
kinematics, will be presented in forthcoming articles. The results of 
Section~\ref{sect:spatdist} have already demonstrated that the 
computation of global \lya\ quantities is a strong function of the 
adopted radius, but nevertheless it is illustrative to define a 
standard radius and compare the galaxies equally in terms of the 
quantities derived within. For this we choose to adopt the frequently
used definition of $2\times r_\mathrm{P20}$, where $r_\mathrm{P20}$ is
the isophotal Petrosian radius, determined to be the isophote at which
the ratio of local to internal surface brightness, $\eta$, reaches 0.2.
We calculate these apertures in the image that transmits \lya\ and the 
FUV continuum, and use the same definitions as in \citet{Hayes2013} and 
Guaita et al (in prep).

\begin{deluxetable*}{clccccccccc}[t!]
\tabletypesize{\scriptsize}
\tablecaption{Line and Continuum properties of the LARS galaxies.\label{tab:lumew}}
\tablehead{
\colhead{LARS} & 
		\colhead{Common name} & 
		\colhead{$z$} & 
		\colhead{$2\times r_\mathrm{P20}$} & 
		\colhead{$L_{\mathrm{Ly}\alpha}$} & 
		\colhead{$L_{FUV}$} & 
		\colhead{$W_{\mathrm{Ly}\alpha}$} & 
		\colhead{$L_{\mathrm{H}\alpha}$} & 
		\colhead{$W_{\mathrm{H}\alpha}$} & 
		\colhead{$L_{\mathrm{H}\beta}$} & 
		\colhead{$W_{\mathrm{H}\beta}$} \\
\colhead{ID} & 
		\colhead{} & 
		\colhead{} & 
		\colhead{kpc} & 
		\colhead{$10^{42}$~cgs.} & 
		\colhead{$10^{40}$~cgs.} &  
		\colhead{\AA} & 
		\colhead{$10^{42}$~cgs.} &
		\colhead{\AA} & 
		\colhead{$10^{42}$cgs.} &
		\colhead{\AA} \\
\colhead{(1)} & 
		\colhead{(2)} & 
		\colhead{(3)} & 
		\colhead{(4)} &  
		\colhead{(5)} & 
		\colhead{(6)} &  
		\colhead{(7)} & 
		\colhead{(8)} &  
		\colhead{(9)} &  
		\colhead{(10)} &  
		\colhead{(11)}}  
\startdata
01 & Mrk\,259	       & 0.028 & 2.39 & 0.85 & 2.52 & 33.0 & 0.63 & 409  & 0.11 & 73.9 \\ 
02 & \nodata         & 0.030 & 2.32 & 0.81 & 0.96 & 81.7 & 0.18 & 313  & 0.08 & 69.4 \\ 
03 & Arp\,238	       & 0.031 & 1.87 & 0.10 & 0.57 & 16.3 & 0.59 & 199  & 0.14 & 26.8 \\ 
04 & \nodata         & 0.033 & 1.83 & 0.00 & 3.02 & 0.00 & 0.60 & 242  & 0.20 & 44.4 \\ 
05 & Mrk\,1486	     & 0.034 & 1.88 & 1.11 & 3.00 & 35.9 & 0.51 & 436  & 0.17 & 66.7 \\ 
06 & KISSR\,2019	   & 0.034 & 1.30 & 0.00 & 0.53 & 0.00 & 0.08 & 166  & 0.04 & 41.2 \\ 
07 & IRAS\,1313+2938 & 0.038 & 1.74 & 1.01 & 2.38 & 40.9 & 0.52 & 434  & 0.15 & 57.2 \\ 
08 & \nodata         & 0.038 & 9.76 & 1.00 & 4.32 & 22.3 & 1.50 & 96.7 & 0.29 & 15.0 \\ 
09 & IRAS\,0820+2816 & 0.047 & 10.1 & 0.33 & 9.47 & 3.31 & 2.61 & 247  & 0.62 & 40.5 \\ 
10 & Mrk\,0061	     & 0.057 & 5.28 & 0.16 & 1.70 & 8.90 & 0.34 & 85.1 & 0.08 & 14.9 \\ 
11 & \nodata         & 0.084 & 16.0 & 1.20 & 15.0 & 7.38 & 1.66 & 65.3 & 0.29 & 10.7 \\ 
12 & SBS\,0934+547   & 0.102 & 3.98 & 0.93 & 9.98 & 8.49 & 1.96 & 418  & 0.54 & 33.2 \\ 
13 & IRAS\,0147+1254 & 0.147 & 7.94 & 0.72 & 10.4 & 6.06 & 2.46 & 195  & 0.63 & 26.0 \\ 
14 & \nodata         & 0.181 & 1.62 & 4.46 & 9.60 & 39.4 & 1.99 & 947  & 0.94 & 120.
\enddata
\tablecomments{Redshifts (col 3) are derived from SDSS. 
All quantities are computed within $2\times r_\mathrm{P20}$ light radii 
in the UV, which are given in col (4). Equivalent widths are given in 
the restframe. No quantities are corrected for extinction.}
\end{deluxetable*}

\begin{deluxetable*}{cccccccccccc}[t!]
\tabletypesize{\scriptsize}
\tablecaption{Inferred properties of LARS galaxies.\label{tab:infprop}}
\tablehead{
\colhead{LARS} & 
		\colhead{$E_{B-V}^{\mathrm {neb.}}$} & 
		\colhead{$E_{B-V}^{\mathrm {stel.}}$} & 
		\colhead{Age} & 
		\colhead{Mass} & 
		\colhead{\lya/\halpha} & 
		\colhead{\halpha/\hbeta} & 
		\colhead{\fesclya} & 
		\colhead{SFR$^{\mathrm{H}\alpha}$} &
		\colhead{SFR$^{\mathrm{FUV}}$} &
		\colhead{SFR$_\mathrm{corr.}^{\mathrm{H}\alpha}$} &
		\colhead{SFR$_\mathrm{corr.}^{\mathrm{FUV}}$} \\
\colhead{ID} & 
		\colhead{mag} & 
		\colhead{mag} &  
		\colhead{Myr} & 
		\colhead{$10^9$M$_\odot$} & 
		\colhead{} & 
		\colhead{} & 
		\colhead{} & 
		\colhead{\msunyr} & 
		\colhead{\msunyr} & 
		\colhead{\msunyr} & 
		\colhead{\msunyr} \\
\colhead{(1)} & 
		\colhead{(2)} & 
		\colhead{(3)} &  
		\colhead{(4)} & 
		\colhead{(5)} & 
		\colhead{(6)} & 
		\colhead{(7)} & 
		\colhead{(8)} & 
		\colhead{(9)} & 
		\colhead{(10)} & 
		\colhead{(11)} &
		\colhead{(12)} } 
\startdata
01 & 0.112 & 0.019 & 7.14 & 6.10 & 1.36   & 3.20 & 0.119  & 4.97 &  2.48  & 6.52 &  3.14 \\ 
02 & 0.000 & 0.005 & 7.95 & 2.35 & 4.53   & 2.77 & 0.521  & 1.41 &  0.95  & 1.41 &  1.01 \\ 
03 & 0.717 & 0.115 & 23.4 & 20.1 & 0.16   & 5.85 & 0.003  & 4.65 &  0.56  & 26.3 &  2.41 \\ 
04 & 0.106 & 0.010 & 13.5 & 12.9 & 0.00   & 3.18 & 0.000  & 4.74 &  2.97  & 6.13 &  3.37 \\ 
05 & 0.148 & 0.007 & 4.71 & 4.27 & 2.16   & 3.32 & 0.174  & 4.06 &  2.94  & 5.81 &  3.23 \\ 
06 & 0.000 & 0.013 & 10.5 & 2.09 & 0.00   & 2.50 & 0.000  & 0.61 &  0.52  & 0.61 &  0.62 \\ 
07 & 0.336 & 0.009 & 7.78 & 4.75 & 1.94   & 4.00 & 0.100  & 4.12 &  2.34  & 9.27 &  2.61 \\ 
08 & 0.469 & 0.058 & 40.4 & 93.3 & 0.67   & 4.57 & 0.025  & 11.8 &  4.24  & 36.8 &  8.81 \\ 
09 & 0.281 & 0.038 & 11.3 & 51.0 & 0.13   & 3.79 & 0.007  & 20.6 &  9.30  & 40.7 &  15.0 \\ 
10 & 0.302 & 0.025 & 39.7 & 21.5 & 0.47   & 3.87 & 0.026  & 2.69 &  1.67  & 5.59 &  2.29 \\ 
11 & 0.348 & 0.033 & 25.6 & 121  & 0.72   & 4.05 & 0.036  & 13.1 &  14.7  & 30.5 &  22.3 \\ 
12 & 0.759 & 0.024 & 7.44 & 7.41 & 0.48   & 6.10 & 0.009  & 15.5 &  9.80  & 97.0 &  13.3 \\ 
13 & 0.497 & 0.048 & 7.19 & 59.2 & 0.29   & 4.70 & 0.010  & 19.5 &  10.2  & 64.8 &  18.8 \\ 
14 & 0.187 & 0.012 & 3.21 & 1.75 & 2.24   & 3.45 & 0.163  & 15.8 &  9.43  & 24.8 &  11.0    
\enddata
\tablecomments{Properties are derived inside the same apertures as 
those given in Table~\ref{tab:lumew}. (2) Nebular \ebv\ is estimated 
from \halpha/\hbeta\ assuming an intrinsic ratio of 2.86 for 10,000~K 
gas. (3 and 4) Stellar \ebv\ and age are estimated from a flux-weighted 
SED fit to the stellar continuum. (5) Mass refers to the stellar mass 
derived by resolved SED fitting.  (9--12) SFRs are calculated from the 
calibration of \citet{Kennicutt1998} and are compiled for comparative
purposes.}
\end{deluxetable*}

\begin{figure*}[t]
\centering 
\includegraphics[scale=0.35]{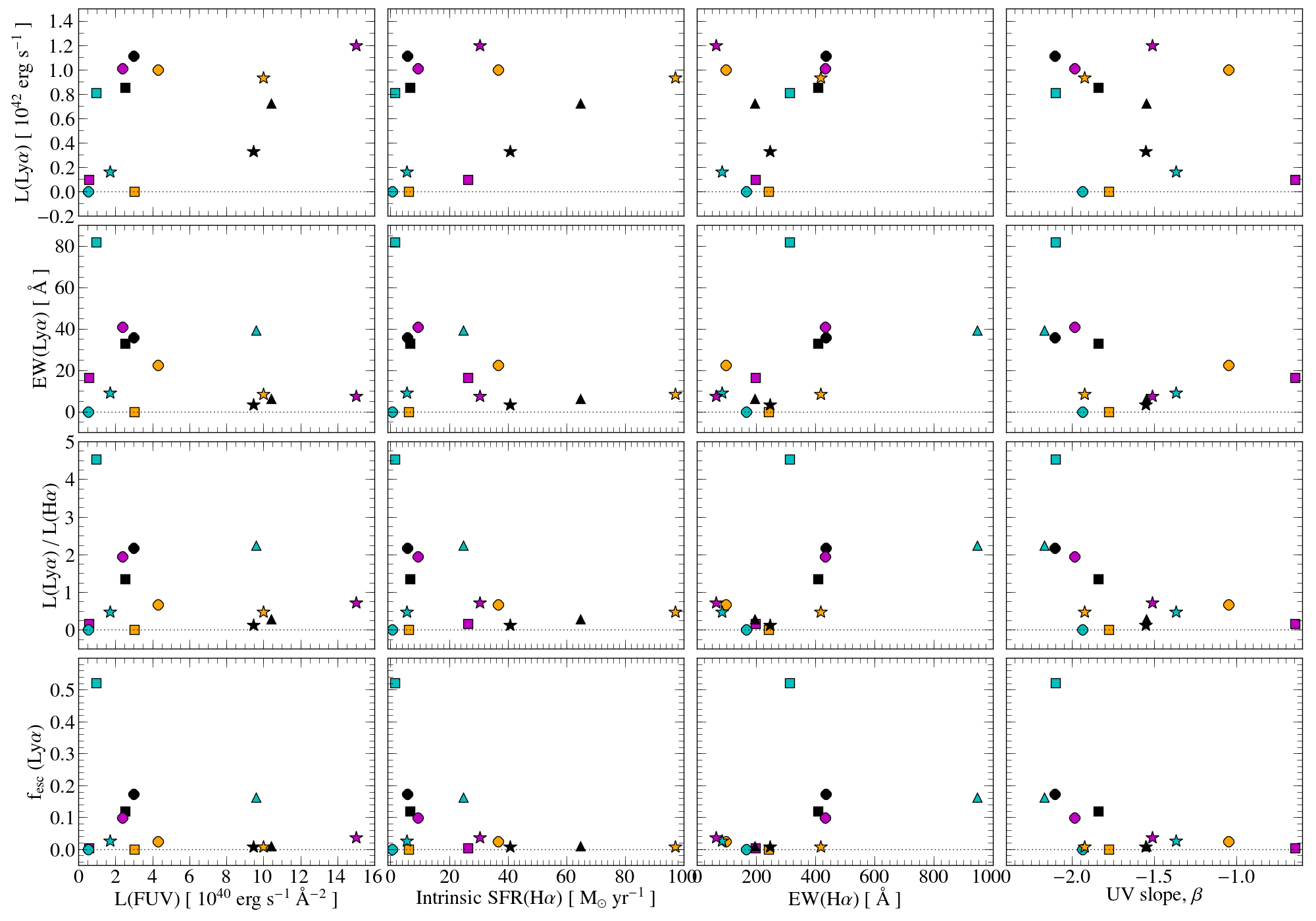}
\caption{Possible correlations between `global' \lya\ properties and 
quantities that can be derived from our HST imaging data. From 
{\em Top} to {\em Bottom} we show \lya\ luminosity, rest-frame EW, 
\lya/\halpha, and \fesclya\ on the ordinate axis. These quantities 
have not been corrected for extinction.  The \emph{upper} plots of 
\lya\ luminosity do not show LARS\,14, which is four times more 
luminous than the next brightest system.  From {\em Left} to 
{\em Right} we show: FUV luminosity, intrinsic (dust-corrected) 
instantaneous SFR derived from \halpha\ and \hbeta, \halpha\ equivalent 
width, and the UV continuum slope $\beta$. 
Symbols are the same as in Figure~\ref{fig:growcurve}.  } 
\label{fig:corrhst}
\end{figure*}

\begin{figure*}[t]
\ContinuedFloat
\centering 
\includegraphics[scale=0.35]{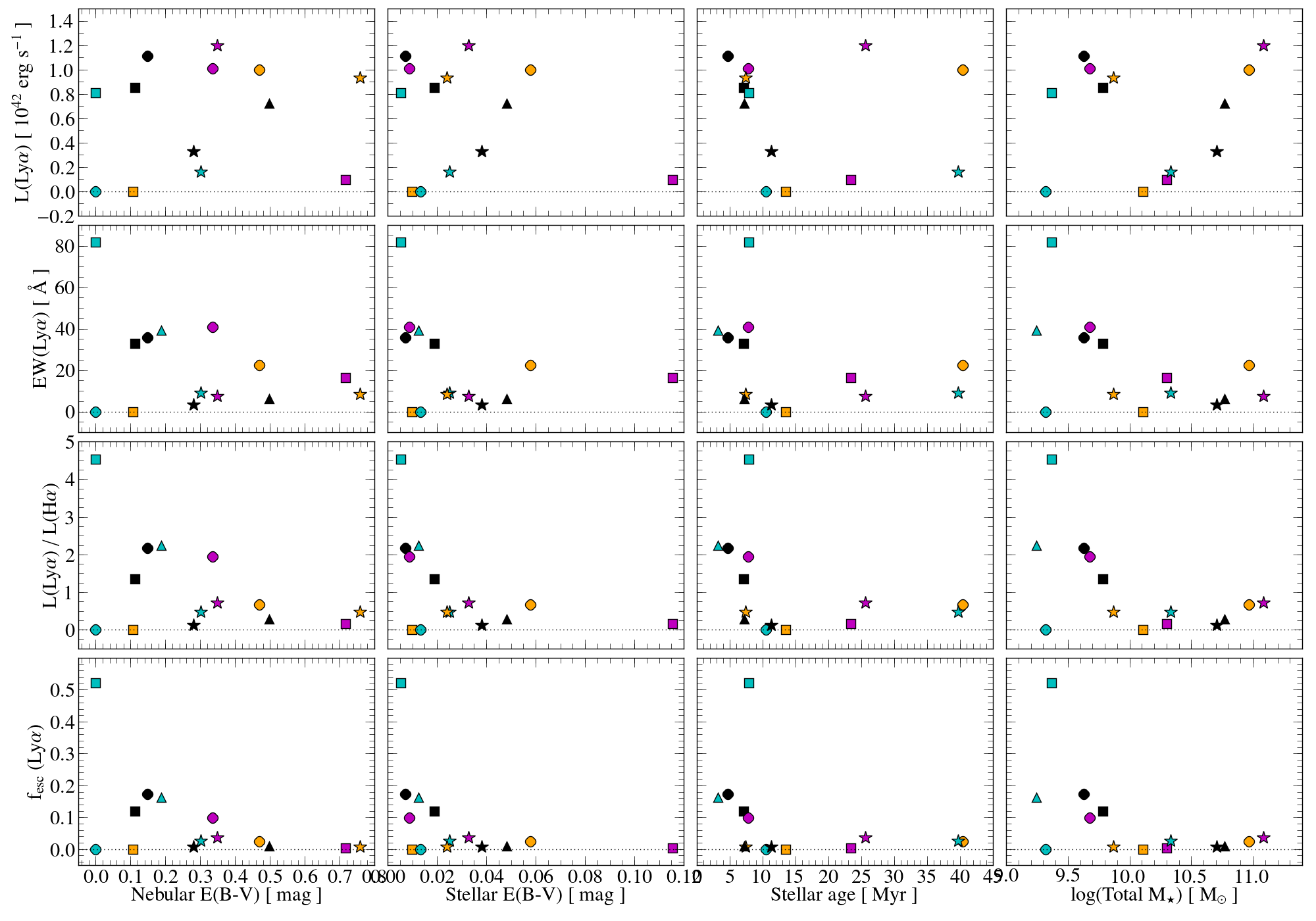}
\caption{ {\bf --- continued ---} 
From {\em Left} to {\em Right}: nebular \ebv\ derived from 
\halpha/\hbeta, stellar \ebv\ derived from the SED fitting, the 
luminosity-weighted stellar age of the burst population derived also 
from the SED, and the total stellar mass (summed over underlying and 
starburst components). } 
\end{figure*}

\begin{figure*}[t]
\centering 
\includegraphics[scale=0.35]{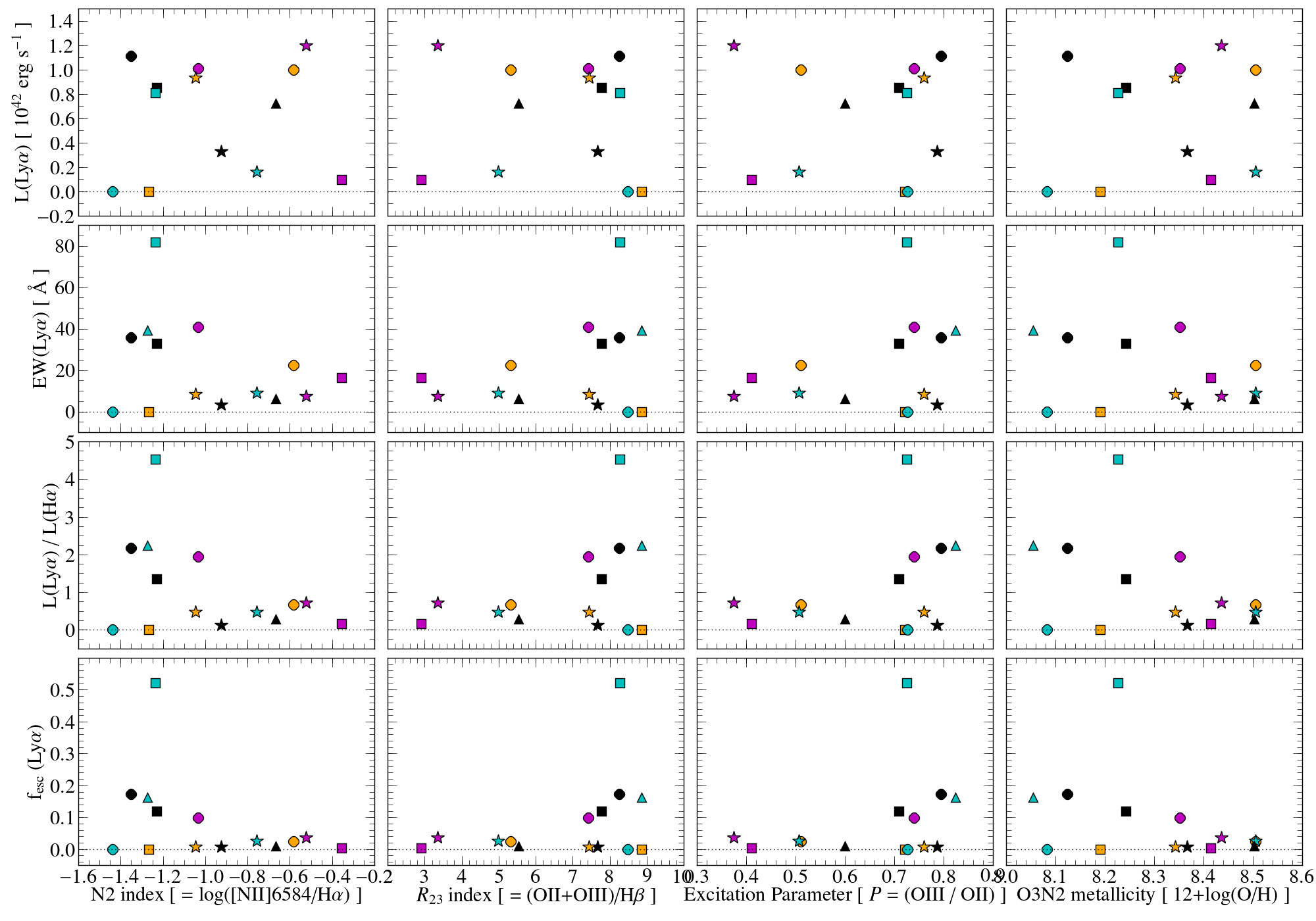}
\caption{Same as Figure~\ref{fig:corrhst} but with properties derived 
from SDSS spectroscopy. As such the quantities have not been 
derived in the same apertures as the information from which the \lya\
quantities have been derived. From {\em Left} to {\em Right} we show:
N2 index, $R_{23}$ index, excitation parameter, and metallicity $Z$ 
computed from the O3N2 index.
Symbols are the same as in Figure~\ref{fig:growcurve}.}
\label{fig:corrsdss}
\end{figure*}

We report a number of observable quantities of the emission lines and 
adjacent continuum in Table~\ref{tab:lumew}. Further quantities derived
from these line strengths and/or SED fitting (same apertures) are 
reported in Table~\ref{tab:infprop}. In images there are various ways of 
computing these integrated quantities. In this paper, fluxes, flux 
densities, and stellar masses are computed by simple summation, and
wherever a ratio of two fluxes is presented the quantities are summed 
individually before dividing. Stellar attenuation and ages are computed 
from the two-dimensional images by taking a flux-weighted average 
(see below). We have also performed total photometry inside the same 
apertures in the raw HST images and run {\tt LaXs} on these fluxes, 
verifying that results are highly consistent. Quantities derived from
the SDSS spectra can be found in Paper I. 

Graphically we elect to compile a number of observable/derivable 
quantities of \lya, that are frequently discussed. These include 
\llya, \ewlya, \lya/\halpha, and \fesclya, which are presented in 
Figure~\ref{fig:corrhst} against HST aperture-derived quantities, 
and in Figure~\ref{fig:corrsdss} against quantities derived from 
optical spectroscopy of the central regions. We compare with the 
following:
\begin{itemize}
\item Far UV continuum luminosity near \lya, \lfuv. \lfuv\ is derived 
from the modeled image of the stellar continuum in the filter that 
transmits \lya\ (F125LP or F140LP). Thus it represents a very small 
extrapolation from the adjacent filter that samples the continuum alone
(F140LP or F150LP). In the unobscured case, \lfuv\ traces the 
average star formation rate over the last few hundred Myr and for 
completeness we also compile the FUV SFR using the 
\citet{Kennicutt1998} calibration, and correct this for dust using the
attenuation measured on the stellar continuum. The FUV 
magnitude, or its corresponding SFR, are among the most frequently 
presented quantities in high-$z$ studies.

\item Instantaneous SFR, SFR$_{\mathrm{H}\alpha}$. This is derived from
the dust-corrected \halpha\ luminosity and adopting the calibration of 
\citet{Kennicutt1998}, which assumes a Salpeter initial mass function
between limits of 0.1 and 100~\msun.  The dust correction is done by 
adopting the \ebv\ measured from the \halpha/\hbeta\ ratio measured in
the same apertures.  The \halpha\ luminosity is also corrected for 
\nII$\lambda\lambda6548$,6584\AA\ emission (allowing for narrowband 
transmission level) using the \nII/\halpha\ ratio measured in the 
SDSS spectrum, and also underlying Balmer absorption from stars (see 
Section~\ref{sect:data}).

\item \halpha\ equivalent width, \ewha. This is computed by by simply 
dividing the continuum-subtracted \halpha\ flux by the total modeled 
continuum in the narrowband filter. 

\item Dust attenuation measured in the nebular gas, \ebvneb. This is 
estimated from \halpha/\hbeta, using the SMC extinction law 
\citep{Prevot1984}. Both lines have been corrected for underlying 
stellar absorption, \halpha\ is corrected for \nII, and the quantity
is fed back to the dust-correction of \halpha\ and computation of 
\fesclya\ (above). 

\item Dust attenuation measured in the stellar population, \ebvstel\ 
from SED fitting. This map is returned by the {\tt LaXs} software on a 
pixel-by-pixel basis. For the computation of the global value we 
compute a flux-weighted average taking the $I-$band for 
the weighting bandpass: $\sum$\ebvstel$L_I$/ $\sum L_I$. Adopting other
filters for weighting has negligible effect on our results. 

\item Age of the stellar population. This is also output by {\tt LaXs}
as a 2D map, and the average is computed in the same way as for \ebvstel.

\item Mass of the stellar population. This is also output by {\tt LaXs}
as a 2D map, and the sum is computed within the apertures. 

\item Nebular indices and metallicity,  $Z$. These properties are 
derived from the SDSS spectra. Because 
the SDSS fibers are of constant size (a standard 3 arcsec diameter 
circle) and the defined HST apertures are variable, these apertures are 
not matched. Nonetheless each one is, or is derived from, a flux 
ratio, and they will provide a reasonable picture of the average 
oxygen abundance/excitation state in the galaxies and are at least 
consistent for every object. Because high-$z$ observations of these 
restframe optical lines are emerging in a piecemeal fashion, we elect
to show: the $R_{23}$ index 
[$\equiv$ (\oII3727+\oIII4959+\oIII5007)/\hbeta]; the N2 index 
[$\equiv$ log( \nII6584/\halpha)]; the excitation parameter, $P$ 
[$\equiv$ \oIII5007/\oII3727]. For a standard strong-line metallicity, 
we adopt the O3N2 method and the empirical calibration of 
\citet{Yin2007}. For about half the sample we can make direct 
$T_\mathrm{e}-$based estimates of the metallicity (Paper I) but as we 
cannot do so for the whole sample, we compare only strong line 
quantities, admitting the systematic uncertainties. 

\end{itemize}

The LARS sample comprises only 14 galaxies, and thus does not reach a 
high level of statistical significance. Nevertheless in sample size it 
is a three-fold improvement upon the existing body of work on low-$z$ 
\lya\ imaging \citep{Ostlin2009}, and the sample is no longer  biased
by the selection of historically interesting `favorites' of the local 
starburst community. With global photometry in the UV and \lya\ we are 
first able to compare our sample with high-$z$ UV- and \lya-selected 
samples, which we will do in Section~\ref{sect:select}. 
Furthermore with a larger number of galaxies than previously studied
and many aperture-matched measurements, we may examine trends between 
\lya\ output and various properties of the galaxies; this we do in 
Section~\ref{sect:individual}. However, before discussing these results
we will comment upon the individual galaxies.

\section{Comments Upon the Individual Targets}\label{sect:individual}

\paragraph{LARS\,01} LARS\,01 (Markarian\,259) is studied in 
detail in Paper I and also enters the sample of \citet{Wofford2013},
so the reader is directed to these publications for 
more extensive information. It has a metallicity of 
$\approx 0.25Z_\odot$ (Paper I; \citealt{Asplund2009}), 
$\sim 0.1$ magnitudes of differential extinction (\ebv) in the gas
phases, has a modest SFR of a few \msunyr, and strong \halpha\ emission
with an EW of $\sim 400$\AA. By these measures it is roughly comparable 
to ESO\,338--IG04 \citep{Hayes2005,Atek2008,Ostlin2009}. Indeed it is a
strong \lya\ emitting galaxy: with an aperture EW of 35\AA\ and 
$>50$\AA\ globally, it would be likely selected as a high-$z$ LAE; a 
luminosity of almost 
$10^{42}$~\ergsec\ suggests that it could be detected by moderately 
deep surveys. Morphologically, it consists of a bright UV star-forming 
center, with an extended tail to the SW: there is a bright hot-spot of 
\lya\ emission that emanates from the major star-forming condensation, 
and flows out in a diffuse fan-like structure preferentially in the NE
direction. In this region the \lya\ EW and \lya/\halpha\ ratio exceed 
500\AA\ and 30, respectively. Such ratios would indicate substantial 
re-scattering of photons into the line-of-sight. Interestingly there is
evidence for nebular 
filamentary structure in both \halpha\ and \hbeta\ that extends in this
direction, which may indicate a bubble that expanded in
this direction before blowing out. This would certainly be suggestive of a
strong outflow of hot gas, which could also mitigate the trapping of 
\lya\ photons and enable them to flow in this direction. The \lya\
curves-of-growth, are all continually rising out to a radius of 
$\approx 8$~kpc and \fesclya\ does not converge: it is likely that the 
total \lya\ output from LARS\,01 is somewhat higher than presented.

\paragraph{LARS\,02} This galaxy shows a metallicity and \halpha\ EW 
roughly similar to LARS\,01, however with a SFR of around 1\msunyr\ it
is among the more dwarf-like objects in our sample. It is 
largely dust-free, and according to all the \lya-related quantities is 
the strongest emitting LARS galaxy: \lya\ EW almost 100\AA, 
\lya/\halpha$=4.5$, \fesclya$=52$~\% inside the 
$2\times r_\mathrm{P20}$ radius. 
Here again we see that the \lya\ EW and \lya/\halpha\ 
ratio locally exceed their recombination values by large factors on kpc
scales, and the curves-of-growth for all the \lya\ quantities rise 
very rapidly. Furthermore, this photometric 
curve-of-growth is rising so steeply that this galaxy emits 75~\% of its 
\lya\ photons by a radius of 3.5~kpc.  
Such high levels of relative \lya\ emission have not been observed before
in targeted observations of local galaxies, although some \lya-selected
nearby galaxies have been seen to exhibit \fesclya\ higher than 
would be expected for their measured attenuations \citep{Atek2009galex}. It
is plausible, although not required, that some mechanical enhancement 
of \lya\ could produce such \fesclya, or that we are either 
underestimating the intrinsic luminosity in \halpha\ (or its dust 
correction by overestimating \hbeta). Future X-ray observations could 
most likely provide the answer. 
LARS\,02 would be selected as \lya\ emitter in deep high-$z$ surveys. 

\paragraph{LARS\,03} Arp\,238 is among the most well-studied 
galaxies in
the LARS sample. The galaxy comprises two merging 
nuclei and with ACS/SBC we were able to point at just one of the two 
cores. Our observation therefore covers only the south-eastern nucleus.
It is a luminous infrared galaxy (LIRG) and with a 
$\log (L_\mathrm{FIR}/L_\odot) = 11.8$ is close to the domain of ultra
luminous infrared galaxies (ULIRGs). Such IR luminous local galaxies 
have never before been imaged in \lya; the most similar galaxy so far 
being NGC\,6090 \citep{Ostlin2009}, which is 0.4 dex fainter in 
$L_\mathrm{FIR}$.   Arp\,238 does show \lya\ in 
emission with an EW of $\sim 16$\AA\ at twice $r_\mathrm{P20}$ and this 
rises to $\approx 40$\AA\ when large apertures are considered. 
Thus it passes the canonical 
definition for high-$z$ narrowband--broadband color selection. However 
the luminosity and escape fraction are small (\fesclya $\approx 1$~\%) and 
it is the simultaneous 
large extinction on the UV continuum that results in the high measured
EW. This suggests that \lya\ and the UV continuum are suppressed by 
roughly the same fraction.
Within the spatial extents that we can probe with the SBC, the \lya\ 
luminosity is low enough for LARS\,03 to evade detection in all but the 
deepest high-$z$ surveys, but the \lya\ luminosity does continue to
increase with radius out to 10~kpc; it is unclear what happens at 
larger radii.

\paragraph{LARS\,04} LARS\,04 shows a highly irregular UV morphology comprising 
an apparent bar and extended tail to the west. \lya\ is emitted on 
small scales throughout all these regions, but in complete annuli a net 
absorption is found at all radii. Maps show that even on small scales the 
\lya\ EW never exceeds expectations and no large scale component of 
emission is apparent -- between the UV bright regions, where diffuse 
emission should easily be detectable, the continuum-subtracted \lya\ 
flux negative, suggesting much neutral gas absorption is ongoing and 
if \lya\ is emitted it must be on very large scales indeed. In terms 
of SFR and dust content it is quite similar to LARS\,01, which emits
copious \lya. 

\paragraph{LARS\,05} Mrk\,1486 is blue highly inclined 
(edge-on) disk galaxy with SFR of around 2~\msunyr. In 
the very central regions -- along the plane of the disk -- \lya\ 
absorption of the stellar continuum is quite strong, and cancels much 
of the emission, resulting in a narrow absorbing band. However, this 
gives way to emission at small projected distances and \lya\ luminosity 
and EW rise rapidly; EW$>500$~\AA\ is reached at a projected distance of 
2.5~kpc from the disk. The large-scale morphology is one of an 
azimuthally symmetric and featureless halo of emission, although 
the aperture growth-curve in \lya\ appears to flatten at a radius of 
4~kpc which would indicate that we are able to capture most of the 
\lya\ emission. At brighter \lya\ isophotes, however, 
the \lya\ surface resembles more more the shape of the UV disk where
\halpha\ filamentary structures that extend in the polar direction are 
visible. This is reminiscent of outflowing gas that has accelerated and 
become Rayleigh-Taylor unstable \citep{MacLow1988,Cooper2008}, and could
explain the \lya\ enhancement in the same directions by permitting 
radiation to stream in this direction before scattering in the halo, 
similar to the bubble to the NE of LARS\,01.  Inside the $2\times r_\mathrm{P20}$ radius 
(1.9~kpc) it has a \lya\ 
EW of 36~\AA, and its total EW of $\sim 45$\AA\ is reached by 
$\approx 4$~kpc. By EW and luminosity LARS\,05
would be selected in deeper surveys as a \lya-emitting galaxy. 

\paragraph{LARS\,06} KISSR\,2019 is the most feebly star-forming
galaxy in the sample by both FUV and \halpha\ measures, and also
exhibits the lowest nebular metallicity ($12+\log(\mathrm{O/H}) =8.08$;
Paper I). It is effectively unreddened according to
\hbeta; the observation of which is of insufficient quality to trace
the extinction over some of the outer of regions where we detect
\halpha.  In the UV and \halpha\ it reveals a morphology
comprising one major star-forming condensation around which no hint of
\lya\ emission is seen. Most wavelengths also reveal a tail of
stars extending to the south, although there is no evidence for this
in \lya\ either. Interestingly the tail is also rather weak in \halpha, 
which suggests that the stars of which it is made up have evolved to
the stage where they remain  UV-bright, but are not hot enough to 
produce strong nebular lines; this may be evidenced also by its age 
of $\approx 10$~Myr.
The galaxy shows strong \lya\ absorption and an almost
complete absence of emission over the surface. 

\paragraph{LARS\,07} Also known as IRAS\,1313+2938 and KISSR\,242, LARS\,07 has already 
been studied in \lya\ spectroscopically with HST/COS 
\citep{France2010,Wofford2013}. Morphologically LARS\,07 resembles LARS\,05 as an
edge-on disk and very compact in \halpha\ and the stellar continuum, 
but there is evidence of some faint nebular gas extending away from the
disk that is also somewhat extinguished as evidenced by \halpha/\hbeta. 
It
also shows a large-scale and low surface brightness \lya\ halo although
this halo is somewhat extended in the direction of the disk. 
Like LARS05 it also shows filamentary structure in \halpha\ that extends 
away from the plane of the galaxy in the polar directions, again reminiscent
of a RT unstable outflow. 
Very high local EW and \lya/\halpha\ are clear in these regions. In \lya\ its 
growth curves follow those of LARS\,01 and 05 closely and it seems that 
we have recovered almost all the \lya\ photons at radii of 4~kpc. 
LARS\,07 is the among the more \lya-luminous galaxies in
the sample, and with a EW of $\approx 50$~\AA\ should be detected in 
most surveys. 

\paragraph{LARS\,08} This is a face-on disk-like galaxy with a heavily 
obscured nucleus, and shows the highest nebular metallicity in the
sample according to strong-line diagnostics (Paper I).  It shows numerous patches of 
UV bright star-forming regions in the disk, that appears to be more 
extended to the west. \lya\ emission is seen coming from the surface of
this disk, but it is not strong compared to the FUV or \halpha, with no
significant super-recombination ratios visible at any position. 
Notably there is no sign of \lya\ emission emanating from the nuclear 
regions, but instead \lya\ is seen only from the regions that are 
brighter in the FUV, possibly as a correlation with the dust content. 
LARS\,08 does show an ever increasing photometric curve-of-growth in its 
luminosity, but this does not translate into its EW or \fesclya, which
both flatten at smaller radii of about 2~kpc. This suggests that
\lya, FUV and \halpha\ all increase roughly similar ways with radius, 
at least for radii that we can probe.  We caution also that this
is one of the two most extended galaxies in the sample, so is likely 
also to be the object in which we have least capability to probe the 
fainter regions. It could be considered a high-$z$ \lya-emitting analog
galaxy by virtue of its luminosity and EW. 

\paragraph{LARS\,09} IRAS\,0820+2816 is a very extended 
late-type starbursting spiral. It is also an NVSS and 5C radio source.
It emits \lya, but at the brightest UV levels is very faint, showing 
mostly \lya\ absorption in the nucleus. A fuzz of \lya\ emission is 
clearly visible but does not behave as an extended halo, and instead 
broadly traces out the optical structure. Its \llya\ curve-of-growth is
remarkable, showing \lya\ in absorption in the central regions, that
gets stronger out to radii of $\sim 3$~kpc. After this radius, 
the curve increases very steeply until the object becomes a global 
\lya\ emitter at $r\sim 9$~kpc.  Only
The growth of EW and integrated \fesclya\ with radius are close to 
straight, linear growths with radius and both grow slowly. 
Inside the $2\times r_\mathrm{P20}$ aperture, it is a net emitter, but 
not a luminous one and it shows a low \ewlya\ of 3\AA\ (and below 15\AA)
at all radii. Note that a field star, coincident with the edge of the 
southern arm as probed by the optical images,  has been masked in 
Figures~\ref{fig:maps} and \ref{fig:ratios}.

\paragraph{LARS\,10} Mrk\,61 has the optical morphology of the
later stage of a merger, showing a UV-bright core that extends over 
$\approx 5$~kpc and an extended tail to the south-east. 
In the central few kpc, it shows a patchy, fine-scale mixture of \lya\ 
emission and absorption, but absorbs in total within $r=2$~kpc.
In this region, \lya\ EWs do not exceed around 50~\AA, and 
\lya/\halpha\ ratios are found in general not to suggest re-scattering 
of photons in this galaxy -- inside $2\times r_\mathrm{P20}$, this 
object shows \ebvneb$\approx 0.3$, and it seems likely that \lya\ 
photons can not escape the central regions. Some re-scattering of \lya\
does occur, however, and at $r \gtrsim 5$~kpc a faint diffuse halo 
structure begins to emerge; the net \ewlya\ within this radius is 
sufficient to LARS\,10 a LAE analog. The rather smooth, flat, curves-of-growth 
suggest that much of the \lya\ has been captured by 
roughly 12~kpc. The galaxy does have quite a high \lya\ EW 
($\approx 30$\AA) but is not luminous enough 
(\llya$\lesssim2\times 10^{41}$~\ergsec) 
to be detected in \lya\ at high redshift 

\paragraph{LARS\,11} This is a highly inclined edge-on disk that 
extends over 40~kpc in the ultraviolet. The whole disk appears to be 
lit with ongoing star formation, as shown by the \halpha\ images; the
total SFR is on the order of 20--30~\msunyr, and its stellar mass 
is the largest in the sample. 
Projected on the sky with a major axis that runs SE to NW, the SE 
half is somewhat more active in star formation. None of the brighter
star-forming regions appear to emit their \lya\ directly, and instead a
faint rim of \lya\ emission is seen running along the lower (more 
southerly) edge of the disk. This may hint at some projection effect 
where we see only \lya\ from the near side.  The \lya\ is clearly 
projected away from the UV disk, and equivalent widths are relatively
high, on the order of 100~\AA. However the \lya/\halpha\ ratios in the
same resolution elements are around 3--5 at most, which would 
not be indicative of strong re-scattering of photons by extended gas 
along this plane of the disk. Globally we find LARS\,11 to be a weakly 
emitting galaxy in \lya\ (EW$\approx 7$\AA) with curves-of-growth that
increase slowly; the total EW is close to 20\AA, and it probably would
become a LAE analog if a larger FOV could be observed.

\paragraph{LARS\,12} SBS\,0934+547  is a UV-bright merging 
system. In front of the brightest star-forming regions it shows strong 
\lya\ absorption. However this changes to emission at $r>2$~kpc, and 
the aperture-integrated \lya\ measurements all increase rapidly 
out to $\approx 10$kpc, after which it roughly converges; 
a diffuse \lya\ halo can be seen in the images 
extending over roughly this area. While very luminous in net \lya, 
the EW never exceeds about 15~\AA\ (5~\AA\ at $2\times r_\mathrm{P20}$) 
and LARS\,12 would not be selected as a LAE if projected to the 
high-$z$ universe. 

\paragraph{LARS\,13} IRAS\,0147+1254 is also a Lyman-break 
analog system \citep{Hoopes2007}, and is also clearly a merging system.
It shows variable dust attenuations in the nebular phase which is 
strongly variable and scattered across the face of the object. \lya\
emission is also rather patchy and varies between absorption and weak
emission.  Unlike LARS\,12, however, this emission never becomes 
particularly strong and LARS\,13 has no substantial halo emission
component. 
Overall, LARS\,13 emits more \lya\ than it 
absorbs, but has both small EW (6 \AA), and \fesclya\ (1 per cent)
in the $2\times r_\mathrm{P20}$ aperture. 

\paragraph{LARS\,14} LARS\,14 is among the most UV luminous galaxies in the 
sample but has a remarkably low metallicity 
($12+\log(\mathrm{O/H}) = 7.8$ from the electron temperature method). 
It also has a moderately high SFR ($\approx 15$~\msunyr), but the 
current episode of star formation is also rather low-mass, making 
LARS\,14 by far the highest specific SFR galaxy in the sample.
Its extremely compact morphology and strong
oxygen lines classify it as a green pea galaxy \citep{Cardamone2009}. 
Diagnostics of \oIII/\hbeta\ vs. \oI/\halpha\ and 
\sII/\halpha\ place it right on the delimiting line between 
starburst and Seyfert systems, however, and inferred gas pressures in 
the central regions are found to be several orders of magnitude higher 
than in `ordinary' starbursts \citep{Overzier2008,Overzier2009}. 
In the \lya\ images it 
shows a very bright emission, centered upon the brightest
star-forming core, and a featureless, symmetric halo of surrounding 
\lya. In no region does it show \lya\ in absorption. Even in the 
central regions it shows \lya\ EW and \lya/\halpha\ that are close to
the recombination values. Inside the $2\times r_\mathrm{P20}$ aperture 
it shows $\approx 4$ times the \lya\ luminosity of the next most 
luminous \lya\ emitter.  LARS\,14 emits more \lya\ photons than the rest 
of the sample combined. Note that in Figure~\ref{fig:growcurve} its luminosity 
has been divided by 10 so that it may be visualized. 

\section{Comparison of LARS with high-redshift UV selections}\label{sect:select}

We now proceed to discuss how the objects in our sample at 
$0.028 < z <0.18$ would be recovered by current high-$z$ ($\gtrsim 2$)
galaxy surveys. 

\subsection{UV continuum selection} 

Based upon their FUV luminosities alone, our galaxies have FUV 
($\lambda\sim 1400$\AA) absolute magnitudes ranging between 
--16.8 and --20.3. For reference the faintest bins in the UV continuum
($\lambda\sim 1700$\AA) LF of \citet{Reddy2009} $z \sim 2$ \emph{BM/BX}
galaxies and $z \sim 3$ LBGs are centered around AB$=-18.3$. Ten of the 14 
LARS galaxies are bright enough to have been detected by these surveys.
The brightest LARS galaxy is just 0.4 magnitudes fainter than \mstar\ 
as determined by the same $z\sim 2$ continuum surveys. However, given 
the evolving LF at higher redshift, the brightest object corresponds to
approximately \mstar\ at $z\approx 6$ and exceeds \mstar\ at $z\sim 7$ 
and 8 \citep{Bouwens2011}. 

\subsection{Lyman-alpha selection} 

Making a direct comparison of our sample against high-$z$ \lya\ 
selection is less straightforward because both flux and EW criteria 
need to be fulfilled. Assuming the canonical EW cut of 20\AA, within 
the chosen apertures, LARS contains six LAE analog galaxies 
(01, 02, 05, 07, 08, and 14). Every one of these objects is 
sufficiently compact (five have aperture radii here between 1.7 and 
2.4~kpc, and only LARS\,08 is notably larger) that this 
emission would be recovered by photometry in apertures
of 1.5 arcsec radius if they were at $z> 2$. We refer to this sub-sample of six galaxies as 
\emph{EW20LAEs}. We note also that among our six \emph{EW20LAEs}, five
have \fesclya\ that exceeds 10~\%, whereas one exhibits a notably 
lower \fesclya\ of just 2.5~\% (LARS\,08). Were we to cut the sample 
by galaxies with \fesclya$>10$~\% we would retain only these five 
galaxies -- there are no high EW galaxies with low \fesclya, which in 
principle could be produced by a simple dust screen. 
We refer to the subsample with \fesclya$>10$~\% as the 
\emph{FESC10LAEs}.

Whether a galaxy would be found as a LAE depends also upon its 
flux and the design of the observation, so contrasting LARS with 
high-$z$ \lya\ surveys will be strongly dependent upon the assumed 
observational parameters and redshift.
At $\langle z\rangle= 0.3$ in the GALEX-selected LAE sample of 
\citet{Cowie2010}, the faintest object has 
\llya=$1.5\times 10^{41}$~\ergsec. LARS contains 10 galaxies brighter
than this in \lya, including all six of our \emph{EW20LAEs}.
The brightest (non AGN) LAE in \citet{Cowie2010} has 
\llya=$2.7\times 10^{42}$~\ergsec, around twice as bright as our brighter
objects LARS\,05 and 07, but only around half the luminosity of 
LARS\,14. 

At high-$z$ the extremely deep $2.67<z<3.75$ spectroscopic 
survey of \citet{Rauch2008} finds galaxies as faint as 
$10^{-18}$~\ergseccm\ in \lya, corresponding to a luminosity of 
$\approx 7\times 10^{40}$~\ergsec: the same LARS galaxies that would 
have been detected by \citet{Cowie2010} would also have been found by 
\citet{Rauch2008}. For a typical `deep' narrowband \lya\ survey we 
adopt the depth of our own $z\sim 2$ survey \citep{Hayes2010}, which 
reached \llya$\approx 3\times 10^{41}$~\ergsec\ -- we would have 
recovered all six of our \emph{EW20LAEs}. Reaching 
\llya=$1.3\times 10^{42}$~\ergsec, the survey of \citet{Guaita2010} 
would only recover LARS\,14 using our $2\times r_\mathrm{P20}$ 
apertures photometry. However extending these apertures to 8~kpc 
($\approx 1$ arcsec), \citet{Guaita2010} would also
recover LARS\,01, 02, and 05 from the \emph{EW20LAEs}. LARS\,12 is 
also sufficiently luminous, but does not have high enough EW to 
survive most survey cuts. 
LARS\,14 is around \lstar\ for $z=3.1$ LAEs 
\citep{Ouchi2008} and the next two LARS galaxies are a factor of 4 fainter. 
Even LARS\,14, however, is too faint to be detected by the $z=5.7$ and 6.5
survey of \citet{Hu2010}, especially given the uncertain impact that 
the IGM would have on the \lya\ throughput from these redshifts. 

When contrasting our results against high-$z$ surveys, however, note 
that one would also need to account for the fact that more neutral gas 
may reside around galaxies with increasing $z$. Consequently, the 
surface brightness may be even more extended, which may act to render a 
given galaxy even less detectable.

\section{What Governs \lya\ Transmission?}\label{sect:corr}

While the sample is small compared to a high-$z$ survey, it is large 
compared to previous low-$z$ HST \lya\ studies and is also far better
selected, and therefore gives us modest statistical power to examine 
correlation of global properties. The 
number of properties that may affect \lya\ emission is also large, 
and some studies (e.g. of neutral gas contents, kinematics and 
covering fractions) are to be deferred until future papers. The HST
imaging data alone, combined with the available smaller aperture 
optical spectroscopy from SDSS, are sufficient to derive a number of 
interesting quantities. 
Regarding \lya\ output, we elect to study four quantities: the 
luminosity, equivalent width, \lya/\halpha\ ratio, and the escape 
fraction \fesclya\ (Equation~\ref{eq:fesc}).  We remind the reader that
all our `global' quantities are computed in apertures of twice the 
Petrosian radius, with $\eta=0.2$. 

In Figures~\ref{fig:corrhst} and \ref{fig:corrsdss}, the 
uppermost  panels always show \lya\ luminosity on the ordinate axis. 
One of the most interesting features of this row of plots is that
in none of them does a systematic trend emerge between 
\llya\ and the quantity presented on the abscissa. Total \lya\ 
luminosity does not correlate with any other quantity that 
we have measured thus far. This in itself is a particularly
interesting result, since even though we know the transport of \lya\ to
be a complex process we would still have expected some degree of 
correlation with most basic quantities like FUV or \halpha\ luminosity,
over which quantities the LARS sample spans a dynamic range of 1.5 dex. 
The \lya\ numbers presented on the ordinate axes of other three rows
of Figures~\ref{fig:corrhst} and \ref{fig:corrsdss} are relative 
quantities -- \lya\ luminosity divided by FUV luminosity, \lya/\halpha,
and \fesclya\ (\lya/intrinsic \lya)  -- and as such these 
trace the 
transmission of \lya\ either absolutely (in the case of \fesclya) or 
relative to other wavelengths (\lya/\halpha\ or \lya\ EW). In many of
these plots a number of relationships and `zones of avoidance' 
begin to emerge. Furthermore they seem to strengthen in the order 
presented: zones of avoidance are better defined in \fesclya\ plots 
than \lya/\halpha\ plots, which in turn are better defined than those
in \ewlya\ plots.  We recall that \ewlya\ on any sightline is a 
function of \{SFH, extinction, scattering\}, and \lya/\halpha\ is 
more simply a function of \{extinction, scattering\}; in contrast 
\fesclya\ attempts to correct also for the obscuration of \halpha. 
Specifically we find that more \lya\ is transferred in galaxies with
lower \lfuv, lower instantaneous SFR, higher \halpha\ equivalent 
width, lower metallicity, lower dust attenuation, and at younger ages. 
Cautioning at the outset that many of these quantities will be correlated,
we now discuss each in turn.

\subsection{FUV luminosity}\label{sect:pars:lfuv} 

\lfuv\ is shown in the left-most column of 
Figure~\ref{fig:corrhst}.
This is a priority high-$z$ observable, and has been much studied 
previously and contrasted with \lya\ \citep{Gronwall2007,Ouchi2008}. 
\citet{Ando2006} presented a plot demonstrating the notable absence of 
strong \lya\ emission (EW) at higher UV luminosities, which has since 
been confirmed in large samples of LBGs \citep{Stark2010,Kornei2010} 
and LAEs \citep{Kashikawa2011} and discussed as a natural consequence 
of extinction increasing with UV magnitude in LBGs, couple with the
associated radiative transfer effects \citep{Verhamme2008}. 

A number of galaxies are clearly 
visible with high \ewlya\ but faint \lfuv, and five of our six
\emph{EW20LAEs} have \lfuv$\lesssim 4\times 10^{40}$\ergsecaa\ 
($M_{1500}=-19.5$ AB), while the sample of non-emitting galaxies extends 
to approximately four times this value. LARS\,14 -- a strongly
\lya\ emitting and UV luminous galaxy -- is the only object that bucks 
the trend somewhat. 
A very similar distribution of points is also present in the 
\lya/\halpha\ and \fesclya\ plots, which demonstrates that this effect 
is not purely one related to the star formation history, but 
is genuinely a result of the relative transmission of \lya. We note 
that the cutoff magnitude suggested by 
\citet{Ando2006} is 1.5 magnitudes brighter than that remarked upon
here, although given the strongly differing selection functions, which
include \ewha, and very different cosmic epochs, it is 
perhaps not surprising to see qualitative but not quantitative 
agreement.

\subsection{Star Formation Rate} \label{sect:corr:sfr}

FUV luminosity, as discussed in Section~\ref{sect:pars:lfuv}, traces 
the (possibly obscured) SFR averaged over the last $\approx 100$~Myr;
we now show the unobscured instantaneous SFR (second column of 
Figure~\ref{fig:corrhst}), which we derive from dust-corrected \halpha,
using the calibration of \citet{Kennicutt1998}. We note that some of 
the SFRs, even the dust corrected ones, listed in 
Table~\ref{tab:infprop} are rather discrepant, and we recall that we 
are observing galaxies that have experienced a burst of massive star 
formation, with a recent significant increase of their SFR.
Indeed note that the ages measured for the stellar population are 
generally very young, and mostly far from the ages at which the 
stationary level on which the SFR calibrations are based has been
reached.

When considering the \lya\ EW we see something that resembles
the \citet{Ando2006} effect, but with \lfuv\ replaced by \halpha\ SFR, as 
expected from the models of \citet{Garel2012}.
Indeed the effect is even more prominent than the comparison against 
\lfuv, with LARS\,14 falling nearer the strongly \lya-emitting but 
less intensely star-forming quadrant. It is clear that higher SFR 
galaxies do not strongly emit \lya: all the \emph{FESC10LAEs} exhibit
SFRs below 30~\msunyr, and galaxies with higher SFR do not emit
much \lya. The upper right part of the diagram
(high SFR, high \fesclya\ or \ewlya) is completely avoided. 

\subsection{Relative Star Formation Intensity} 

In the third column of Figure~\ref{fig:corrhst} we show 
how \lya\ quantities correlate with \ewha\ which, 
since \halpha\ traces SFR and the $R-$band luminosity is a rough tracer
of stellar mass, is an observable that strongly correlates with the 
instantaneous specific SFR (sSFR).  Certainly no correlation is seen 
here, but five of the six \emph{EW20LAEs} and all of the 
\emph{FESC10LAEs} have \ewha\ above 300~\AA.

We have previously shown at $z\approx 2$ that when galaxy 
selection criteria extend down to very low \ewha\ ($\gtrsim 20$\AA),
only around 10~\% of the same sample are recovered by \lya\ 
selection using the same \ewha\ threshold \citep{Hayes2010}. At 
$z\sim 0.3$ \citet{Cowie2011} remarked that \lya\
emission from sources with \ewha\ below 100 \AA\ is rare (roughly 
1~\%) but increases with increasing \ewha\ (30~\% at EW$> 100$\AA; 
60~\% at EW$>250$\AA). 

\ewha$>100$\AA\ is a condition of the LARS selection function 
(Paper I), but we note that these estimates of \ewha\ were drawn from 
3 arcsec circular apertures, and are not global quantities. The 
1 arcsec slitlets employed by 
\citet{Cowie2011} are not global either, but their galaxies have the
advantage of occupying a factor of 5 smaller physical sampling than ours
(calculated from the median redshift of the two samples). The 
\ewha\ we use in this paper 
has been recomputed inside aperture matched HST images, and in a few 
cases \ewha\ is reduced below the 100\AA\ used for selection -- the 
minimum \ewha\ in the sample is now 65\AA\ -- but nevertheless the 
\ewha\ of LARS galaxies is still somewhat high compared to the limits 
used by both \citet{Cowie2011} and 
\citet{Hayes2010}. We therefore would expect a large fraction of 
LARS galaxies to show \lya\ emission, which is exactly what we see.

\subsection{Ultraviolet Continuum Slopes} 

After UV luminosity, probably the next most commonly studied high-$z$
observable is the restframe color of the UV continuum 
\citep{Nilsson2009survey,Guaita2010,Blanc2011}, where it is frequently
invoked as a proxy for stellar dust attenuation. The correlation 
between the throughput \lya\ radiation and the UV continuum slope 
$\beta$ is one of the strongest in the sample, and is
shown in the fourth column of Figure~\ref{fig:corrhst}. The six reddest 
galaxies ($\beta > -1.7$) suppress more than 97~\% of their \lya\ 
radiation, while much spread in \ewlya\ and \fesclya\ are seen at blue
UV slopes.
All objects with \fesclya$>10$~\%  all show $\beta$ slopes bluer
than --1.8. As we will discuss in the coming sub-section, this is 
likely because of the correlation between the UV slope and dust 
content.

\subsection{Stellar and Nebular Dust Attenuation} 

Being among the easier quantities to measure in individual galaxies, the effect 
of dust content on \lya\ emission has been studied extensively 
\citep[non-exhaustively:][]{Atek2008,Atek2009galex,Scarlata2009,
Finkelstein2009dust,Pentericci2009,Finkelstein2011dust,Kornei2010,
Hayes2010, Ono2010,Cowie2011,Nakajima2012nb}. The convergent bottom line 
is that measures of the relative \lya\ output decrease as 
dust content increases, although the spread in these relationships is, 
as usual for \lya, large because of the additional properties that 
govern the
\lya\ transport \citep[see ][for a detailed discussion]{Hayes2011evol}. 
The LARS sample gives us the opportunity to study the effect of both 
attenuation as probed by the interstellar emission lines \halpha\ and 
\hbeta\ and also on the stellar continuum from a full fit of the SED. 
We do this in the fifth and sixth columns of Figure~\ref{fig:corrhst}.

The plot of \ewlya\ vs nebular \ebv\ shows little correlation:
The galaxy with the highest \ewlya\ is effectively dust free, but 
the remaining five \emph{EW20LAEs} show \ebvneb\ up to 0.5 
magnitudes. However a comparison of \ewlya\ and dust content will 
always be complicated by the fact that intrinsic equivalent widths 
are a 
function of the star-formation history, and that the stellar and 
nebular radiation may be subject to different dust contents. 
Indeed when 
we compare instead the dust contents with \lya/\halpha\ ratio and 
\fesclya, a more traditional picture emerges: the \emph{EW20LAEs} with
higher dust content tend to be among the ones with lower \fesclya, and
the zone-of-avoidance on the plot at high-\fesclya\ and high-\ebvneb\
becomes quite prominent:
all the \emph{FESC10} LAEs are found at lower dust content, whereas no
galaxies with \ebvneb$>0.35$ show \fesclya\ above a few percent. 

\begin{figure}[t]
\centering 
\includegraphics[scale=0.7, clip=true, trim=0mm 0mm 0mm 0mm]{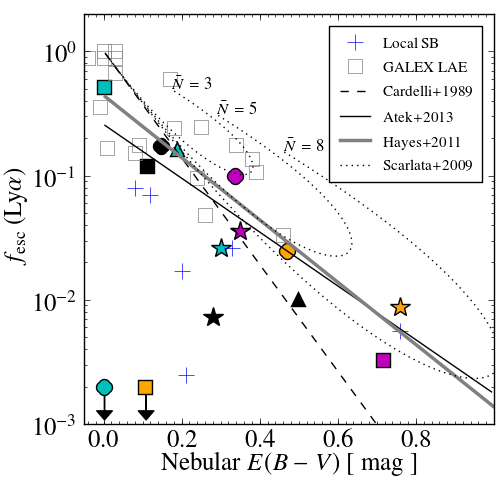}
\caption{\fesclya\ vs. nebular dust attenuation \ebv\ (always derived 
from \halpha/\hbeta) for LARS and various local galaxy samples. This is
the same as Figure~\ref{fig:corrhst} column 5 row 4, but in logarithmic 
space. Globally absorbing galaxies are represented by upper limits, 
arbitrarily set to \fesclya of 0.2 percent. We now include the six 
hand-picked galaxies at $0.009 < z < 0.028$ from our local \lya\ 
imaging pilot study \citep{Ostlin2009} with narrowband Balmer line 
observations \citep{Atek2008}, and 21 objects at $0.18 < z < 0.35$ 
selected on \lya\ strength from GALEX slitless spectroscopy and 
followed up by \citet{Atek2009galex}. The dotted line shows the \fesclya\ 
that would be expected for a pure dust-screen based upon the
\cite{Cardelli1989} extinction law. The solid gray line shows the 
best fitting curve to the \fesclya--\ebvstel\ points assembled in $z=2-3$
galaxies in \citet{Hayes2011evol}, and rescaled to \ebvneb\ using the factor 
of 2.2 from \citet{Calzetti2000}. The solid black line represents a 
similar fit to \fesclya--\ebvneb\ derived for $z\sim 0.25$ LAEs (Atek et
al., submitted). The black dotted lines show loci of points that would
be expected for Poissonianly distributed clumps of dust, as implemented
by \citet{Scarlata2009}. 
Symbols for LARS galaxies are the same as in Figure~\ref{fig:growcurve}.} 
\label{fig:corr:ebvn}
\end{figure}

In Figure~\ref{fig:corr:ebvn} we show a more detailed plot of \fesclya\ vs. 
\ebvneb, together with points from our previous studies 
\citep{Atek2008,Atek2009galex}. This shows how \ebvneb\ impacts \fesclya\ in 
samples for which we are able to match apertures, and the logarithmic 
scaling of the \fesclya\ axis illustrates the large dispersion inherent
in the relationship, even when dust content is low. We have set globally
\lya- absorbing
galaxies LARS\,04 and 06 to an arbitrary upper limit, below the 
lowest \fesclya\ emitter. It is remarkable that even with such a 
small sample, \fesclya\ spans two orders of magnitude at low \ebvneb, 
while galaxies with dust contents a factor of 5 higher are clearly 
emitters. For illustration, we plot as the dashed black line the 
\fesclya\ expected from pure dust-screen extinction, assuming 
the \citet{Cardelli1989} Galactic curve. All the galaxies at lower dust
content lie below this curve, where \lya\ is preferentially suppressed
by \hi\ scattering. However there is also a group of three LARS galaxies
and several from the \lya-selected samples that lie above this curve. 
In solid black and gray lines, 
respectively, 
we plot the empirical \fesclya-\ebv\ relationships of Atek et al 
(submitted), which was derived at $\langle z \rangle=0.25$ from 
\halpha/\hbeta, and of \citet{Hayes2011evol} which was derived at 
$z=2-3$ from SED fitting to the stellar continuum, and scaling 
\ebvstel\ to \ebvneb\ using the factor of 2.2 suggested by 
\citet{Calzetti2000}. With the dotted lines we plot the global dust 
attenuation expected from the models of \citet{Natta1984}, which have 
been advocated to explain the observed line ratios for LAEs
by \citet{Scarlata2009}; see also \citet{Calzetti1994} and 
\citet{Charlot2000} for more discussion of these geometries. 
Here $\bar{N}$ Poissonianly distributed clumps lie along the 
line-of-sight, and \ebvneb\ is computed from the optical depth of all
clumps.  These three latter parameterizations are able, with reasonable 
accuracy, to capture the upper envelope of points in the LARS 
sample. Eight galaxies lie along these relationships, including 
all the \emph{FESC10LAEs}. Importantly, all the six
\emph{EW20LAEs} lie along the empirical lines, which were derived in 
samples that were selected with precisely that selection function. 
Unlike the Hayes et al. and Atek et al. empirical prescriptions for 
\lya, however, the clumpy dust models \citep{Natta1984} were 
developed with concerns completely independent from \lya, and
only later applied to the GALEX-selected objects. The 
broad agreement between the models, particularly at higher levels of 
attenuation, may hint at their reality.  It is interesting also 
that in the \fesclya--\ebvneb\ plane the upper envelope of our 
UV-selected sample behaves very similarly to the \lya-selected samples,
at both low- and high-$z$.

In the UV-selected LARS sample, and also the mixed sample-selection
of \citet{Atek2008} and \citet{Ostlin2009} there is a sequence of 
points that falls away from this line. Here it is clear that \lya\ must
be preferentially attenuated compared to other hydrogen line radiation
as \hi\ increases the dust absorption probability or scatters \lya\ to
very large radii where it cannot be detected. For example LARS\,09 only
becomes a net \lya\ emitter when large apertures are considered and it 
is possible that both LARS\,04 and 06 would do the same if a very wide 
field UV  observations could be obtained. In \citet{Hayes2013} we 
discuss how \lya\ 
becomes systematically more extended compared to UV and \halpha\ as dust
content decreases. It is likely that in defining our apertures based 
upon UV sizes we are underestimating \lya\ fluxes in a way that the 
underestimate is larger in the less dusty systems. Correcting for this
would increase \fesclya\ more in lower \ebv\ galaxies, and could in 
principle reduce the apparent dispersion in this figure. If we could 
assemble a very large UV-selected sample with global \lya\ information 
we would expect the lower region of the plot at lower \fesclya\ and 
\ebvneb\ to be filled in. Based upon current information derived from 
both \lya\ and UV selection, we see no evidence for stronger \lya\ 
emission than described by the envelopes of our defined relationships, 
and no need for a finely tuned `scattering+shielding' radiation 
transport \citep{Neufeld1991}.

\subsection{Stellar Age and Mass}\label{sect:ages}

We show stellar age and mass in columns (7) and (8)
of Figure~\ref{fig:corrhst}.
Most observational studies of the stellar populations of \lya\ galaxies
suggest that they are relatively young and low-mass systems
\citep{Ono2010,Nilsson2011,Aquaviva2012}, although see 
\citet{Finkelstein2009dust} for counterexamples. In 
addition, several theoretical studies have also predicted \lya\ strength
to follow evolutionary sequences
\citep{Tenorio-Tagle1999,Thommes2005,Mao2007}, although details vary.
In contrast, \citet{Pentericci2009,Pentericci2010} find no correlation 
of \lya\ EW with stellar age among LBGs, and \citet{Verhamme2008} have 
also noted that trends with age should not be strong in LBGs because 
star formation proceeds close to equilibrium, where the 
\lya\ EW has saturated.
An obvious question, therefore, is whether we see any age-dependent 
trends among galaxies in the LARS sample. 

Looking at our six \emph{EW20LAEs}, it seems that five of them have a 
preference for very young ages, while the other (the lowest \ewlya\
of the subsample) exhibits the oldest stellar population in the sample. 
However when cast by \fesclya, all the seven oldest galaxies transmit 
below 3~\% of their intrinsic \lya\ radiation, while all 
\emph{FESC10LAEs} have ages below 10~Myr. It is curious that this 
timescale corresponds roughly to the ionizing lifespan of an SSP, and
while it makes sense for \ewlya\ to follow such a trend the same would
not necessarily be predicted for \fesclya. Scenarios could be invoked 
whereby a certain evolutionary time is needed to produce the dust to 
suppress the \lya\ emission, yet the oldest galaxies are not the most 
dusty (Table~\ref{tab:infprop}). The result is also largely in tension
with the scenario of \citet{Tenorio-Tagle1999} which requires a certain
timescale for kinetic feedback to accelerate the neutral gas. 

At high stellar masses there is also a notable deficiency of strong
\lya-emitters. A declining relationship between all the relative 
\lya\ quantities and the mass is clearly seen, and is most apparent 
in the figure of \fesclya. There are seven galaxies with a total stellar 
mass below $10^{10}$\msun, and five of them are the \emph{FESC10LAEs}; 
obviously, every galaxy above this mass limit transmits below 3\% of its 
\lya. These results are certainly in qualitative 
agreement with the hypothesis that \lya-selection would find lower-mass
systems, and radiative transport predictions for \lya\ transmission 
\citep[e.g.][]{Laursen2009,Yajima2012fesc}. Current thinking suggests 
that this phenomenon is due to the increased gas and dust content of 
more massive galaxies -- from direct \hi\ observations we will soon 
empirically demonstrate this to be the case within the LARS sample 
(Pardy et al, in preparation).

\subsection{Nebular Metallicity and Excitation} 

Finally we show some properties derived from SDSS 
spectroscopy, relating to nebular abundances and the excitation 
parameter, in Figure~\ref{fig:corrsdss}.
\citet{Charlot1993} presented an anticorrelation between \lya\ EW and 
metallicity, derived from IUE spectroscopy and ground-based 
observations. Even at the time, it was easy to find local dwarf 
starbursts that would outlie this relatively narrow distribution by 
roughly 2 dex \citep{Kunth1994,Thuan1997}.  Indeed the \ewlya--$Z$
relationship became weaker after the development of new data-reduction 
tools and acquisition of aperture-matched supplementary data 
\citep{Giavalisco1996}. 
However more recently and working in a sample that was both larger 
and much more cleanly selected, \citet{Cowie2011} showed that indeed 
\lya-emitting galaxies have metallicities systematically lower than 
UV-selected galaxies of similar continuum magnitude that show no \lya\
emission. Note here that in contrast to \citet{Giavalisco1996}
the \lya\ and optical apertures are not matched, but the UV does
encompass all the \lya. Our case is similar -- the \lya\ 
fluxes may be close to global, but the metallicities are derived from 
smaller SDSS fibers. Piecemeal results at high-$z$ have also found 
strongly \lya-emitting galaxies to be of systematically low metallicity 
\citep{Fosbury2003,Erb2010,Finkelstein2011hetdex,Nakajima2012nb,Guaita2013},
and also with high ionization parameter \citep{Nakajima2012ion}. 

From the SDSS spectra we examine four quantities that are frequently
used to quantify nebular properties at low-$z$, and are now emerging
as diagnostics of galaxies at $z=2-3$. These include the $R_{23}$ index 
[$\equiv$ (\oII3727+\oIII4959+\oIII5007)/\hbeta]; the N2 index 
[$\equiv$ log( \nII6584/\halpha)]; the excitation parameter, $P$ 
[$\equiv$ \oIII5007/\oII3727]; and a strong-line metallicity, for 
which we adopt the O3N2 method and the calibration of \citet{Yin2007}. 
All these quantities, and the spectroscopic fluxes from
which they were derived, can be found in \"Ostlin et al (2013). 
With the large current investment in NIR followup of high-$z$ galaxies 
and new high-multiplexing NIR spectrographs soon arriving, we can 
expect a lot more such measurements to be published in the coming 
years. 

Trends are visible between \ewlya\ and all four of the above listed 
nebular quantities, with strongly \lya-emitting galaxies exhibiting 
lower N2, and metallicity, and higher $R_{23}$ and $P$. Once again, 
trends appear stronger when cast as \fesclya, and all \emph{FESC10LAEs}
are found to exhibit low \nII/\halpha\ ratios, high oxygen/\hbeta\ 
ratios, and high \oIII/\oII\ ratios. All of these three quantities are
correlated with each-other and also with the nebular metallicity, but 
curiously the trend with O3N2-determined oxygen abundance is the 
weakest in Figure~\ref{fig:corrsdss}. While at least plausibility may
be invoked for a relationship with $Z$ -- e.g. metal abundance 
correlates with dust abundance and suppresses \lya\ -- it is not clear 
why the relationships with $R_{23}$, $P$, and N2 should be tighter.
Taken at face-value, it appears that a hotter stellar population is 
needed not only to produce \lya\ photons but also to facilitate their 
escape. 

\subsection{Summary} 

To summarize this section, we note briefly that we find no
relationship between total \lya\ luminosity and any of the secondary 
galaxy properties we have tested here. When considering the 
\lya\ EW, however, some trends begin to emerge, but they are seen more 
clearly when examining direct nebular relative measurements for \lya:
\lya/\halpha\ and \fesclya. We find \lya\ transmission to be higher 
at lower \lfuv, lower intrinsic instantaneous SFR, higher \ewha, 
bluer UV continuum colors, lower dust attenuation, younger ages, lower
stellar masses, lower N2 and metallicity, and higher $R_{23}$ and 
excitation parameter. Of course these are just observational 
considerations -- we do not claim that all of these 
effects form a direct causal relation nor that these quantities are 
uncorrelated. Clearly many of them will be.

\section{Summary and Conclusions}\label{sect:conc}

The \emph{Lyman alpha Reference Sample} (LARS) comprises 14 galaxies at 
redshift between 0.028 and 0.18 in which the dominant source of 
ionization is determined to be star formation 
(\"Ostlin et al 2013; Paper I).  In this article we have 
presented individual images obtained with the Hubble Space Telescope in
the far ultraviolet continuum, \halpha, and \lya, and maps of the ratios 
of \halpha/\hbeta, \lya/\halpha, and \lya/FUV (i.e. the \lya\ equivalent 
width).  From the intensity maps we have produced radial light profiles in 
\lya, \halpha, and the FUV, and photometric aperture curves-of-growth for \lya, 
FUV, \ewlya, and \fesclya. We have defined standardized apertures, and 
computed a number of global properties of the sample, including 
aperture-matched quantities that describe the \lya\ output and also the
properties of the nebular gas and massive stellar population. We find 
that:

\begin{itemize}

\item{the morphology of \lya\ is usually very different from the 
\halpha\ and FUV morphologies, and that many features seen in \halpha\ 
are not visible in \lya. We interpret this, together with the large, 
extended \lya\ haloes that we have presented previously 
\citep{Hayes2013}, as the scattering of \lya\ photons in the neutral 
interstellar medium that surrounds the star-forming regions. This is 
supported by the observed line ratios that exceed the canonical case 
B values by factors of more than 3.}

\item{radial profiles in \lya\ are flatter than those of the UV 
continuum, and exhibit S{\'e}rsic indices that are systematically 
lower in \lya\ than the UV.  Moreover, radial profiles in 
\ewlya\ and \lya/\halpha\ often exceed their intrinsic recombination 
values by large factors, even in complete radial annuli. These two 
observations the spatial redistribution of \lya\ to large radii as a 
result of \hi\ scattering. }

\item{photometric growth-curves rise more slowly in \lya\ than
in the UV and \halpha. We caution that in some cases there may be 
unpredictable, and possibly substantial aperture-dependencies and 
systematic effects involved in the measurement of these quantities in 
high-$z$ samples, at least with `normal' apertures of 1--2 arcsec. 
These cases, however, do 
not appear to be in the majority in our sample, at least within the 
limits of the available field-of-view. }

\item{some regions of locally enhanced \lya\ emission coincide 
with filamentary structures in the nebular gas. This could easily be 
explained by outflows that become Rayleigh-Taylor unstable and 
fragment, thereby increasing the \lya\ flow in these directions because
of a reduced covering fraction and increased velocity offset in the 
neutral gas.}

\item{ten of our 14 galaxies are analogous in luminosity to high-$z$ 
Lyman-break galaxies, and six could be detected by deep \lya\ 
narrowband surveys that select objects with \ewlya\ above 20~\AA. We 
find five galaxies to have high \lya\ escape fractions (above 10~\%), 
with the remaining nine showing \fesclya\ below 3~\%. Several of these 
strongly emitting 
galaxies are indeed bright, showing EWs above 60\AA\ in two cases and 
in one case -- LARS\,02 -- a \fesclya\ of 75~\% at radius of just a few
kpc. Such values have not previously been reported in nearby galaxies, 
and appear also to be rare at high-$z$. This object may be a very 
interesting laboratory for detailed study.  }

\item{\lya\ throughput (EW, \halpha\ ratio, \fesclya) is 
systematically higher in galaxies of faint FUV magnitude, lower 
star formation rate, higher \halpha\ EW, bluer UV colors, lower
extinction, lower mass, and nebular quantities that suggest more 
intense UV radiation fields. In contrast, we have not yet found 
\lya\ luminosity to correlate strongly with any of the quantities 
we have measured. }

\end{itemize}

\acknowledgments
M.H. received support from Agence Nationale de la recherche bearing the reference ANR-09-BLAN-0234-01, and also acknowledges support from the Swedish research council (VR) and the Swedish National Space Board (SNSB).
G.\"O. is a Swedish Royal Academy of Sciences research fellow supported by a grant from Knut and Alice Wallenberg foundation, and also acknowledges support from the Swedish research council (VR) and the Swedish National Space Board (SNSB).
A.V. benefits from the fellowship `Boursi\`ere d'excellence de l'Universit\'e de Gen\`eve'.
I.O. was financed through the Sciex fellowship by the Rectors' Conference of Swiss Universities. 
H.A. and D.K. are supported by the Centre National d'\'Etudes Spatiales (CNES) and the Programme National de Cosmologie et Galaxies (PNCG).
P.L. acknowledges support from the ERC-StG grant EGGS-278202.
H.O.F. acknowledges financial support from CONACYT grant 129204, Spanish FPI grant BES-2006-13489, and was also financed through a postdoctoral UNAM grant.
H.O.F. and J.M.M.H. are partially funded by Spanish MICINN grants CSD2006-00070 (CONSOLIDER GTC), AYA2010-21887-C04-02 (ESTALLIDOS) and AYA2012-39362-C02-01
.

{\it Facilities:} \facility{HST (ACS,WFC3)}.

\bibliographystyle{apj}

\begin{thebibliography}{}
\expandafter\ifx\csname natexlab\endcsname\relax\def\natexlab#1{#1}\fi

\bibitem[{{Acquaviva} {et~al.}(2012){Acquaviva}, {Vargas}, {Gawiser}, \&
  {Guaita}}]{Aquaviva2012}
{Acquaviva}, V., {Vargas}, C., {Gawiser}, E., \& {Guaita}, L. 2012, \apjl, 751,
  L26

\bibitem[{{Ando} {et~al.}(2006){Ando}, {Ohta}, {Iwata}, {Akiyama}, {Aoki}, \&
  {Tamura}}]{Ando2006}
{Ando}, M., {Ohta}, K., {Iwata}, I., {et~al.} 2006, \apjl, 645, L9

\bibitem[{{Asplund} {et~al.}(2009){Asplund}, {Grevesse}, {Sauval}, \&
  {Scott}}]{Asplund2009}
{Asplund}, M., {Grevesse}, N., {Sauval}, A.~J., \& {Scott}, P. 2009, \araa, 47,
  481

\bibitem[{{Atek} {et~al.}(2008){Atek}, {Kunth}, {Hayes}, {{\"O}stlin}, \&
  {Mas-Hesse}}]{Atek2008}
{Atek}, H., {Kunth}, D., {Hayes}, M., {{\"O}stlin}, G., \& {Mas-Hesse}, J.~M.
  2008, \aap, 488, 491

\bibitem[{{Atek} {et~al.}(2009){Atek}, {Kunth}, {Schaerer}, {Hayes},
  {Deharveng}, {{\"O}stlin}, \& {Mas-Hesse}}]{Atek2009galex}
{Atek}, H., {Kunth}, D., {Schaerer}, D., {et~al.} 2009, \aap, 506, L1

\bibitem[{{Bertin} \& {Arnouts}(1996)}]{Bertin1996}
{Bertin}, E., \& {Arnouts}, S. 1996, \aaps, 117, 393

\bibitem[{{Blanc} {et~al.}(2011){Blanc}, {Adams}, {Gebhardt}, {Hill}, {Drory},
  {Hao}, {Bender}, {Ciardullo}, {Finkelstein}, {Fry}, {Gawiser}, {Gronwall},
  {Hopp}, {Jeong}, {Kelzenberg}, {Komatsu}, {MacQueen}, {Murphy}, {Roth},
  {Schneider}, \& {Tufts}}]{Blanc2011}
{Blanc}, G.~A., {Adams}, J.~J., {Gebhardt}, K., {et~al.} 2011, \apj, 736, 31

\bibitem[{{Bouwens} {et~al.}(2011){Bouwens}, {Illingworth}, {Oesch},
  {Labb{\'e}}, {Trenti}, {van Dokkum}, {Franx}, {Stiavelli}, {Carollo},
  {Magee}, \& {Gonzalez}}]{Bouwens2011}
{Bouwens}, R.~J., {Illingworth}, G.~D., {Oesch}, P.~A., {et~al.} 2011, \apj,
  737, 90

\bibitem[{{Calzetti} {et~al.}(2000){Calzetti}, {Armus}, {Bohlin}, {Kinney},
  {Koornneef}, \& {Storchi-Bergmann}}]{Calzetti2000}
{Calzetti}, D., {Armus}, L., {Bohlin}, R.~C., {et~al.} 2000, \apj, 533, 682

\bibitem[{{Calzetti} {et~al.}(1994){Calzetti}, {Kinney}, \&
  {Storchi-Bergmann}}]{Calzetti1994}
{Calzetti}, D., {Kinney}, A.~L., \& {Storchi-Bergmann}, T. 1994, \apj, 429, 582

\bibitem[{{Cappellari} \& {Copin}(2003)}]{Cappellari2003}
{Cappellari}, M., \& {Copin}, Y. 2003, \mnras, 342, 345

\bibitem[{{Cardamone} {et~al.}(2009){Cardamone}, {Schawinski}, {Sarzi},
  {Bamford}, {Bennert}, {Urry}, {Lintott}, {Keel}, {Parejko}, {Nichol},
  {Thomas}, {Andreescu}, {Murray}, {Raddick}, {Slosar}, {Szalay}, \&
  {Vandenberg}}]{Cardamone2009}
{Cardamone}, C., {Schawinski}, K., {Sarzi}, M., {et~al.} 2009, \mnras, 399,
  1191

\bibitem[{{Cardelli} {et~al.}(1989){Cardelli}, {Clayton}, \&
  {Mathis}}]{Cardelli1989}
{Cardelli}, J.~A., {Clayton}, G.~C., \& {Mathis}, J.~S. 1989, \apj, 345, 245

\bibitem[{{Charlot} \& {Fall}(1993)}]{Charlot1993}
{Charlot}, S., \& {Fall}, S.~M. 1993, \apj, 415, 580

\bibitem[{{Charlot} \& {Fall}(2000)}]{Charlot2000}
---. 2000, \apj, 539, 718

\bibitem[{{Cooper} {et~al.}(2008){Cooper}, {Bicknell}, {Sutherland}, \&
  {Bland-Hawthorn}}]{Cooper2008}
{Cooper}, J.~L., {Bicknell}, G.~V., {Sutherland}, R.~S., \& {Bland-Hawthorn},
  J. 2008, \apj, 674, 157

\bibitem[{{Cowie} {et~al.}(2010){Cowie}, {Barger}, \& {Hu}}]{Cowie2010}
{Cowie}, L.~L., {Barger}, A.~J., \& {Hu}, E.~M. 2010, \apj, 711, 928

\bibitem[{{Cowie} {et~al.}(2011){Cowie}, {Barger}, \& {Hu}}]{Cowie2011}
---. 2011, \apj, 738, 136

\bibitem[{{Diehl} \& {Statler}(2006)}]{Diehl2006}
{Diehl}, S., \& {Statler}, T.~S. 2006, \mnras, 368, 497

\bibitem[{{Dijkstra} \& {Jeeson-Daniel}(2013)}]{Dijkstra2013}
{Dijkstra}, M., \& {Jeeson-Daniel}, A. 2013, ArXiv e-prints, arXiv:1305.3613

\bibitem[{{Dijkstra} {et~al.}(2007){Dijkstra}, {Wyithe}, \&
  {Haiman}}]{Dijkstra2007}
{Dijkstra}, M., {Wyithe}, J.~S.~B., \& {Haiman}, Z. 2007, \mnras, 379, 253

\bibitem[{{Erb} {et~al.}(2010){Erb}, {Pettini}, {Shapley}, {Steidel}, {Law}, \&
  {Reddy}}]{Erb2010}
{Erb}, D.~K., {Pettini}, M., {Shapley}, A.~E., {et~al.} 2010, \apj, 719, 1168

\bibitem[{{Feldmeier} {et~al.}(2013){Feldmeier}, {Hagen}, {Ciardullo},
  {Gronwall}, {Gawiser}, {Guaita}, {Hagen}, {Bond}, {Acquaviva}, {Blanc},
  {Orsi}, \& {Kurczynski}}]{Feldmeier2013}
{Feldmeier}, J., {Hagen}, A., {Ciardullo}, R., {et~al.} 2013, ArXiv e-prints,
  arXiv:1301.0462

\bibitem[{{Finkelstein} {et~al.}(2011{\natexlab{a}}){Finkelstein}, {Cohen},
  {Moustakas}, {Malhotra}, {Rhoads}, \& {Papovich}}]{Finkelstein2011dust}
{Finkelstein}, S.~L., {Cohen}, S.~H., {Moustakas}, J., {et~al.}
  2011{\natexlab{a}}, \apj, 733, 117

\bibitem[{{Finkelstein} {et~al.}(2009){Finkelstein}, {Rhoads}, {Malhotra}, \&
  {Grogin}}]{Finkelstein2009dust}
{Finkelstein}, S.~L., {Rhoads}, J.~E., {Malhotra}, S., \& {Grogin}, N. 2009,
  \apj, 691, 465

\bibitem[{{Finkelstein} {et~al.}(2011{\natexlab{b}}){Finkelstein}, {Hill},
  {Gebhardt}, {Adams}, {Blanc}, {Papovich}, {Ciardullo}, {Drory}, {Gawiser},
  {Gronwall}, {Schneider}, \& {Tran}}]{Finkelstein2011hetdex}
{Finkelstein}, S.~L., {Hill}, G.~J., {Gebhardt}, K., {et~al.}
  2011{\natexlab{b}}, \apj, 729, 140

\bibitem[{{Fosbury} {et~al.}(2003){Fosbury}, {Villar-Mart{\'{\i}}n},
  {Humphrey}, {Lombardi}, {Rosati}, {Stern}, {Hook}, {Holden}, {Stanford},
  {Squires}, {Rauch}, \& {Sargent}}]{Fosbury2003}
{Fosbury}, R.~A.~E., {Villar-Mart{\'{\i}}n}, M., {Humphrey}, A., {et~al.} 2003,
  \apj, 596, 797

\bibitem[{{France} {et~al.}(2010){France}, {Nell}, {Green}, \&
  {Leitherer}}]{France2010}
{France}, K., {Nell}, N., {Green}, J.~C., \& {Leitherer}, C. 2010, \apjl, 722,
  L80

\bibitem[{{Fynbo} {et~al.}(2001){Fynbo}, {M{\"o}ller}, \&
  {Thomsen}}]{Fynbo2001}
{Fynbo}, J.~U., {M{\"o}ller}, P., \& {Thomsen}, B. 2001, \aap, 374, 443

\bibitem[{{Garel} {et~al.}(2012){Garel}, {Blaizot}, {Guiderdoni}, {Schaerer},
  {Verhamme}, \& {Hayes}}]{Garel2012}
{Garel}, T., {Blaizot}, J., {Guiderdoni}, B., {et~al.} 2012, \mnras, 422, 310

\bibitem[{{Giavalisco} {et~al.}(1996){Giavalisco}, {Koratkar}, \&
  {Calzetti}}]{Giavalisco1996}
{Giavalisco}, M., {Koratkar}, A., \& {Calzetti}, D. 1996, \apj, 466, 831

\bibitem[{{Gonz{\'a}lez Delgado} {et~al.}(1999){Gonz{\'a}lez Delgado},
  {Leitherer}, \& {Heckman}}]{Gonzalez-Delgado1999}
{Gonz{\'a}lez Delgado}, R.~M., {Leitherer}, C., \& {Heckman}, T.~M. 1999,
  \apjs, 125, 489

\bibitem[{{Gronwall} {et~al.}(2007){Gronwall}, {Ciardullo}, {Hickey},
  {Gawiser}, {Feldmeier}, {van Dokkum}, {Urry}, {Herrera}, {Lehmer}, {Infante},
  {Orsi}, {Marchesini}, {Blanc}, {Francke}, {Lira}, \&
  {Treister}}]{Gronwall2007}
{Gronwall}, C., {Ciardullo}, R., {Hickey}, T., {et~al.} 2007, \apj, 667, 79

\bibitem[{{Guaita} {et~al.}(2013){Guaita}, {Francke}, {Gawiser}, {Bauer},
  {Hayes}, {{\"O}stlin}, \& {Padilla}}]{Guaita2013}
{Guaita}, L., {Francke}, H., {Gawiser}, E., {et~al.} 2013, \aap, 551, A93

\bibitem[{{Guaita} {et~al.}(2010){Guaita}, {Gawiser}, {Padilla}, {Francke},
  {Bond}, {Gronwall}, {Ciardullo}, {Feldmeier}, {Sinawa}, {Blanc}, \&
  {Virani}}]{Guaita2010}
{Guaita}, L., {Gawiser}, E., {Padilla}, N., {et~al.} 2010, \apj, 714, 255

\bibitem[{{Haiman} \& {Spaans}(1999)}]{Haiman1999}
{Haiman}, Z., \& {Spaans}, M. 1999, \apj, 518, 138

\bibitem[{{Hayes} {et~al.}(2007){Hayes}, {{\"O}stlin}, {Atek}, {Kunth},
  {Mas-Hesse}, {Leitherer}, {Jim{\'e}nez-Bail{\'o}n}, \& {Adamo}}]{Hayes2007}
{Hayes}, M., {{\"O}stlin}, G., {Atek}, H., {et~al.} 2007, \mnras, 382, 1465

\bibitem[{{Hayes} {et~al.}(2009){Hayes}, {{\"O}stlin}, {Mas-Hesse}, \&
  {Kunth}}]{Hayes2009}
{Hayes}, M., {{\"O}stlin}, G., {Mas-Hesse}, J.~M., \& {Kunth}, D. 2009, \aj,
  138, 911

\bibitem[{{Hayes} {et~al.}(2005){Hayes}, {{\"O}stlin}, {Mas-Hesse}, {Kunth},
  {Leitherer}, \& {Petrosian}}]{Hayes2005}
{Hayes}, M., {{\"O}stlin}, G., {Mas-Hesse}, J.~M., {et~al.} 2005, \aap, 438, 71

\bibitem[{{Hayes} {et~al.}(2011){Hayes}, {Schaerer}, {{\"O}stlin}, {Mas-Hesse},
  {Atek}, \& {Kunth}}]{Hayes2011evol}
{Hayes}, M., {Schaerer}, D., {{\"O}stlin}, G., {et~al.} 2011, \apj, 730, 8

\bibitem[{{Hayes} {et~al.}(2010){Hayes}, {{\"O}stlin}, {Schaerer}, {Mas-Hesse},
  {Leitherer}, {Atek}, {Kunth}, {Verhamme}, {de Barros}, \&
  {Melinder}}]{Hayes2010}
{Hayes}, M., {{\"O}stlin}, G., {Schaerer}, D., {et~al.} 2010, \nat, 464, 562

\bibitem[{{Hayes} {et~al.}(2013){Hayes}, {{\"O}stlin}, {Schaerer}, {Verhamme},
  {Mas-Hesse}, {Adamo}, {Atek}, {Cannon}, {Duval}, {Guaita}, {Herenz}, {Kunth},
  {Laursen}, {Melinder}, {Orlitov{\'a}}, {Ot{\'{\i}}-Floranes}, \&
  {Sandberg}}]{Hayes2013}
---. 2013, \apjl, 765, L27

\bibitem[{{Heckman} {et~al.}(2011){Heckman}, {Borthakur}, {Overzier},
  {Kauffmann}, {Basu-Zych}, {Leitherer}, {Sembach}, {Martin}, {Rich},
  {Schiminovich}, \& {Seibert}}]{Heckman2011}
{Heckman}, T.~M., {Borthakur}, S., {Overzier}, R., {et~al.} 2011, \apj, 730, 5

\bibitem[{{Hoopes} {et~al.}(2007){Hoopes}, {Heckman}, {Salim}, {Seibert},
  {Tremonti}, {Schiminovich}, {Rich}, {Martin}, {Charlot}, {Kauffmann},
  {Forster}, {Friedman}, {Morrissey}, {Neff}, {Small}, {Wyder}, {Bianchi},
  {Donas}, {Lee}, {Madore}, {Milliard}, {Szalay}, {Welsh}, \&
  {Yi}}]{Hoopes2007}
{Hoopes}, C.~G., {Heckman}, T.~M., {Salim}, S., {et~al.} 2007, \apjs, 173, 441

\bibitem[{{Hopkins} \& {Beacom}(2006)}]{Hopkins2006}
{Hopkins}, A.~M., \& {Beacom}, J.~F. 2006, \apj, 651, 142

\bibitem[{{Hu} {et~al.}(2010){Hu}, {Cowie}, {Barger}, {Capak}, {Kakazu}, \&
  {Trouille}}]{Hu2010}
{Hu}, E.~M., {Cowie}, L.~L., {Barger}, A.~J., {et~al.} 2010, \apj, 725, 394

\bibitem[{{Hummer} \& {Storey}(1987)}]{Hummer1987}
{Hummer}, D.~G., \& {Storey}, P.~J. 1987, \mnras, 224, 801

\bibitem[{{Kashikawa} {et~al.}(2006){Kashikawa}, {Shimasaku}, {Malkan}, {Doi},
  {Matsuda}, {Ouchi}, {Taniguchi}, {Ly}, {Nagao}, {Iye}, {Motohara},
  {Murayama}, {Murozono}, {Nariai}, {Ohta}, {Okamura}, {Sasaki}, {Shioya}, \&
  {Umemura}}]{Kashikawa2006}
{Kashikawa}, N., {Shimasaku}, K., {Malkan}, M.~A., {et~al.} 2006, \apj, 648, 7

\bibitem[{{Kashikawa} {et~al.}(2011){Kashikawa}, {Shimasaku}, {Matsuda},
  {Egami}, {Jiang}, {Nagao}, {Ouchi}, {Malkan}, {Hattori}, {Ota}, {Taniguchi},
  {Okamura}, {Ly}, {Iye}, {Furusawa}, {Shioya}, {Shibuya}, {Ishizaki}, \&
  {Toshikawa}}]{Kashikawa2011}
{Kashikawa}, N., {Shimasaku}, K., {Matsuda}, Y., {et~al.} 2011, \apj, 734, 119

\bibitem[{{Kennicutt}(1998)}]{Kennicutt1998}
{Kennicutt}, Jr., R.~C. 1998, \araa, 36, 189

\bibitem[{{Kornei} {et~al.}(2010){Kornei}, {Shapley}, {Erb}, {Steidel},
  {Reddy}, {Pettini}, \& {Bogosavljevi{\'c}}}]{Kornei2010}
{Kornei}, K.~A., {Shapley}, A.~E., {Erb}, D.~K., {et~al.} 2010, \apj, 711, 693

\bibitem[{{Kulas} {et~al.}(2012){Kulas}, {Shapley}, {Kollmeier}, {Zheng},
  {Steidel}, \& {Hainline}}]{Kulas2012}
{Kulas}, K.~R., {Shapley}, A.~E., {Kollmeier}, J.~A., {et~al.} 2012, \apj, 745,
  33

\bibitem[{{Kunth} {et~al.}(1994){Kunth}, {Lequeux}, {Sargent}, \&
  {Viallefond}}]{Kunth1994}
{Kunth}, D., {Lequeux}, J., {Sargent}, W.~L.~W., \& {Viallefond}, F. 1994,
  \aap, 282, 709

\bibitem[{{Kunth} {et~al.}(1998){Kunth}, {Mas-Hesse}, {Terlevich}, {Terlevich},
  {Lequeux}, \& {Fall}}]{Kunth1998}
{Kunth}, D., {Mas-Hesse}, J.~M., {Terlevich}, E., {et~al.} 1998, \aap, 334, 11

\bibitem[{{Laursen} {et~al.}(2009){Laursen}, {Sommer-Larsen}, \&
  {Andersen}}]{Laursen2009}
{Laursen}, P., {Sommer-Larsen}, J., \& {Andersen}, A.~C. 2009, \apj, 704, 1640

\bibitem[{{Lidman} {et~al.}(2012){Lidman}, {Hayes}, {Jones}, {Schaerer},
  {Westra}, {Tapken}, {Meisenheimer}, \& {Verhamme}}]{Lidman2012}
{Lidman}, C., {Hayes}, M., {Jones}, D.~H., {et~al.} 2012, \mnras, 420, 1946

\bibitem[{{Mac Low} \& {McCray}(1988)}]{MacLow1988}
{Mac Low}, M.-M., \& {McCray}, R. 1988, \apj, 324, 776

\bibitem[{{Malhotra} \& {Rhoads}(2004)}]{Malhotra2004}
{Malhotra}, S., \& {Rhoads}, J.~E. 2004, \apjl, 617, L5

\bibitem[{{Mao} {et~al.}(2007){Mao}, {Lapi}, {Granato}, {de Zotti}, \&
  {Danese}}]{Mao2007}
{Mao}, J., {Lapi}, A., {Granato}, G.~L., {de Zotti}, G., \& {Danese}, L. 2007,
  \apj, 667, 655

\bibitem[{{Mas-Hesse} {et~al.}(2003){Mas-Hesse}, {Kunth}, {Tenorio-Tagle},
  {Leitherer}, {Terlevich}, \& {Terlevich}}]{Mas-Hesse2003}
{Mas-Hesse}, J.~M., {Kunth}, D., {Tenorio-Tagle}, G., {et~al.} 2003, \apj, 598,
  858

\bibitem[{{Matsuda} {et~al.}(2012){Matsuda}, {Yamada}, {Hayashino}, {Yamauchi},
  {Nakamura}, {Morimoto}, {Ouchi}, {Ono}, {Umemura}, \& {Mori}}]{Matsuda2012}
{Matsuda}, Y., {Yamada}, T., {Hayashino}, T., {et~al.} 2012, \mnras, 425, 878

\bibitem[{{Nakajima} {et~al.}(2012{\natexlab{a}}){Nakajima}, {Ouchi},
  {Shimasaku}, {Hashimoto}, {Ono}, \& {Lee}}]{Nakajima2012ion}
{Nakajima}, K., {Ouchi}, M., {Shimasaku}, K., {et~al.} 2012{\natexlab{a}},
  ArXiv e-prints, arXiv:1208.3260

\bibitem[{{Nakajima} {et~al.}(2012{\natexlab{b}}){Nakajima}, {Ouchi},
  {Shimasaku}, {Ono}, {Lee}, {Foucaud}, {Ly}, {Dale}, {Salim}, {Finn},
  {Almaini}, \& {Okamura}}]{Nakajima2012nb}
---. 2012{\natexlab{b}}, \apj, 745, 12

\bibitem[{{Natta} \& {Panagia}(1984)}]{Natta1984}
{Natta}, A., \& {Panagia}, N. 1984, \apj, 287, 228

\bibitem[{{Neufeld}(1991)}]{Neufeld1991}
{Neufeld}, D.~A. 1991, \apjl, 370, L85

\bibitem[{{Nilsson} {et~al.}(2011){Nilsson}, {{\"O}stlin}, {M{\o}ller},
  {M{\"o}ller-Nilsson}, {Tapken}, {Freudling}, \& {Fynbo}}]{Nilsson2011}
{Nilsson}, K.~K., {{\"O}stlin}, G., {M{\o}ller}, P., {et~al.} 2011, \aap, 529,
  A9

\bibitem[{{Nilsson} {et~al.}(2009){Nilsson}, {Tapken}, {M{\o}ller},
  {Freudling}, {Fynbo}, {Meisenheimer}, {Laursen}, \&
  {{\"O}stlin}}]{Nilsson2009survey}
{Nilsson}, K.~K., {Tapken}, C., {M{\o}ller}, P., {et~al.} 2009, \aap, 498, 13

\bibitem[{{Ono} {et~al.}(2010){Ono}, {Ouchi}, {Shimasaku}, {Akiyama}, {Dunlop},
  {Farrah}, {Lee}, {McLure}, {Okamura}, \& {Yoshida}}]{Ono2010}
{Ono}, Y., {Ouchi}, M., {Shimasaku}, K., {et~al.} 2010, \mnras, 402, 1580

\bibitem[{{Osterbrock}(1989)}]{Osterbrock1989}
{Osterbrock}, D.~E. 1989, {Astrophysics of gaseous nebulae and active galactic
  nuclei}

\bibitem[{{{\"O}stlin} {et~al.}(2009){{\"O}stlin}, {Hayes}, {Kunth},
  {Mas-Hesse}, {Leitherer}, {Petrosian}, \& {Atek}}]{Ostlin2009}
{{\"O}stlin}, G., {Hayes}, M., {Kunth}, D., {et~al.} 2009, \aj, 138, 923

\bibitem[{{Ot{\'{\i}}-Floranes} {et~al.}(2012){Ot{\'{\i}}-Floranes},
  {Mas-Hesse}, {Jim{\'e}nez-Bail{\'o}n}, {Schaerer}, {Hayes}, {{\"O}stlin},
  {Atek}, \& {Kunth}}]{Oti-Florannes2012}
{Ot{\'{\i}}-Floranes}, H., {Mas-Hesse}, J.~M., {Jim{\'e}nez-Bail{\'o}n}, E.,
  {et~al.} 2012, \aap, 546, A65

\bibitem[{{Ouchi} {et~al.}(2008){Ouchi}, {Shimasaku}, {Akiyama}, {Simpson},
  {Saito}, {Ueda}, {Furusawa}, {Sekiguchi}, {Yamada}, {Kodama}, {Kashikawa},
  {Okamura}, {Iye}, {Takata}, {Yoshida}, \& {Yoshida}}]{Ouchi2008}
{Ouchi}, M., {Shimasaku}, K., {Akiyama}, M., {et~al.} 2008, \apjs, 176, 301

\bibitem[{{Overzier} {et~al.}(2008){Overzier}, {Heckman}, {Kauffmann},
  {Seibert}, {Rich}, {Basu-Zych}, {Lotz}, {Aloisi}, {Charlot}, {Hoopes},
  {Martin}, {Schiminovich}, \& {Madore}}]{Overzier2008}
{Overzier}, R.~A., {Heckman}, T.~M., {Kauffmann}, G., {et~al.} 2008, \apj, 677,
  37

\bibitem[{{Overzier} {et~al.}(2009){Overzier}, {Heckman}, {Tremonti}, {Armus},
  {Basu-Zych}, {Gon{\c c}alves}, {Rich}, {Martin}, {Ptak}, {Schiminovich},
  {Ford}, {Madore}, \& {Seibert}}]{Overzier2009}
{Overzier}, R.~A., {Heckman}, T.~M., {Tremonti}, C., {et~al.} 2009, \apj, 706,
  203

\bibitem[{{Partridge} \& {Peebles}(1967)}]{Partridge1967}
{Partridge}, R.~B., \& {Peebles}, P.~J.~E. 1967, \apj, 147, 868

\bibitem[{{Pentericci} {et~al.}(2009){Pentericci}, {Grazian}, {Fontana},
  {Castellano}, {Giallongo}, {Salimbeni}, \& {Santini}}]{Pentericci2009}
{Pentericci}, L., {Grazian}, A., {Fontana}, A., {et~al.} 2009, \aap, 494, 553

\bibitem[{{Pentericci} {et~al.}(2010){Pentericci}, {Grazian}, {Scarlata},
  {Fontana}, {Castellano}, {Giallongo}, \& {Vanzella}}]{Pentericci2010}
{Pentericci}, L., {Grazian}, A., {Scarlata}, C., {et~al.} 2010, \aap, 514, A64+

\bibitem[{{Petrosian}(1976)}]{Petrosian1976}
{Petrosian}, V. 1976, \apjl, 209, L1

\bibitem[{{Prevot} {et~al.}(1984){Prevot}, {Lequeux}, {Prevot}, {Maurice}, \&
  {Rocca-Volmerange}}]{Prevot1984}
{Prevot}, M.~L., {Lequeux}, J., {Prevot}, L., {Maurice}, E., \&
  {Rocca-Volmerange}, B. 1984, \aap, 132, 389

\bibitem[{{Quider} {et~al.}(2009){Quider}, {Pettini}, {Shapley}, \&
  {Steidel}}]{Quider2009}
{Quider}, A.~M., {Pettini}, M., {Shapley}, A.~E., \& {Steidel}, C.~C. 2009,
  \mnras, 398, 1263

\bibitem[{{Raiter} {et~al.}(2010){Raiter}, {Schaerer}, \&
  {Fosbury}}]{Raiter2010}
{Raiter}, A., {Schaerer}, D., \& {Fosbury}, R.~A.~E. 2010, \aap, 523, A64

\bibitem[{{Rauch} {et~al.}(2008){Rauch}, {Haehnelt}, {Bunker}, {Becker},
  {Marleau}, {Graham}, {Cristiani}, {Jarvis}, {Lacey}, {Morris}, {Peroux},
  {R{\"o}ttgering}, \& {Theuns}}]{Rauch2008}
{Rauch}, M., {Haehnelt}, M., {Bunker}, A., {et~al.} 2008, \apj, 681, 856

\bibitem[{{Reddy} \& {Steidel}(2009)}]{Reddy2009}
{Reddy}, N.~A., \& {Steidel}, C.~C. 2009, \apj, 692, 778

\bibitem[{{Rhoads} {et~al.}(2003){Rhoads}, {Dey}, {Malhotra}, {Stern},
  {Spinrad}, {Jannuzi}, {Dawson}, {Brown}, \& {Landes}}]{Rhoads2003}
{Rhoads}, J.~E., {Dey}, A., {Malhotra}, S., {et~al.} 2003, \aj, 125, 1006

\bibitem[{{Santos}(2004)}]{Santos2004}
{Santos}, M.~R. 2004, \mnras, 349, 1137

\bibitem[{{Scarlata} {et~al.}(2009){Scarlata}, {Colbert}, {Teplitz}, {Panagia},
  {Hayes}, {Siana}, {Rau}, {Francis}, {Caon}, {Pizzella}, \&
  {Bridge}}]{Scarlata2009}
{Scarlata}, C., {Colbert}, J., {Teplitz}, H.~I., {et~al.} 2009, \apjl, 704, L98

\bibitem[{{Schaerer}(2003)}]{Schaerer2003}
{Schaerer}, D. 2003, \aap, 397, 527

\bibitem[{{Schaerer} {et~al.}(2011){Schaerer}, {de Barros}, \&
  {Stark}}]{Schaerer2011frac}
{Schaerer}, D., {de Barros}, S., \& {Stark}, D.~P. 2011, \aap, 536, A72

\bibitem[{{Shapley} {et~al.}(2003){Shapley}, {Steidel}, {Pettini}, \&
  {Adelberger}}]{Shapley2003}
{Shapley}, A.~E., {Steidel}, C.~C., {Pettini}, M., \& {Adelberger}, K.~L. 2003,
  \apj, 588, 65

\bibitem[{{Shapley} {et~al.}(2006){Shapley}, {Steidel}, {Pettini},
  {Adelberger}, \& {Erb}}]{Shapley2006}
{Shapley}, A.~E., {Steidel}, C.~C., {Pettini}, M., {Adelberger}, K.~L., \&
  {Erb}, D.~K. 2006, \apj, 651, 688

\bibitem[{{Shimasaku} {et~al.}(2006){Shimasaku}, {Kashikawa}, {Doi}, {Ly},
  {Malkan}, {Matsuda}, {Ouchi}, {Hayashino}, {Iye}, {Motohara}, {Murayama},
  {Nagao}, {Ohta}, {Okamura}, {Sasaki}, {Shioya}, \&
  {Taniguchi}}]{Shimasaku2006}
{Shimasaku}, K., {Kashikawa}, N., {Doi}, M., {et~al.} 2006, \pasj, 58, 313

\bibitem[{{Stark} {et~al.}(2010){Stark}, {Ellis}, {Chiu}, {Ouchi}, \&
  {Bunker}}]{Stark2010}
{Stark}, D.~P., {Ellis}, R.~S., {Chiu}, K., {Ouchi}, M., \& {Bunker}, A. 2010,
  \mnras, 408, 1628

\bibitem[{{Stark} {et~al.}(2011){Stark}, {Ellis}, \& {Ouchi}}]{Stark2011}
{Stark}, D.~P., {Ellis}, R.~S., \& {Ouchi}, M. 2011, \apjl, 728, L2

\bibitem[{{Steidel} {et~al.}(2003){Steidel}, {Adelberger}, {Shapley},
  {Pettini}, {Dickinson}, \& {Giavalisco}}]{Steidel2003}
{Steidel}, C.~C., {Adelberger}, K.~L., {Shapley}, A.~E., {et~al.} 2003, \apj,
  592, 728

\bibitem[{{Steidel} {et~al.}(2011){Steidel}, {Bogosavljevi{\'c}}, {Shapley},
  {Kollmeier}, {Reddy}, {Erb}, \& {Pettini}}]{Steidel2011}
{Steidel}, C.~C., {Bogosavljevi{\'c}}, M., {Shapley}, A.~E., {et~al.} 2011,
  \apj, 736, 160

\bibitem[{{Tapken} {et~al.}(2007){Tapken}, {Appenzeller}, {Noll}, {Richling},
  {Heidt}, {Meink{\"o}hn}, \& {Mehlert}}]{Tapken2007}
{Tapken}, C., {Appenzeller}, I., {Noll}, S., {et~al.} 2007, \aap, 467, 63

\bibitem[{{Tenorio-Tagle} {et~al.}(1999){Tenorio-Tagle}, {Silich}, {Kunth},
  {Terlevich}, \& {Terlevich}}]{Tenorio-Tagle1999}
{Tenorio-Tagle}, G., {Silich}, S.~A., {Kunth}, D., {Terlevich}, E., \&
  {Terlevich}, R. 1999, \mnras, 309, 332

\bibitem[{{Thommes} \& {Meisenheimer}(2005)}]{Thommes2005}
{Thommes}, E., \& {Meisenheimer}, K. 2005, \aap, 430, 877

\bibitem[{{Thuan} \& {Izotov}(1997)}]{Thuan1997}
{Thuan}, T.~X., \& {Izotov}, Y.~I. 1997, \apj, 489, 623

\bibitem[{{Valls-Gabaud}(1993)}]{Valls-Gabaud1993}
{Valls-Gabaud}, D. 1993, \apj, 419, 7

\bibitem[{{Vanzella} {et~al.}(2010){Vanzella}, {Grazian}, {Hayes},
  {Pentericci}, {Schaerer}, {Dickinson}, {Cristiani}, {Giavalisco}, {Verhamme},
  {Nonino}, \& {Rosati}}]{Vanzella2010}
{Vanzella}, E., {Grazian}, A., {Hayes}, M., {et~al.} 2010, \aap, 513, A20

\bibitem[{{Verhamme} {et~al.}(2008){Verhamme}, {Schaerer}, {Atek}, \&
  {Tapken}}]{Verhamme2008}
{Verhamme}, A., {Schaerer}, D., {Atek}, H., \& {Tapken}, C. 2008, \aap, 491, 89

\bibitem[{{Wofford} {et~al.}(2013){Wofford}, {Leitherer}, \&
  {Salzer}}]{Wofford2013}
{Wofford}, A., {Leitherer}, C., \& {Salzer}, J. 2013, \apj, 765, 118

\bibitem[{{Yajima} {et~al.}(2012){Yajima}, {Li}, {Zhu}, {Abel}, {Gronwall}, \&
  {Ciardullo}}]{Yajima2012fesc}
{Yajima}, H., {Li}, Y., {Zhu}, Q., {et~al.} 2012, ArXiv e-prints,
  arXiv:1209.5842

\bibitem[{{Yamada} {et~al.}(2012){Yamada}, {Matsuda}, {Kousai}, {Hayashino},
  {Morimoto}, \& {Umemura}}]{Yamada2012}
{Yamada}, T., {Matsuda}, Y., {Kousai}, K., {et~al.} 2012, \apj, 751, 29

\bibitem[{{Yin} {et~al.}(2007){Yin}, {Liang}, {Hammer}, {Brinchmann}, {Zhang},
  {Deng}, \& {Flores}}]{Yin2007}
{Yin}, S.~Y., {Liang}, Y.~C., {Hammer}, F., {et~al.} 2007, \aap, 462, 535

\end{thebibliography}

\clearpage

\end{document}